# The Market Consequences of Perceived Strategic Generosity: An Empirical Examination of NFT Charity Fundraisers

Chen Liang[1], Murat Tunc[2], Gordon Burtch[3]

## Abstract

Crypto donations now represent a significant fraction of charitable giving worldwide. Nonfungible token (NFT) charity fundraisers, which involve the sale of NFTs of artistic works with the proceeds donated to philanthropic causes, have emerged as a novel development in this space. A unique aspect of NFT charity fundraisers is the significant potential for donors to reap financial gains from the rising value of purchased NFTs. Questions may arise about donors' motivations in these charity fundraisers, potentially resulting in a negative social image. NFT charity fundraisers thus offer a unique opportunity to understand the economic consequences of a donor's social image. We investigate these effects in the context of a large NFT charity fundraiser, leveraging random variation in transaction processing times on the blockchain to identify the causal effect of purchasing a charity NFT on a donor's later market outcomes. Further, we demonstrate a clear pattern of heterogeneity based on an individual's decision to relist (versus hold) the purchased charity NFT (a sign of perceived strategic generosity) and based on an individual's social exposure within the NFT marketplace. We show that charity-NFT 're-listers' experience significant market penalties, with an estimated 15.3% decrease in the prices they can command for other NFTs in their portfolio. This negative effect is particularly pronounced among those who are more socially exposed. Two controlled online experiments (one incentive-compatible and one scenario-based) corroborate our findings, demonstrating that the re-listing of a charity NFT for sale at a profit leads onlookers to perceive the initial donation as strategic generosity and reduces their willingness to purchase NFTs from the donor. Our study underscores the growing importance of digital visibility and traceability, features that characterize crypto-philanthropy and online philanthropy more broadly.

---

[1] University of Connecticut
[2] Tilburg University
[3] Boston University

## 1. Introduction

Blockchain technologies, particularly cryptocurrency, have had a disruptive influence on financial markets over the past 15 years. This disruption has recently extended to philanthropy (Tan and Tan 2022). Crypto donations now comprise a substantial fraction of charitable donations around the world, and many of the world's best-known charitable organizations now accept cryptocurrency donations. Within the broader crypto philanthropy space, charity fundraisers involving artistic work minted as non-fungible tokens, or NFTs (Kanellopoulos et al. 2025), have gained prominence.[4] Several NFT charity fundraisers have seen massive success over the last few years, organized by the likes of Bill Murray,[5] Sotheby's,[6] and Taco Bell,[7] and numerous online platforms have even emerged that cater specifically to this phenomenon, e.g., DoinGud[8] and DigitalArt4Climate.[9]

NFT charity fundraisers are typically organized to support a particular philanthropic organization or pursuit. Each fundraiser focuses on artwork that is commissioned or contributed for sale to 'donors,' such that the resulting proceeds can be provided to the beneficiary organization. A unique aspect of these fundraisers, and crypto philanthropy more broadly, is that donors have the potential to benefit financially from their ostensibly prosocial 'donations' (Tan and Tan 2022). For example, donors in NFT charity fundraisers may reap financial gains because both the purchased NFTs and their underlying cryptocurrencies can experience rapid increases in value. This dynamic implies that, although individuals may be motivated by altruism, they may also be driven by the prospect of personal gain. Thus, depending on the perceptions and inferences onlookers hold or draw, respectively, about an individual's motivations for purchasing an NFT, positive or negative consequences may arise for that individual's social image.

---

[4] https://cointelegraph.com/news/nft-philanthropy-demonstrates-new-ways-of-giving-back
[5] https://www.coindesk.com/business/2022/08/31/beer-with-bill-murray-nft-sells-for-185k-in-eth-at-charity-auction/
[6] https://cointelegraph.com/news/sotheby-s-metaverse-announces-latest-and-largest-nft-charity-auction
[7] https://charitydigital.org.uk/topics/topics/nfts-for-good-how-charities-can-fundraise-with-nfts-8899
[8] https://doingud.com/
[9] https://digitalart4climate.space/

Philanthropy involving NFT charity fundraisers provides a unique context for studying social image, distinct from traditional charitable giving. While charitable donations in some offline settings also allow donors to acquire an asset, e.g., a piece of art in a charity art auction, which may later be resold, NFT fundraisers differ in two important ways. First, NFTs may experience rapid and significant changes in value, creating the potential for donors to turn a quick profit. Second, the transparency of blockchain-based marketplaces means that a donor's decision to resell a charity NFT, and to attempt to do so at a profit, may be highly visible to third parties. Imbued with the ability to see when an NFT relisting has taken place, and aware of the potential for donors to engage in opportunistic behavior by flipping a charity NFT at a profit, onlookers may infer strategic behavior on the part of a re-listing donor and respond negatively, reducing their willingness to transact with the donor.

Thus, we suggest here that the combination of traceability and speed that characterizes NFT markets has the potential to amplify the role of reputation and perceived motivations ascribed to donors in NFT charity fundraisers. Onlookers are not only aware of the possibility that donors may hold profit-seeking motives, but can also observe direct signs that donors have actually taken steps to execute on those motives, may heighten the reputational stakes of participating in NFT charity auctions. This is doubly important in NFT markets because the highly subjective and volatile valuation of NFTs means that social perceptions of donors' motivations—whether philanthropic or opportunistic—have significant potential to influence onlookers' willingness to pay for an NFT, potentially affecting market outcomes for NFT donors in a meaningful way.

Extensive research in economics and psychology has documented evidence that social image concerns shape a variety of behaviors (Bursztyn and Jensen 2017), including numerous prosocial activities, such as volunteering (Carpenter and Myers 2010), blood donation (Lacetera and Macis 2010), and charitable contributions (Bénabou and Tirole 2006; Kafashan et al. 2014; Simpson and Willer 2015). Research argues and shows that individuals are more likely to engage in prosocial behavior when it improves their social image (Lacetera and Macis 2010). Conversely, work also shows that individuals are less likely to engage in prosocial behavior when potential benefits are absent or undermined by the publicly observable presence of extrinsic incentives, e.g., financial compensation (Bénabou and Tirole 2006).

Though a great deal of research has explored how image concerns affect prosocial behavior, little work has examined the perspective and perception of onlookers (Berman and Silver 2022); that is, whether, when, and to what extent onlookers' perceptions that an individual's prosocial behavior reflects generosity versus the mere appearance of generosity translate to economic benefits or penalties, respectively. We adopt a multimethod approach to address that gap. Drawing on qualitative insights from an expert interview with a veteran NFT collector/donor and integrating these with scholarly literature, we identify three unique features of blockchain-based charity that create conditions for economic punishment of perceived strategic generosity: *the potential for personal gain under charitable guise*, *transparent transaction history as reputational signal*, and *market punishment for attributed strategic motives*. We then test our proposed mechanism through a combination of observational and experimental studies: an observational econometric study establishes external validity by documenting real-world market consequences, a preliminary incentive-compatible experiment clarifies underlying motives and behavioral premises, and a scenario-based online experiment confirms causality through randomized manipulation.

Our main observational study is situated in the context of a large NFT charity fundraiser, focusing on a fundraiser organized in support of Ukraine amid its military conflict with Russia. We collect and analyze a dataset spanning the 14 months surrounding the fundraiser, which took place on February 26, 2022, examining NFT-listing and sale activities associated with a group of individuals who attempted to purchase an NFT offered as part of the fundraiser. We contrast the market activities and outcomes of those who successfully obtained a charity NFT (our treated, 'generous' group) with those who failed (our control group), examining how market responses vary depending on successful acquirers' re-listing behavior. To obtain causal estimates, our research design exploits the fact that, conditional on bid timing and gas limit, variation in blockchain transaction processing times is plausibly exogenous, generating quasi-random variation in NFT purchase outcomes. Because the fundraiser offered only a limited number of NFTs for sale, these uncontrollable delays led some interested individuals to ultimately fail in their attempt to purchase an NFT, while others succeeded.

We first demonstrate that purchasing a charity NFT has a significant negative association with later market outcomes, on average. However, exploring heterogeneity in those effects depending on whether an individual's initial generosity is followed by behavior that may be perceived as revealing that the donation was in fact strategic, i.e., whether the donor held versus re-listed their charity NFT for sale, we demonstrate that the average effect is driven primarily by individuals who choose to re-list. That is, whereas individuals who held their charity NFTs actually experienced a small benefit in their market outcomes, on average, those who re-listed their NFT for sale experienced a systematic penalty in the prices they were able to command for the other NFTs in their portfolio (an estimated 15.3% decline). These penalties are systematically larger among re-listing donors whose behavior was observable to a larger audience, i.e., individuals who had more followers in the marketplace. Our results are robust to a variety of specifications and matching techniques, alternative estimation strategies (matrix completion and instrumental variable regression), and a set of placebo tests.

To assess the generalizability of our results, we collected data from two additional NFT collections (another NFT charity auction, as well as a non-charity NFT collection). We replicate our main findings in the second charity NFT collection but observe no such effects in the non-charity NFT collection, consistent with the idea that these penalties are tied, in particular, to the charitable nature of the NFT fundraiser.

Finally, to gain direct evidence for the hypothesized mechanism, we conducted two online experiments on Prolific (one incentive-compatible and one scenario-based), screening for participants with NFT experience. The preliminary incentive-compatible experiment simulated an actual NFT purchase decision and provides support for the notion that onlookers in NFT marketplaces actively scrutinize donor behavior and trade histories, enabling them to detect actions such as relisting charity NFTs. The main scenario-based experiment corroborates our observational findings, revealing that re-listing a charity NFT for sale at a profit reduces onlookers' interest in transacting with the holder of a recently purchased charity NFT. Moreover, the negative impact of re-listing on onlookers' willingness to transact with an NFT donor is mediated by onlookers' perception of strategic generosity associated with the re-listing behavior. Taken together, these insights enrich our narrative by illustrating how reputationally driven, market-level penalties can arise in response to perceived strategic motives in blockchain-based philanthropy.

Our study contributes to the literature on social image and pro-social contribution, as well as the literature on crypto philanthropy. Our work offers unique insights into the dynamics of social penalties and ostracism that arise in response to perceived strategic generosity. As recent work observes, field evidence for the role of social image in prosocial contributions is generally hard to come by, which has resulted in a heavy focus on laboratory experiments in the literature (Exley 2018). NFT charity fundraisers provide a research context that uniquely lends itself to the study of social image, in large part because all transactions, involved parties, and associated prices, are publicly observable, auditable, and traceable on the blockchain via smart contracts (Cong et al. 2023).

Our study also contributes to a broader literature on the role of social image concerns in prosocial behavior, particularly in the context of charitable giving and donation. Existing studies on the potential negative *social consequences* for individuals whose prosocial behavior is perceived as strategic generosity often rely on *surveys or lab experiments* (Critcher and Dunning 2011; Lin-Healy and Small 2013; Newman and Cain 2014). Our findings expand on this by empirically demonstrating the extent to which the negative social consequences that donors experience translate to negative *economic consequences in real markets*. Our work also contributes to the nascent literature on crypto philanthropy, exploring a unique social dynamic that can arise as a result of features enabled by NFTs and blockchain technology, namely the public visibility, traceability, and rapidity of transactions. We demonstrate that these affordances operationalize social monitoring of strategic generosity and enable collective sanctioning at scale, even within pseudonymous environments.

Finally, in terms of practical implications, our findings suggest that donors to NFT charity fundraisers may face positive or negative economic consequences in terms of the prices for their NFTs in subsequent transactions, depending on the motivational attributions that onlookers make. This finding has implications for the behavior and participation of potential donors and may have implications for the long-run success of NFT charity fundraisers. To the extent donors recognize and experience economic gains or penalties, their participation may rise or decline in the long run. Further, to the extent perceived strategic generosity is penalized in these markets, individuals may begin to behave systematically less strategically than in other, less transparent contexts. Our study also highlights the importance of considering the role of social image concerns in charitable giving, not only from a donor's standpoint, but also from the perspective of third-party onlookers. While donors may be motivated by both altruistic and strategic considerations, and concerns about third-party perceptions may influence their choices about whether and how much to contribute, our work demonstrates that *onlookers' actual perceptions* are also important to consider, and that donors and platforms should take care to manage those perceptions.

## 2. Literature Review

### 2.1. Image Concerns & Prosocial Behavior

The literature concerning the impact of social image on prosocial behavior offers a nuanced view of how individuals' actions are shaped by the prospect of social recognition and material rewards. The research by Lacetera and Macis (2010) sheds light on the intricacies of blood donation, highlighting how symbolic prizes can enhance donation activity by serving as markers of positive social recognition, especially when these acknowledgments are made public. Conversely, Carpenter and Myers (2010) highlight that the benefit of making incentives publicly observable appears to invert when those incentives are monetary in nature. Those authors leverage data on volunteer firefighters to show that, while monetary incentives do motivate greater participation among these firefighters, their efficacy diminishes when the incentives are made public, as the firefighters are aware that the mere presence of publicly observable financial incentives undermines the social image benefit that a volunteer obtains from volunteering, by shifting onlookers' perceptions of the volunteer's motives. Bénabou and Tirole (2006) provide a theoretical framework that formalizes the interplay between altruism, material incentives, transparency, and social image. Consistent with the noted empirical findings, their model predicts that material or reputational incentives can lead to a reduction, or even a complete displacement, of intrinsic prosocial motivations.

Recently, this literature has shifted toward understanding how third-party observers perceive and judge an individual's prosocial actions. For example, Berman et al. (2015) explore the consequences of bragging about one's prosocial activities. Those authors demonstrate that bragging can enhance an individual's perceived generosity if the highlighted activities were not previously known to the individual's audience. However, bragging can have a negative consequence if the highlighted activities were already common knowledge. Further, Bliege Bird et al. (2018) show that engaging in observable prosocial actions can help attract new transaction partners in a marketplace, due to reputational gains. That said, Berman and Silver (2022) observe, as part of a broader discussion, that prosocial acts may engender greater skepticism from onlookers in competitive contexts, if onlookers perceive that the ostensibly prosocial activities are in fact strategic, intended to foster trust or more favorable evaluations.

The findings and discussion highlighted recently in this literature are particularly relevant in blockchain-based philanthropy and charity NFT platforms. In these settings, charity and financial compensation are broadly observable to all participants in the marketplace, due to the transparent nature of the blockchain. Strategic displays of generosity are thus very likely to influence individuals' ability to transact in the marketplace (Bliege Bird et al. 2018). That said, these effects may also fail to manifest, as prior work notes the importance of prevailing norms in a marketplace or community (Berman and Silver 2022). To the extent that strategic behavior is common, e.g., reselling NFTs, it might be perceived as acceptable and thus uninformative about an individual's social character.

## 2.2. Crypto Philanthropy & NFT Charity Fundraisers

Cryptocurrencies have disrupted financial markets broadly (Catalini et al. 2022), and the market for philanthropy is no exception (Tan and Tan 2022). NFTs, or Non-Fungible Tokens, are digital assets stored on a blockchain, a decentralized and distributed digital ledger technology. Unlike traditional cryptocurrencies, NFTs are unique (hence the term 'non-fungible'), with each NFT tied to a particular asset, whether physical, e.g., real estate, or digital, e.g., artistic images, videos, or audio clips. The ownership of an NFT is recorded on the blockchain, which provides proof of ownership and ensures that the asset is unique and cannot be duplicated or altered. This makes NFTs attractive for creators of digital art, collectibles, and other unique digital items who want to protect the authenticity and value of their creations. NFTs are bought and sold on digital marketplaces, e.g., OpenSea, with the value of each NFT being determined by supply and demand. As with any market, the value of NFTs can fluctuate depending on a variety of factors such as popularity, rarity, and perceived value. While NFTs have been around for several years, they have gained significant attention and popularity of late due to several high-profile sales of NFT-based digital art, reaching millions of dollars.

A small literature in IS has begun to examine aspects of NFTs, and their implications for traditional markets. For example, Halaburda et al. (2022) conceptually discuss the nature of NFTs, the novel functionalities they afford, and new types of markets they may enable, while Kanellopoulos et al. (2025) examine the effect of the introduction of NFT collectibles on the demand for traditional collectibles in the context of basketball trading cards. However, by far, NFTs' greatest area of adoption to date has been in the realm of art (Kugler 2021). Many NFT markets have begun to host auction fundraisers to support charitable causes. Several dedicated platforms have even emerged that are tailored for this purpose. In these charity fundraisers, a limited supply of NFT artwork is first contributed (minted) by artists, and then sold, with all or a portion of the proceeds being allocated toward a philanthropic organization.

While there is a small prior literature on charity fundraisers (Carpenter et al. 2008; 2010; Elfenbein and McManus 2010; Leszczyc and Rothkopf 2010), addressing why people participate, the efficacy of different fundraiser formats, and showing that individuals are willing to pay more for the same item when their payment is linked to charity, the prior literature has focused little on the role of financial returns to donors, or on the perception of such returns by third parties. This is likely because donors' financial benefits, to the extent they do exist, are only likely to manifest over the long term, and are generally not visible to onlookers. NFT charity fundraisers are different, of course, because there is a significant potential for short-run financial gains, due to the highly speculative, volatile nature of NFT value (Zaucha and Agur 2022), and because the role of the blockchain makes resale and subsequent transactions highly visible to others in the market. Transparency and visibility can increase trust between donors and onlookers (Weiss and Obermeier, 2021) or expose self-interested giving and suspicious activities (Chao and Fisher 2022;

Tahmasbi and Fuchsberger 2022). In turn, this creates the potential for social image to significantly influence an ostensibly prosocial donor's ability to conduct future transactions in the marketplace.

Onlookers' perception that another's generosity is strategic implies an interpretation of giving behavior as motivated by self-interest rather than genuine altruism. That onlookers may hold such perceptions is supported by literature dealing with charitable donation and social psychology. Lin-Healy and Small (2013) note that when charitable behavior results in both personal and charitable benefits, onlookers may be suspicious of the actor's true motives. Critcher and Dunning (2011) observe that people often cynically reinterpret prosocial behavior as being motivated by self-interest, even in the face of evidence to the contrary. Newman and Cain (2014) argue that when donors' prosocial behaviors yield both personal and charitable benefits, onlookers may be suspicious of their true motives. They posit that tainted altruism — prosocial behavior that conveys ulterior motives — can be viewed more negatively than no altruism at all. These studies collectively support that publicly visible post-donation behavior (i.e., quick resales) may yield perceptions of strategic intent, as observers in the NFT market interpret the actions. However, prior research on the social consequences of perceived strategic generosity has predominantly relied on surveys or lab experiments; evidence from real-world, economically relevant contexts remains scant. Our research addresses this gap by documenting the impact of perceived strategic generosity on individuals' subsequent NFT sales performance, leveraging data from multiple charity NFT collections that involve large samples of actual economic participants.

## 3. Theoretical Development and Research Roadmap

The literature reviewed above establishes that image concerns shape prosocial behavior and that crypto philanthropy represents a novel and rapidly growing domain of charitable giving. However, existing research has yet to fully examine how the unique technological and social features of blockchain-based charity, particularly NFT fundraisers, create distinct conditions for reputational dynamics and market consequences. While prior work on image concerns and prosocial behavior has primarily relied on laboratory experiments or traditional charitable contexts, the NFT charity setting introduces novel elements:

the potential for financial gain from donated assets, radical transaction transparency, and pseudonymous yet socially interconnected market participants. To bridge this gap between established theory and an emerging empirical context, we conducted an expert interview to gain practitioner insights into the mechanisms that may link charity NFT behavior to market outcomes. These qualitative insights, combined with the scholarly literature, inform our identification of the distinctive features of blockchain-based charity and guide our multi-method empirical approach.

### 3.1 Expert Interview

To develop a contextualized understanding of how image concerns manifest in NFT charity settings, we conducted a semi-structured interview with a veteran NFT collector who has been active in NFT marketplaces for over five years. The interviewee is a confirmed donor to multiple charity NFT campaigns (including RELI3F UKR, https://rarible.com/reli3f-ukr) and operates an NFT-focused social media account with over 15,000 followers on X (formerly Twitter). These qualitative insights inform the theoretical development that follows and enrich our understanding of how onlookers perceive and respond to charity NFT owners' behavior in blockchain-based philanthropy.

The interview was structured around several core themes: the respondent's background, motivations for purchasing charity NFTs, decisions around holding versus reselling, the role of social image, the importance of transparency, community reactions to donor behaviors, and perceptions of arbitrage and strategic activity within NFT markets. Selected excerpts from this interview are presented below, with the full interview transcript provided in Appendix A.

During the conversation, our interviewee spoke at length about the mechanisms by which perceived strategic generosity on the part of other participants can influence their decisions to make purchases from those individuals. He explained that he systematically examines sellers' complete transaction histories, for multiple reasons. The interviewee explained that he does this *"[f]or a number of reasons. One is for safety. You don't want to buy from a scammer that stole NFTs from somebody else."* Beyond security concerns, *"provenance matters, so that means who has owned it before you. If that's a well-known person in the space, for example, you want to buy from them."*

Most relevant, the interviewee stated that he specifically looks for sellers who *"do well for the community and for the broader society; I would rather buy from them than from somebody else."* Accessing this information is relatively easy, due to the transparency of blockchain technology: *"It's very easy. The whole chain of transactions are always visible on marketplaces like OpenSea. You can immediately see who owns certain NFT and who bought it from who, who minted it first and then the various transactions."* When asked how he would view someone who had purchased and quickly resold a charity NFT for profit, the interviewee stated: *"This would affect my willingness [to buy from them] in the sense of negatively, yes."* He further explained that such behavior becomes part of lasting community knowledge, noting that charity NFT flippers *"are called out by the community also publicly as sellers... So the community generally views this negatively."* These quotes illustrate the chain of events underlying the mechanism: before purchasing, onlookers review transaction histories for safety, provenance, and ethical considerations; interpret resale decisions, especially those involving charity NFTs, as signals of underlying strategic motivations; and translate these perceptions into economic decisions.

### 3.2. Unique Features of Blockchain-Based Charity

Drawing on both the fresh qualitative insights from our interview, as well as the scholarly literature in crypto philanthropy, crowdfunding, and prosocial behavior, we identify three distinctive structural features that distinguish blockchain-enabled charity from both traditional charitable giving and conventional NFT markets. These features, summarized in Table 1, collectively shape user behavior and market dynamics in ways that create a distinctive context for studying prosocial behavior. First, the ability to resell charity NFTs creates potential for personal gain under the guise of charitable intent. Second, transparent on-chain transaction histories function as reputational signals in otherwise pseudonymous environments. Third, when market participants attribute strategic motives to ostensibly prosocial actors, they impose economic punishment in subsequent transactions. Together, these features establish the theoretical and empirical foundation for understanding how deviations from expected prosocial norms can trigger both reputational damage and measurable market-level consequences. Detailed discussion, supporting literature, interview evidence, and elaboration on each of these features are provided in Appendix B.

**Table 1. Unique Features of Blockchain-Based Charity**

| Unique Feature | Description | Mechanism | Theoretical Foundation | Interview Evidence | Supporting Literature |
|---|---|---|---|---|---|
| Personal Gain Under Charitable Guise | Charity NFTs enable public signaling of generosity while retaining financial upside through resale | Donors can appear altruistic while pursuing profit, creating conditions for *strategic generosity*, i.e., behavior imitating prosocial intent but ultimately self-interested. Relisting may shift attributed motives from genuine altruism to strategic self-interest. | ▪ *Impure altruism*: giving reflects both genuine concern and personal benefits (Andreoni 1989).<br>▪ *Tainted altruism*: prosocial behavior with self-benefit evaluated more negatively than inaction (Newman & Cain 2014).<br>▪ *Strategic generosity*: prosocial behavior performed to cultivate reputation (Gambetta & Przepiorka 2014). | ▪ "You're willing to pay up for being a good person."<br>▪ "Obviously, it signals that people are not really there for the cause, but basically for the financial gain." | Andreoni (1989); Gambetta & Przepiorka (2014); Newman & Cain (2014) |
| Transparent Transaction History as Reputational Signal | All ownership and trading activity is publicly recorded on-chain, enabling inspection of seller behavior | In pseudonymous markets lacking identity markers, transaction history serves as the primary trust signal. Participants use observable patterns (relisting, holding duration) to infer genuine versus strategic altruism. Individual behaviors aggregate into network structures where reputation shapes trading relationships. Effects persist over time. | ▪ *Social signaling*: contributor visibility signals credibility (Burtch et al. 2016).<br>▪ *Credibility assessment*: observers process observable actions to evaluate motivations (Liu et al. 2018).<br>▪ *Network formation*: traders form clusters based on shared interests and values (Nadini et al. 2021).<br>▪ *Reputation persistence*: charitable behavior serves as lasting quality signal (Elfenbein et al. 2012; Schamp et al. 2022). | ▪ Reviews wallets to avoid "scammers"; prefers "people that do well for the community."<br>▪ Observes "whether that person resells... somebody with purely financial motives, they might put it up for sale soon after." "Some people are really known for being flippers." | Burtch et al. (2015, 2016); Liu et al. (2018); Elfenbein et al. (2012); Lee et al. (2023); Hur et al. (2023); Nadini et al. (2021); Schamp et al. (2022) |
| Market Punishment for Attributed Strategic Motives | Perceived strategic generosity triggers real economic penalties in downstream transactions | When observers perceive ulterior motives, they impose market-level sanctions. Buyers sharply distinguish profiting from regular NFTs versus charity-linked assets. This extends lab-based findings to document real-world financial consequences. | ▪ *Social penalties*: mixed motives reduce trust and moral evaluations (Berman et al. 2015).<br>▪ *Withholding credit*: onlookers withhold charitable credit when self-interest is perceived (Lin-Healy & Small 2012).<br>▪ *CSR skepticism*: prosocial efforts backfire when profit motives are salient (Yoon et al. 2006; Forehand & Grier 2003; Bhattacharjee et al. 2017). | ▪ "Their reputation goes down and I would be less willing to engage with them later on."<br>▪ "I would look much less favorably on people that make money from charity NFTs... I don't mind people making money on regular NFTs, but I would mind it with charity NFTs." | Berman et al. (2015); Lin-Healy & Small (2012); Yoon et al. (2006); Forehand & Grier 2003; Bhattacharjee et al. (2017) |

*Note. CSR denotes corporate social responsibility.*

### 3.3. Research Roadmap: A Multi-Method Approach

Our theoretical framework proposes that relisting charity NFTs signals strategic generosity, which, due to blockchain transparency, becomes observable to other market participants who then impose economic punishment. Testing this framework presents empirical challenges: the decision to purchase charity NFTs is endogenous, and individuals who own charity NFTs may differ systematically from those who do not in ways that also affect their market outcomes. To address these challenges, we employ a multi-method research design combining observational data analysis with experimental methods, each offering distinct advantages in establishing causal evidence.

*Main Observational Study (Section 4).* Our main empirical analysis examines a large charity NFT campaign (RELI3F UKR) using NFT activity data from multiple marketplaces. To identify the causal effect of charity NFT purchase on subsequent market outcomes, we exploit quasi-random variation in blockchain transaction processing times among bids placed in the same minute for treatment assignment and conduct a difference-in-differences (DID) analysis. This strategy compares individuals who successfully purchased charity NFTs with those who attempted to purchase but failed due to exogenous processing delays. We then examine heterogeneity based on relisting behavior and social exposure to test our proposed mechanisms.

**Table 2. Research Roadmap: Multi-Method Evidence Strategy**

| Section | Study | Objective | Method |
|---|---|---|---|
| Section 4 | Main observational study | Establish causal effect of charity NFT purchase on market outcomes | DID analysis, exploiting random variation in blockchain transaction processing times; Secondary analyses examining relisting behavior, social exposure, and ruling out alternative explanations |
| Section 5.3 | Replication and falsification tests | Assess generalizability and confirm charity-specific effects | Replication with alternative charity NFT collection (SHAQ); Falsification with non-charity collection (Porsche) |
| Section 6.1 | Preliminary experiment | Establish behavioral premises underlying proposed mechanism | Incentive-compatible online experiment with real economic stakes; Open-ended qualitative responses analyzed via LLM-assisted coding |
| Section 6.2 | Main scenario-based experiment | Directly test causal mechanism linking relisting to economic punishment | Scenario-based experiment manipulating relisting behavior (higher price / lower price / no relist); Mediation analysis |

*Replication Study (Section 5.3).* We replicate our main findings using data from an additional charity NFT campaign (SHAQ) to demonstrate the generalizability of our results to charity NFTs, and contrast these findings with a non-charity NFT campaign (Porsche) to confirm that the observed effects are specific to the charity context.

*Experimental Studies (Sections 6.1 and 6.2)*. To directly test the psychological mechanism linking relisting behavior to economic punishment, we conduct two online experiments. The first is an incentive-compatible experiment with real economic stakes, providing direct evidence that portfolio inspection is common practice among NFT buyers and capturing participants' qualitative perceptions of charity NFT resale. Building on these premises, our second experiment manipulates relisting behavior and measures its effects on perceived strategic generosity and willingness to transact, providing direct evidence for the causal chain linking relisting to motive attribution to economic punishment.

This multi-method approach offers several advantages. Our interview evidence validates the core assumption that perceived strategic generosity results in market penalties, providing support for the proposed mechanism linking observable resale behavior to reputational and economic consequences. The observational studies provide evidence of real-world economic consequences with high external validity, while the experimental studies establish the psychological mechanisms with high internal validity. Together, they offer converging evidence for our theoretical framework. We now turn to the main observational study.

## 4. Research Context and Observational Study

### 4.1. Study Context and Data

Our study utilizes a dataset of NFT transactions linked to individuals[10] (unique wallets) who attempted to purchase NFTs offered as part of a charity fundraiser aimed at supporting Ukraine in its conflict with Russia. We collected this data via two APIs: the Etherscan API, which provides bid information, and the Rarible API, which provides data on individuals' other NFT-related activities across various NFT marketplaces.

[10] Users on NFT platforms create profiles and have these profiles linked to their wallet addresses. This setup offers a system of pseudonymity. While these profiles are not directly linked to real-world identities (preserving privacy), these digital identities accumulate reputations and social capital within the NFT community over time. As a result, NFT charities create a unique context where social image concerns can arise even in a pseudonymous environment.

We constructed a panel dataset spanning 14 months, from August 2021 to September 2022.[11] Our panel thus includes seven months before and after the date of the NFT charity fundraiser, February 26, 2022. Our sample includes 2,836 individuals who successfully purchased a charity NFT, and 2,856 individuals who tried and failed.[12] Table 3 provides detailed definitions alongside the descriptive statistics for each variable.

**Table 3. Descriptive Statistics of the Full Sample**

| Variable | Definition | Obs | Mean | SD | Min | Max |
|---|---|---|---|---|---|---|
| Treated | A dummy variable; =1 since the month when the individual successfully purchases a charity NFT and onward, and equals zero otherwise | 72,475 | 0.313 | 0.464 | 0.000 | 1.000 |
| LnNumSales | Number of NFTs the individual sold in that month (logged) | 72,475 | 1.344 | 1.532 | 0.000 | 8.809 |
| LnTotRev | The total revenue associated with NFTs sold by the individual in that month (logged) | 72,475 | 0.787 | 1.111 | 0.000 | 7.044 |
| LnTenure | The number of months since the inception of the individual's Ethereum wallet (logged) | 72,475 | 1.899 | 0.790 | 0.000 | 4.190 |
| LnNumListings | Number of NFTs the individual listed in that month (logged) | 72,475 | 1.491 | 1.769 | 0.000 | 10.174 |

*Note: We add the value of 1 to all continuous variables related to NFT listings and sales prior to performing the log transformation. When measuring individuals' behavior concerning NFT listings and sales, we exclude any activities pertaining to the charity NFTs under study.*

Our analysis focuses on the RELI3F UKR NFT collection,[13] which includes 200 units (each) of 37 NFTs, provided by 37 influential artists. All 7,400 copies of these NFTs were sold within 5 minutes (see Figure 1), and all proceeds from the initial sale of each NFT went to support humanitarian relief efforts in Ukraine. The individuals who were able to acquire one of these NFTs when they were first auctioned off, on February 26, 2022, serve as our treatment group.[14] Our control group consists of those individuals who attempted to acquire one of the NFTs, yet were unable to do so due to i) the limited number of units on offer, and ii) random variation in transaction processing times on the blockchain.[15] This aspect of our research design thus benefits from the unique transparency of the blockchain (Li et al. 2024), which documents all bids and purchase attempts, even those that fail to complete.

[11] Given that NFT sales are relatively low-frequency events, we selected a 7-month post-treatment observation period for our analysis. This extended timeframe allows us to capture the long-term impact of perceived strategic generosity and evaluate how the effects evolve over time. The results remain highly consistent even when using a shorter observation window (e.g., 3 months before and after the fundraiser).

[12] Some individuals may submit multiple bids and can be unsuccessful in some of their attempts, yet successful in others. Further, some individuals may initially fail to acquire an NFT, yet manage to obtain one later, via the secondary market, i.e., purchasing an NFT from an individual who obtained the NFT as part of the original fundraiser. We omit these individuals from our sample to ensure mutual exclusivity between those who did versus did not obtain an NFT as part of the charity fundraiser.

[13] https://rarible.com/reli3f-ukr

[14] The magnitude of contributions could be heterogeneous across donors in charity fundraisers. However, the charity fundraiser we study in this paper has a fixed rate of 0.05 ETH. Hence, every user donates the same amount to the charity, which ensures a degree of homogeneity in donations and allows us to focus on post-donation behavior.

To identify donors who may be perceived as strategic in their generosity, we explore heterogeneity in the re-listing behavior of NFT-acquiring individuals following the initial purchase. Relisting and reselling a charity NFT are publicly observable (see Figure 2) and may be taken as a sign by others that the individual's initial purchase was strategic and self-interested. With this idea in mind, we contrast subsequent market outcomes between individuals who held versus relisted their charity NFT.

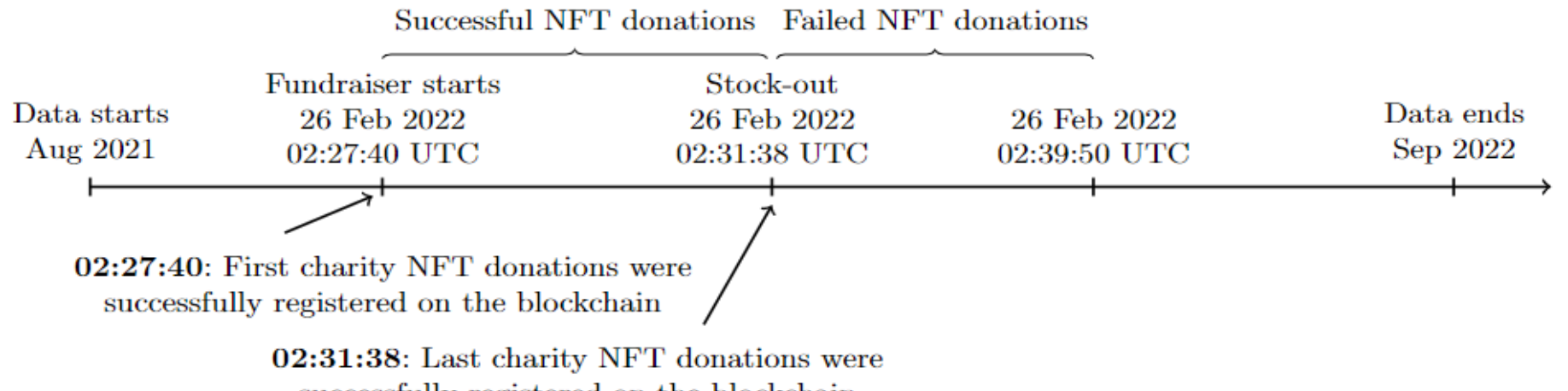

**Figure 1. Timeline of the RELI3F UKR NFT Charity Fundraiser**

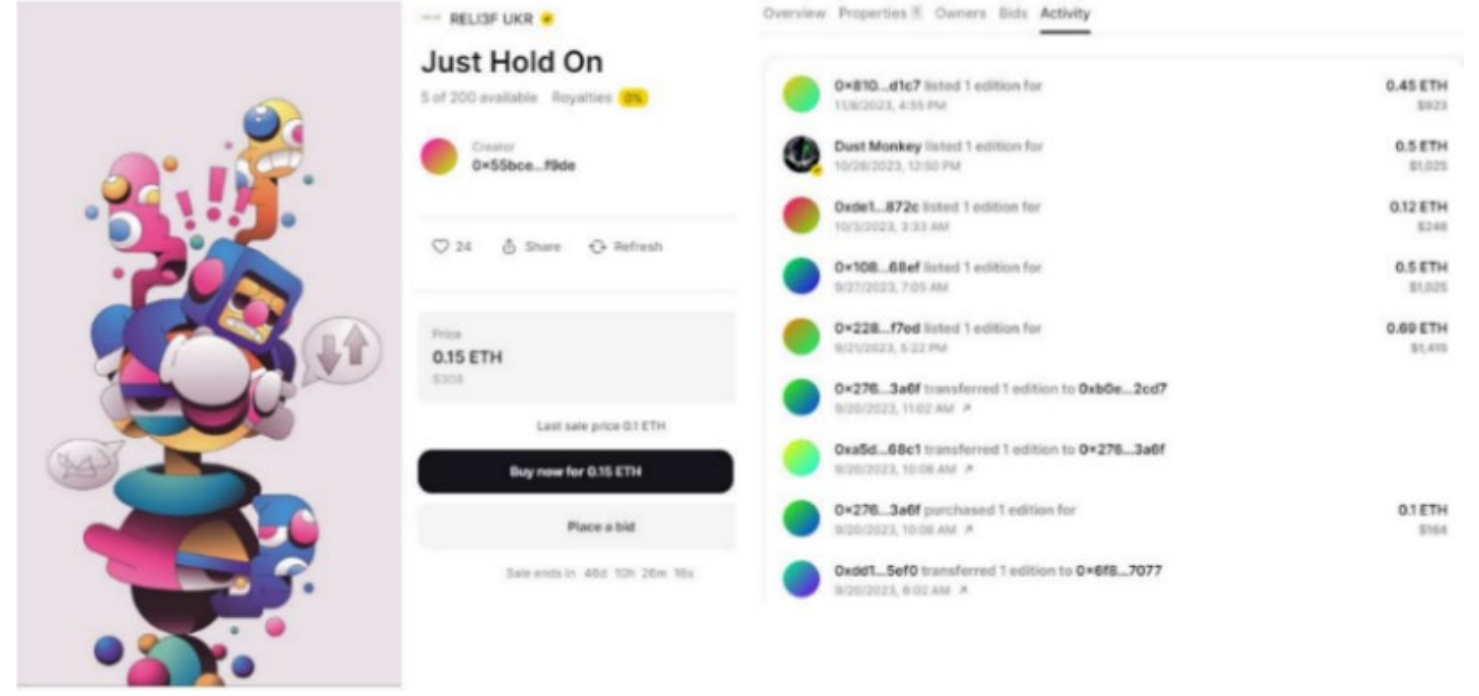

**Figure 2. Screenshot Illustrating a Sample Charity NFT from the RELI3F UKR Collection**

*Note. As shown in the Activity tab, all the previous transactions and the sale listings are publicly visible and traceable.*

### 4.2. Research Design

To establish the causal effect of purchasing a charity NFT on market outcomes related to the other NFTs in an individual's portfolio, we would ideally randomize individuals to purchase events. However, such an experiment is obviously infeasible for several reasons. We therefore employ an alternative research design, leveraging quasi-random variation in blockchain transaction processing times conditional on the timing of individuals' purchase attempts and their specified gas limit. Given that all charity NFTs were sold at the same price within 5 minutes of one another,[16] small variations in transaction processing time had a large impact on whether individuals' purchase attempts were completed successfully. Accordingly, we can divide individuals who attempted to purchase charity NFTs into two groups: one composed of individuals who successfully acquired the charity NFTs and one composed of individuals who attempted yet failed, due to a stock-out.[17] Leveraging these two groups, we estimate a DID design (Angrist and Pischke 2008; Burtch et al. 2018) to recover the effect of successful donation (charity NFT purchase) on the sales performance associated with other NFTs in an individual's portfolio. Subsequently, we estimate moderated DID specifications, to understand heterogeneity in the effects depending on whether an individual relists the charity NFT for sale, or holds it, as well as heterogeneity in other characteristics, e.g., the extent of an individual's social exposure in the marketplace.

[15] In our study, the population of interest is defined as individuals who attempted to donate to the focal NFT charity. This definition ensures that the population is similar in their willingness to donate to this charity. While it is possible that other individuals also browsed through the focal NFT charity campaign, they are not included in our sample because they did not attempt to donate.

#### 4.2.1. Matching

To further ensure comparability between the treated and control groups, we use coarsened exact matching, or CEM (Blackwell et al. 2009), constructing a matched sample based on the timing of purchase attempt, [18] the gas limit associated with the transaction (whether it is higher than, equal to, or lower than the modal gas limit),[19] and pre-treatment (pre-charity) NFT-related activities (i.e., purchases, sales, mints, listings,

---

[16] All the bids for charity NFTs were submitted within a five-minute window around the stock-out timestamp.

[17] Successful and unsuccessful bids are systematically different in their visibility on the NFT marketplaces. Successful bids for charity NFTs are visible on the NFT marketplace, while unsuccessful bids are only recorded on the blockchain ledger as failed transactions but are not displayed on the NFT marketplace interface. This significant difference in visibility on the NFT marketplace reduces the likelihood of unsuccessful bidders being perceived similarly to successful buyers. While it is plausible to navigate the blockchain and access unsuccessful bid information, given the computational complexity and the amount of effort it requires, this suggests that unsuccessful bidders are highly unlikely to be recognized by onlookers and, as a result, are not likely to be influenced by their failed purchase. This supports the validity of using these unsuccessful bidders as a control group.

[18] Bidding time refers to the time when the potential donors made a request to purchase the charity NFTs.

[19] The processing time for transactions on the Ethereum blockchain can be anywhere between 15 seconds and 5 minutes and depends on various factors (e.g., the gas limit, the congestion of the Ethereum network) (see more at https://legacy.ethgasstation.info/blog/ethereum-transaction-how-long/). To account for variations in users' willingness to pay higher gas fees, we match the treatment and control groups based on the gas limits users set. To account for network traffic, we match users based on the time their bids were submitted, as bids placed at similar times should experience comparable levels of

and transfers). After CEM matching, all the covariates are well-balanced between the treatment and control groups after matching (Appendix C). The CEM procedure yields a matched sample of 720 control individuals and 854 treatment individuals (1,574 total) with closely comparable characteristics.

**4.2.2. Econometric Specification**

Our primary regression specification is as per Equation 1:

$$Y_{it} = \alpha + \beta \cdot Treated_{it} + \theta \cdot X_{it} + \gamma_i + \delta_t + \varepsilon_{it} \quad (1)$$

The dependent variable, $Y_{it}$, represents the sales performance of other (non-charity) NFTs held by individual $i$ in month $t$,[20] including both the volume of NFT sales and total revenue (any reference henceforth to NFT sales performance refers to sales outcomes associated with these other, non-charity NFTs; the qualifier "non-charity" will be omitted for brevity and clarity). The independent variable of interest, $Treated_{it}$,[21] is a binary variable that is set to 1 in the month when individual $i$ successfully acquired a charity NFT, and each month subsequent, and is otherwise set to 0. Coefficient $\beta$ captures the impact of charity NFT acquisition on an individual's NFT sales performance. $X_{it}$ represents a vector of time-varying controls, including the tenure,[22] measured in months, of individual $i$'s Ethereum wallet, as well as individual $i$'s activities related to NFT listings and sales in the preceding month. Further, $\gamma_i$ denotes the individual-specific fixed effects and $\delta_t$ represents the month-specific fixed effects. $\varepsilon_{it}$ is an idiosyncratic error. When estimating our regressions, we cluster standard errors at the individual (wallet) level. It is worth noting that some of the dynamic controls may in fact be 'bad' controls (Angrist and Pischke 2008), as they may be impacted by the treatment. Accordingly, we mainly focus on estimations that omit the lagged controls; we consider estimations incorporating these controls only as a robustness check.

---

network congestion. In this way, although processing time is not entirely random across users, we can still leverage the remaining quasi-natural variation in processing time—factors beyond users' control—for identification.

[20] To ensure comparability between the treatment and control groups, we restrict our analysis to the sales performance of their NFTs, excluding those associated with the focal NFT charity fundraiser.

[21] Explicitly, $Treated_{it} = TreatGroup_i \cdot After_t$. Here, $TreatGroup_i$ denotes whether user $i$ belongs to the treatment group, while $After_t$ is equal to 1 after the start of the focal charity NFT fundraiser. Note that the main effects of $Treatment_group_i$ and $After_t$ are unidentified because they are inherently subsumed by the user-specific and time-specific fixed effects, respectively.

[22] The coefficient of tenure is identifiable because it is log-transformed and thus not subsumed by fixed effects. As shown below, our results are highly consistent, irrespective of tenure control.

Following our main estimations, we also report the results of a dynamic DID specification that simultaneously serves to i) evaluate the parallel trend assumption of the DID design, in the pre-treatment period, and ii) provide an indication of post-treatment effect dynamics. We achieve this by replacing the treatment indicator with dummies reflecting chronological distance in months from the time of charity NFT purchase (Autor 2003; Angrist and Pischke 2008; Burtch et al. 2018). Specifically, we replace the treatment dummy, $Treated_{it}$ in Equation (1) with a set of relative time dummies, yielding Equation 2:

$$Y_{it} = \alpha + \beta_1 \cdot Treated_{it}^{-7} + \beta_2 \cdot Treated_{it}^{-6} + \cdots + \beta_{13} \cdot Treated_{it}^{+6} + \beta_{14} \cdot Treated_{it}^{+7}$$

$$+\theta \cdot X_{it} + \gamma_i + \delta_t + \varepsilon_{it} \quad (2)$$

where $Treated_{it}^{-\pi}(Treated_{it}^{\pi})$ equals 1 for individual $i$ in the $\pi$-th month before (after) charity NFT purchase ($\pi \in [-7,7]$). Following prior literature (e.g., Autor 2003; Burtch et al. 2018), we designate the relative month (-1) as the omitted reference period, and the dummies are all coded as 0 for observations associated with control individuals. The coefficients that we estimate associated with pre-treatment dummies are not expected to differ from zero; were they to deviate from zero, this would reflect a violation of the parallel trend assumption. That said, we expect coefficients associated with post-treatment dummies to deviate from zero, as these dummies capture changes in the treatment group, post-treatment. The post-treatment relative time dummies thus allow us to investigate how the treatment effect evolves over time.

Following the dynamic specification, we consider heterogeneity in the treatment effect based on whether individuals relist their charity NFT for resale, which signals strategic generosity, versus holding it. We achieve this by estimating moderation models that interact treatment with relisting behavior and social exposure in the NFT community.

## 4.3. Empirical Results

### 4.3.1. Main Results: Average Treatment Effect

We begin by estimating the effect of charity NFT acquisition on individuals' NFT sales performance. Considering Table 4, we find that charity NFT purchase has no significant impact on an individual's volume of NFT sales. However, it does have a significant negative impact on an individual's total NFT revenue (approximately -6.29%, calculated as 1- exp(-0.065)). This finding suggests that charity NFT purchase adversely influences individuals' subsequent NFT sales performance, on average, lowering the price an individual is able to command in the NFT market. While this result is superficially surprising, in subsequent sections, we demonstrate that this average negative effect is driven by the adverse outcomes experienced by purchasers who eventually relisted the charity NFT for sale.

We visualize the dynamic difference-in-differences coefficients in Figure 3, with detailed results in Appendix C. As Figure 3 shows, all the pre-treatment relative time coefficients are not significantly different from zero, consistent with the parallel trend assumption. While we do observe a negative treatment effect on an individual's subsequent volume of NFT sales, the effect is transient. By contrast, the negative impact on total revenue persists, remaining statistically significant for at least 7 months post-treatment. This sustained price effect reflects market punishment: strategic generosity signals reduce willingness to pay for donors' other NFTs, an effect directly captured by declines in transaction prices rather than volume.

**Table 4. Average Treatment Effect**

| | LnNumSales<br>(1) | LnTotRev<br>(2) | LnNumSales<br>(3) | LnTotRev<br>(4) | LnNumSales<br>(5) | LnTotRev<br>(6) |
|---|---|---|---|---|---|---|
| Treated | -0.060 | -0.066** | -0.058 | -0.065** | -0.018 | -0.064** |
| | (0.050) | (0.027) | (0.047) | (0.025) | (0.051) | (0.032) |
| LnTenure | | | 0.478*** | 0.343*** | 0.215*** | 0.287*** |
| | | | (0.051) | (0.031) | (0.075) | (0.050) |
| LagLnNumSales | | | | | 0.193*** | 0.067*** |
| | | | | | (0.030) | (0.017) |
| LagLnNumListings | | | | | 0.136*** | 0.084*** |
| | | | | | (0.023) | (0.014) |
| Individual FE | Yes | Yes | Yes | Yes | Yes | Yes |
| Year-month FE | Yes | Yes | Yes | Yes | Yes | Yes |
| Observations | 16,780 | 16,780 | 16,780 | 16,780 | 15,206 | 15,206 |
| Num of Individuals | 1,574 | 1,574 | 1,574 | 1,574 | 1,574 | 1,574 |
| R-squared | 0.479 | 0.451 | 0.489 | 0.468 | 0.557 | 0.523 |

*Note: Robust standard errors clustered at the individual level. * p<0.1, ** p<0.05, *** p<0.01.*

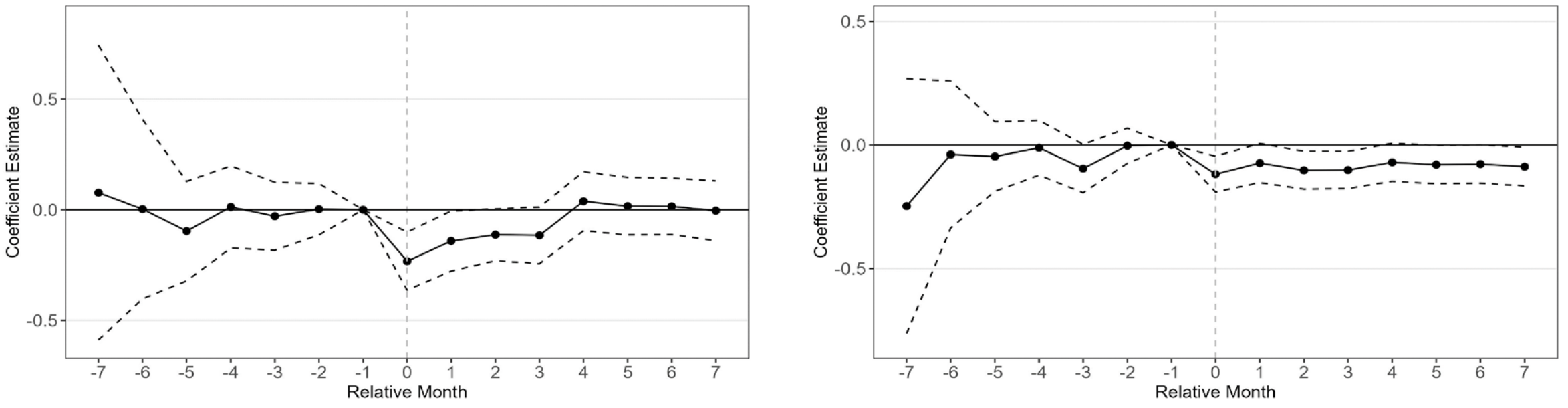

**Figure 3. Relative Time Estimates (Left = Sales Volume Effect; Right = Total Revenue Effect)**

*Notes: The dashed vertical line denotes the treatment time. Error bars represent 95% confidence intervals based on standard errors clustered by individual.*

#### 4.3.2. Perceptions of Strategic Generosity: Heterogeneous Treatment Effects

To gain a deeper understanding of why charity NFT purchase has a negative impact on individuals' subsequent NFT sales performance, on average, we next examine how the treatment effect varies depending on whether the purchasing individual seeks to resell or hold the charity NFT. We incorporate a moderator variable, $ListCharityNFT3Days_i$, denoting whether an individual lists the charity NFT for resale within three days following its acquisition. Our decision to focus on a 3-day window is driven by the fact that, among individuals who opted to list their charity NFTs for resale, the vast majority (93.2%) did so within three days. Note that we later show our results to be robust to this choice; our findings are consistent when employing an alternative measure, $ListCharityNFT_i$, reflecting whether an individual has ever listed the charity NFT for resale over the extent of our observation period.

As shown in Table 5, we find that individuals are only negatively affected by the charity NFT purchase if they also sought to resell it. Specifically, this results in a 15.3% decrease in their total NFT revenue, calculated as (1-exp(-0.205+0.039)). For reference, the median total revenue among untreated observations was 0.257 ETH, or approximately $714 USD at the time of the campaign. This implies a median monthly loss of $109 per individual who relisted UKR NFTs. The final two columns show that purchasing charity NFTs without resale attempts improves sales outcomes, consistent with social image concerns and market punishment of strategic generosity.

**Table 5. Heterogeneous Treatment Effects (HTE) by Intention to Resell the Charity NFTs**

| | LnNumSales<br>(1) | LnTotRev<br>(2) | LnNumSales<br>(3) | LnTotRev<br>(4) | LnNumSales<br>(5) | LnTotRev<br>(6) |
|---|---|---|---|---|---|---|
| Treated | 0.071 | 0.039 | 0.063 | 0.033 | 0.180*** | 0.079** |
| | (0.054) | (0.028) | (0.051) | (0.026) | (0.054) | (0.034) |
| Treated×ListCharityNFT3Days | -0.255*** | -0.205*** | -0.235*** | -0.191*** | -0.382*** | -0.276*** |
| | (0.061) | (0.035) | (0.059) | (0.034) | (0.062) | (0.042) |
| LnTenure | | | 0.473*** | 0.339*** | 0.198*** | 0.275*** |
| | | | (0.051) | (0.031) | (0.075) | (0.050) |
| LagLnNumSales | | | | | 0.193*** | 0.067*** |
| | | | | | (0.030) | (0.017) |
| LagLnNumListings | | | | | 0.136*** | 0.084*** |
| | | | | | (0.023) | (0.014) |
| Individual FE | Yes | Yes | Yes | Yes | Yes | Yes |
| Year-month FE | Yes | Yes | Yes | Yes | Yes | Yes |
| Observations | 16,780 | 16,780 | 16,780 | 16,780 | 15,206 | 15,206 |
| Num of Individuals | 1,574 | 1,574 | 1,574 | 1,574 | 1,574 | 1,574 |
| R-squared | 0.481 | 0.455 | 0.491 | 0.471 | 0.560 | 0.528 |

*Note: Robust standard errors clustered at the individual level. * p<0.1, ** p<0.05, *** p<0.01.*

Additionally, we find evidence that market punishment is more pronounced when individuals list charity NFTs at a higher price (above the original sale price). Detailed results are reported in Appendix D.

## 4.4. Secondary Analyses

Our main analysis establishes that purchasing charity NFTs negatively affects subsequent market outcomes, with effects concentrated among those who relist. We now present three secondary analyses that provide further evidence for the proposed mechanism linking relisting behavior to economic punishment through perceived strategic generosity. First, we disaggregate the treatment by replacing our moderated specification with a vector of treatment dummies reflecting individuals' progression through three distinct states: charity NFT purchase, charity NFT relisting, and charity NFT resale. This finer-grained analysis allows us to pinpoint where in the behavioral sequence the economic penalty emerges. Second, we consider variation in the impact of various treatment statuses depending on an individual's extent of social exposure. The logic of this test is that the negative effect of relisting a charity NFT should be stronger in the presence of greater public awareness if loss of social image truly drives the effect. To test this idea, we introduce another moderating factor, capturing the size of an individual's follower base. Third, by focusing on NFTs that individuals had already listed for sale prior to the treatment, we hold the supply side constant and ensure that any observed decline in sales performance reflects changes in buyer demand rather than shifts in individuals' listing behavior. To do so, we undertake two analyses: i) we repeat our estimations limiting our analysis to other NFTs that individuals had listed prior to the NFT charity fundraiser, which they had not modified, and ii) we re-estimate our models conditioning on the volume of NFTs an individual had listed in a given month, as well as associated listing prices. Fourth, and last, we report the results of a different estimation strategy, namely a listing-level survival analysis that demonstrates that perceived strategic generosity increases the time it takes for individuals' other listed NFTs to sell, consistent with reduced buyer demand rather than changes in listing behavior.

### 4.4.1. Progressive Treatment Status

We construct three time-varying treatment status indicators that capture individuals' behavior around a purchased charity NFT, specifically, whether they have procured the NFT ($TreatedNoCharityListing$), whether they have subsequently listed the charity NFT for resale ($TreatedCharityListingNoSold$), and whether they have successfully resold the charity NFT ($TreatedCharitySold$).[23] Taking our control individuals as reference, we investigate the degree to which effects on sales outcomes manifest under each status. Our expectation is that, if market punishment of perceived strategic generosity is truly driving our results, the negative effect on individuals' NFT sale prices should become more pronounced at the point of charity NFT relisting and resale.

Our results, reported in Table 6, support this expectation. The coefficient of $TreatedNoCharityListing$ is positive in most models and becomes statistically significant when controlling for individuals' listing and selling behavior from the preceding month. This suggests that charity NFT purchases, absent indications of perceived strategic generosity, have no adverse effects on purchasers' NFT sales performance. In contrast, the coefficients of $TreatedCharityListingNoSold$ and $TreatedCharitySold$ are significantly negative for total NFT revenue.

**Table 6. Progressive Treatment Status Effects**

| | LnNumSales | LnTotRev | LnNumSales | LnTotRev | LnNumSales | LnTotRev |
|---|---|---|---|---|---|---|
| | (1) | (2) | (3) | (4) | (5) | (6) |
| TreatedNoCharityListing | 0.072 | 0.039 | 0.063 | 0.033 | 0.189*** | 0.083** |
| | (0.053) | (0.028) | (0.050) | (0.026) | (0.053) | (0.033) |
| TreatedCharityListingNoSold | -0.162*** | -0.143*** | -0.156*** | -0.139*** | -0.184*** | -0.177*** |
| | (0.063) | (0.036) | (0.060) | (0.034) | (0.064) | (0.043) |
| TreatedCharitySold | -0.379* | -0.384*** | -0.290 | -0.321*** | -0.384* | -0.385*** |
| | (0.213) | (0.114) | (0.204) | (0.114) | (0.203) | (0.126) |
| LnTenure | | | 0.471*** | 0.336*** | 0.190** | 0.268*** |
| | | | (0.051) | (0.031) | (0.074) | (0.050) |
| LagLnNumSales | | | | | 0.192*** | 0.066*** |
| | | | | | (0.030) | (0.017) |
| LagLnNumListings | | | | | 0.136*** | 0.085*** |
| | | | | | (0.023) | (0.014) |
| Individual FE | Yes | Yes | Yes | Yes | Yes | Yes |
| Year-month FE | Yes | Yes | Yes | Yes | Yes | Yes |
| Observations | 16,780 | 16,780 | 16,780 | 16,780 | 15,206 | 15,206 |
| Num of Individuals | 1,574 | 1,574 | 1,574 | 1,574 | 1,574 | 1,574 |
| R-squared | 0.481 | 0.455 | 0.491 | 0.471 | 0.560 | 0.528 |

[23] There are a few cases in which individuals sold charity NFTs through accepting buyer-initiated bids before listing them for resale on the market. However, this behavior was observed in a minor fraction of the cases (precisely 5.88% of all individuals who sold their charity NFTs). Given this low frequency, for analytical simplicity, we designate both scenarios—whether individuals relisted and then resold the charity NFTs or resold them without relisting—as the treatment status $TreatedCharitySold$.

*Note: Robust standard errors clustered at the individual level.* * $p<0.1$, ** $p<0.05$, *** $p<0.01$.

#### 4.4.2. Social Exposure

We now investigate whether the effect varies with an individual's level of social exposure, i.e., awareness within the NFT community. Considering the size of an individual's follower base as a proxy for social exposure, we evaluate the extent to which the impacts of the three previously discussed treatment statuses vary with increases in an individual's social exposure. Detailed results are reported in Appendix E for brevity. We find that the negative effect of perceived strategic generosity is more pronounced for individuals who have a larger follower base. Results are consistent when we use another measure of social exposure, the popularity of the charity NFT, as indicated by the number of likes the NFT received. Consistently, the negative effect of perceived strategic generosity is more pronounced for those who relisted more popular charity NFTs. These participants, who tend to have higher pre-treatment revenue, face larger absolute losses, both in terms of reputation and financial damage (Appendix E).

#### 4.4.3. Ruling Out Individuals' NFT Listing Behavior as an Alternative Explanation

A possible alternative explanation for our results is that treated individuals may change their propensity to list non-charity NFTs for sale, or they may change their listing prices. To address this possibility, we perform two additional analyses. First, we repeat our estimations, focusing strictly on other NFTs that an individual had listed and not modified prior to the NFT charity fundraiser. We find that a negative treatment effect only arose for individuals who purchased a charity NFT and sought to resell it. Second, we re-estimate our regressions conditioning on the volume of NFTs an individual had listed in a given month, as well as their listing prices. Our findings suggest that the negative treatment effect persists, ruling out individuals' listing behavior as an explanation for our findings. Detailed results are provided in Appendix F for brevity.

Additionally, we employ an alternative estimation approach using a Cox proportional hazard model (Cox 1972) that models the hazard of other NFTs being sold as a function of strategic generosity signals. Results reported in Appendix G corroborate our main findings, showing that perceived strategic generosity reduces the hazard of sale—NFT listings take longer to sell for these individuals.

## 5. Robustness Checks for the Observational Study

Having established the effects of interest, we now implement an extensive set of robustness checks to support the validity of our findings. First, we employ the Matrix Completion (MC) estimator as an alternate causal inference strategy, imputing counterfactual outcomes for treated observations and subsequently recovering an average treatment effect on the treated (ATT). Second, instead of relying on the parallel trend assumption, which underpins the DID estimation, we pivot our identification strategy towards an alternative approach: an Instrumental Variable (IV) analysis (Appendix H). Third, to demonstrate generalizability and robustness, we replicate our analysis using data from a different charity NFT collection (Appendix I) and conduct a falsification analysis using a non-charity NFT collection (Appendix J). Fourth, to address potential spurious causality concerns, we conduct a placebo test featuring randomized treatment assignments. This helps gauge the likelihood of our results being driven by random disturbances rather than the primary treatment (i.e., the charity NFT purchase) (Appendix K). We conduct a series of additional robustness checks summarized in Table 7, all of which yield consistent results. Due to space constraints, detailed results appear in the online appendices.

**Table 7. Summary of Robustness Checks**

| **Robustness Check** | **Section** |
|---|---|
| *1) Analyses to alleviate concerns regarding potential violations of identification assumptions* | |
| Instrumental variable | Section 5.1; |
| Matrix completion | Section 5.2; Appendix H |
| *2) Analyses to demonstrate generalizability and safeguard against spurious correlations* | |
| Replication with a different charity NFT collection | Section 5.3; Appendix I |
| Falsification test with a non-charity NFT collection | Section 5.3; Appendix J |
| Placebo test | Appendix K |
| *3) Analyses to assess the robustness of our results to various samples and model specifications* | |
| Alternative dependent variable | Appendix L |
| Alternative samples | Appendix M |
| Alternative matching method | Appendix N |
| Alternative time windows | Appendix O |
| Excluding potential suspicious traders | Appendix P |
| Accounting for potential confounding effects of other charity NFTs | Appendix Q |
| Excluding individuals with insufficient balance for purchase | Appendix S |

## 5.1. Matrix Completion

We also explore robustness to an alternative estimator, namely the matrix completion (MC) method of Athey et al. (2021), which yields an ATT estimate. We implement this estimation on the CEM sample. The dynamic treatment effects are presented in Figure 4, where we see that all pre-treatment estimates are approximately zero. Post-treatment estimates, however, are once again pronounced and negative, particularly with respect to the sale price outcome.

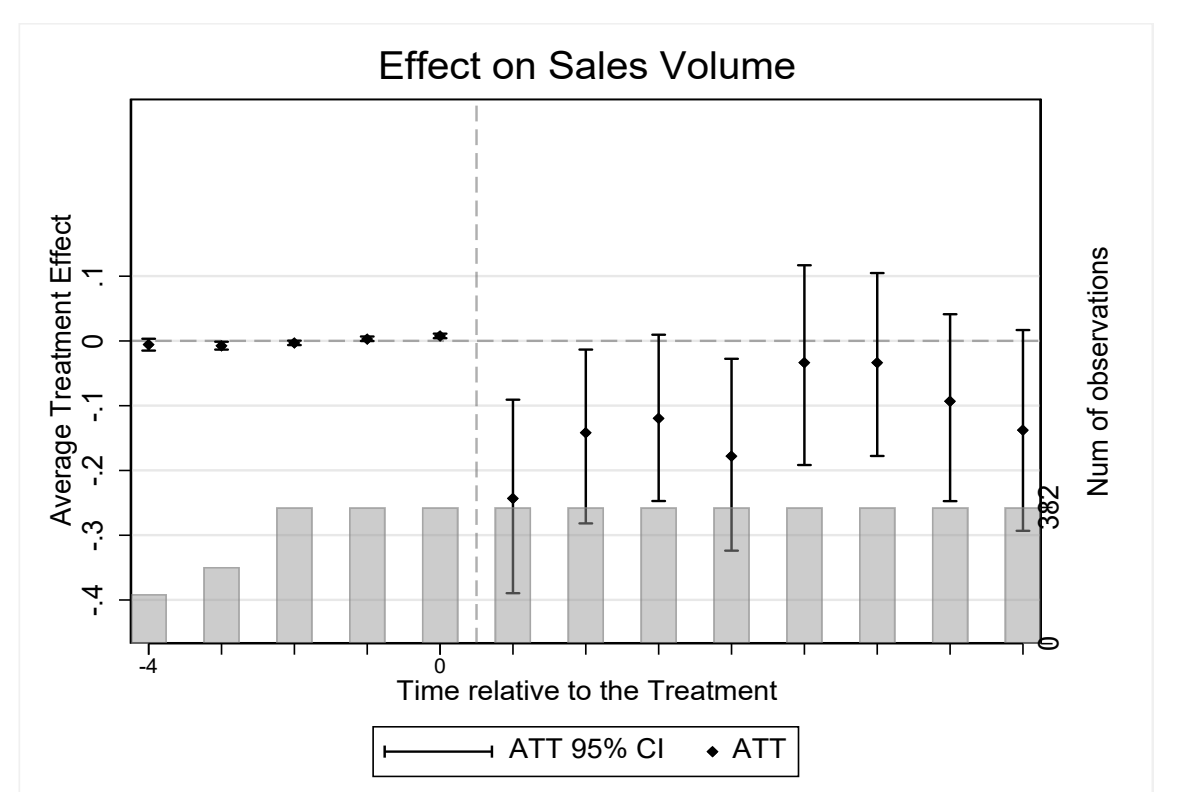

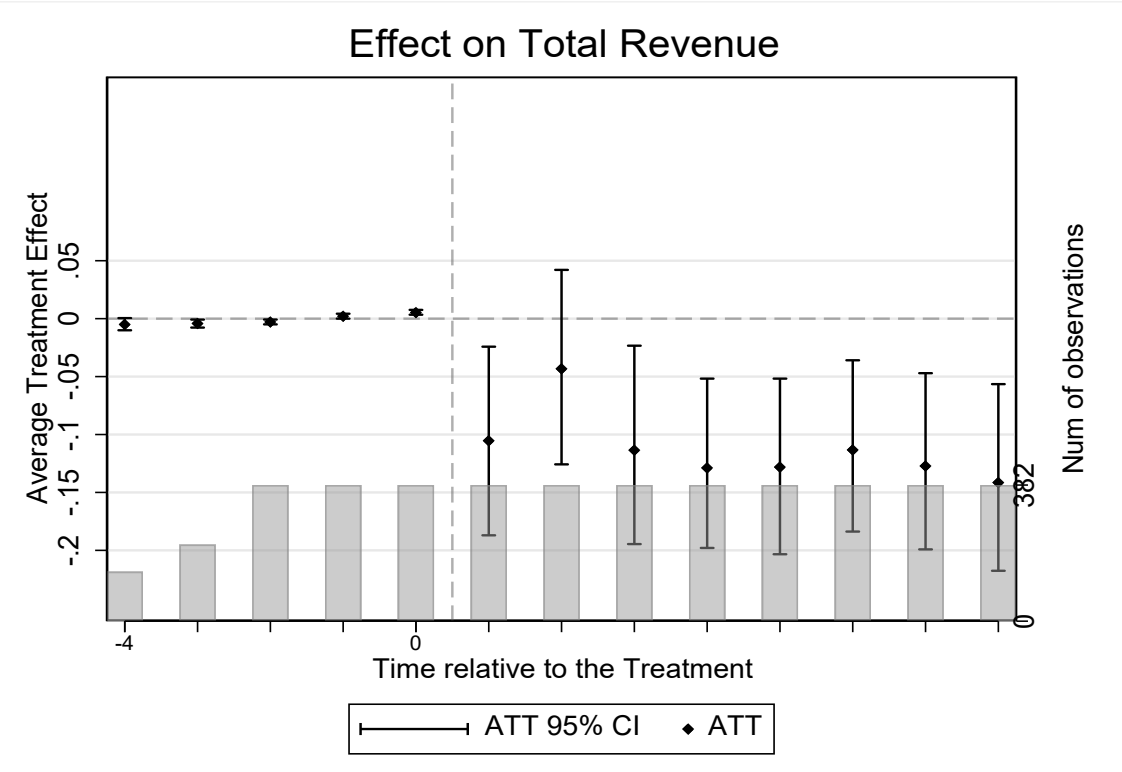

**Figure 4. Dynamic Treatment Effects Based on the MC Method**

Notes: Both plots presented were produced utilizing the "fect" package in Stata. Error bars denote bootstrapped 95% confidence intervals. Results are highly consistent when we include time-varying controls (i.e., tenure in months, lagged terms of individual activities related to NFT listings and sales) or when the full sample is considered.

## 5.2. Instrumental Variable

We can also exploit the random variation in transaction processing times to employ a different estimation approach, namely IV regression. In our matched sample, individuals in both the treatment and control groups submitted their purchase requests at a similar time and set the gas limit at the same level. However, the small differences in the processing time of these similar bids, which are beyond the individuals' control, may play a crucial role in determining whether their bids for charity NFTs were confirmed before such NFTs sold out. The processing time can serve as a valid IV for two important reasons. First, the processing time is highly related to individuals' treatment status, as it plays a critical role in determining whether an individual successfully purchases the charity NFT, satisfying the *relevance* requirement.[24] Second, when using control individuals who are similar in various aspects, such as bid submission time, gas fee limits, pre-treatment NFT activity, and Ethereum experience, as the counterfactual, the processing time of the charity NFT bid would only influence future sales of other NFTs by determining whether the individual successfully obtained the charity NFT. This meets the *exclusion restriction* condition, as the processing time itself has no direct effect on the individuals' future sales of other NFTs. Thus, we use the log-transformed processing time as an instrument for the treatment group dummy and use its interaction with the *After* dummy (which is set to 1 after the NFT charity fundraiser) as an instrument for the *Treated* dummy (Wooldridge 2010, *p.*154) to estimate the impact of successfully purchasing a charity NFT on the sale prices of other NFTs in the individuals' portfolios.

[24] While processing time is related to treatment assignment, it may not be perfectly correlated. There could be cases where users with faster processing times still fail to secure a charity NFT due to other factors, e.g., bid submission time, supply of NFTs, etc.

Detailed results in Appendix H confirm that longer processing times reduce NFT acquisition likelihood, as expected. The Kleibergen-Paap F statistic exceeds Stock-Yogo critical values (Stock and Yogo 2005), indicating instrument strength. Second-stage results replicate our main findings: treatment significantly reduces revenue from the sale of other NFTs.

### 5.3. Replication and Falsification Tests

To assess the generalizability and boundary conditions of our findings, we conducted two additional analyses using additional archival datasets. Specifically, we conducted (1) a replication study based on the *SHAQ Gives Back* charity NFT collection, and (2) a falsification test, repeating our analysis in relation to the *Porsche NFT* collection, a luxury-branded NFT series that had no charitable aspect.

The *SHAQ Gives Back* collection serves as a conceptually relevant replication. This fundraiser, organized by the Shaquille O'Neal Foundation, aims to support underprivileged youth across the United States by providing essential resources, including toys, clothing, and meals. The Foundation's mission emphasizes creating pathways for underserved youth to reach their full potential. This replication analysis enables us to investigate whether our primary results extend beyond crisis-driven or war-related fundraisers to more routine and diverse forms of altruistic fundraising. Full details are provided in Appendix I. Consistent with our main case (UKR NFTs), we observe negative downstream market performance effects for donors who engage in rapid resale of purchased charity NFTs from the SHAQ collection. These results suggest that the market punishment tied to perceived strategic generosity generalizes across different charitable domains and donor types. In contrast, our falsification test, conducted using the *Porsche NFT* collection, a high-profile non-charity project, resulted in no evidence that quick resale behavior had any negative effect on the market performance of NFT purchasers (details are reported in Appendix J). This null result is expected and reinforces the boundary conditions of our findings: the market penalties we observe appear specific to charity NFTs, where altruistic motives are normatively expected.

Collectively, our various analyses tell a consistent story. The replication and falsification tests help delineate the boundary conditions for our findings and bolster the external validity of the proposed mechanism within common charity NFT contexts. In the next section, we offer complementary insights into the underlying perceptions held by actual NFT participants through two experiments that establish the causal impact of reselling charity NFTs on subsequent market punishment.

### 5.4. Other Robustness Checks

We conduct a battery of additional robustness checks to ensure the validity of our causal inferences and to eliminate the possibility of spurious relationships. First, in addition to analyzing total sales volume and total revenue, we examine whether our findings hold when using an alternative dependent variable: the average transaction price of NFTs sold in a month (Appendix L). Second, to assess whether our findings are robust to alternative sample configurations, we employ both the DID estimation and the relative time model across two distinct samples: the full sample and a narrowly defined matched sample concentrating on individuals who attempted to purchase charity NFTs within a tight three-minute bandwidth surrounding the stock-out timestamp (Appendix M). Third, as a pivot from the CEM approach, we adopt the PSM method to match individuals in the treatment and control groups (Appendix N). Fourth, we consider alternative time windows for our estimations (Appendix O). Fifth, to address concerns about potential bot activity, we reanalyze our data after excluding traders who exhibit abnormally high NFT transfers prior to the treatment (Appendix P). Sixth, to address potential confounding effects from other charity NFT transactions, we either focus our analysis on non-charity NFT sales or exclude individuals who have previously owned other charity NFTs (Appendix Q). Finally, we relax our sample restrictions by including individuals regardless of whether they had sufficient balance to purchase charity NFTs (Appendix R). Across all robustness tests, individuals who purchased and relisted charity NFTs consistently experience reduced NFT sales performance. These findings collectively support market penalization of perceived strategic generosity.

## 6. Mechanism Exploration

To directly test the psychological mechanism linking relisting behavior to economic punishment, we conduct two online experiments. We first conduct a preliminary incentive-compatible experiment with real economic stakes, establishing that portfolio inspection is common practice among NFT buyers and capturing participants' qualitative perceptions of charity NFT resale. Building on these foundational insights, our main experiment manipulates relisting behavior and measures its effects on perceived strategic generosity and willingness to transact.

### 6.1. Preliminary Experiment: Portfolio Inspection Behavior

To validate that portfolio inspection is common practice among NFT buyers and to understand how onlookers perceive charity NFT relisting behavior, we conduct an incentive-compatible online experiment on Prolific (https://www.prolific.com/). As part of the design, participants were informed that they would be required to select one NFT for potential purchase at the end of the experiment, introducing real economic stakes. They were also informed that, due to budget constraints, 20% would be randomly selected to receive their chosen NFT, which would be purchased with actual funds and transferred to them. At the end of the experiment, selected participants were informed that, due to institutional restrictions on research spending, we could not make the purchase directly. Instead, they were invited to buy the NFT themselves and were offered a $5 Amazon gift card via email to offset the cost. To assess inspection behavior, participants were shown a (non-charity) NFT and could either proceed directly to purchase or first view the seller's complete portfolio and trading history. We find that *72.2%* of participants chose to examine the seller's portfolio, providing direct evidence that portfolio inspection is a common and deliberate practice among NFT buyers. We also analyze participants' responses to an open-ended question about their perceptions of charity NFT relisting behavior. Following prior studies (Brynjolfsson et al., 2025), we use large language model (LLM)–assisted qualitative coding to analyze these responses. Five primary topic clusters emerge: *ethical concerns* (60%), *greed and self-interest* (56%), *opportunism* (54%), *disappointment and distrust* (51%), and *market dynamics and acceptance* (19%), validating our theoretical argument (see Appendix S for details).

## 6.2. Main Experiment: Relisting Behavior and Perceived Strategic Generosity

To further investigate the mechanism underlying the negative impact on the sales performance of other NFTs in treated individuals' portfolios, we seek more direct measures of perceptions and responses among onlookers. We conduct a scenario-based online experiment on Prolific. This controlled experiment aims to quantify the influence of observing individuals who purchase charity NFTs and subsequently relist them for sale. We aim to understand how this behavior affects participants' perceptions of charity NFT owners and any changes in participants' willingness to purchase NFTs from these owners.

### 6.2.1. Experiment Design and Data Collection

To ensure participants possess prior NFT trading experience, we restrict recruitment to individuals residing in the United States who have previously minted or owned an NFT. As summarized in Figure 5, the experimental process consists of five steps. In Step 1, participants were prompted to imagine encountering a non-charity NFT of interest to them. In Step 2, they were asked to explore other NFTs in the seller's portfolio, where they discovered a SHAQ GIVES BACK charity NFT[25] that the seller had recently purchased as part of a charity offering. Subsequently, participants were randomly assigned to an experimental condition, one of: (a) the *HighRelistingPrice* group, where the seller relists the charity NFT at a price higher than its original purchase price; (b) the *LowRelistingPrice* group, where the seller relists the charity NFT at a price lower than its original purchase price; or (c) the *NoRelisting* group, where the seller does not relist the charity NFT for sale. Details regarding the procedure and design of our experiment are provided in Appendix T.

To assess participants' willingness to engage in transactions with the charity NFT owner, a follow-up scenario was introduced in Step 3. Participants were instructed to imagine browsing the collection page and discovering another copy of the same (non-charity) NFT that initially caught their interest, held by another seller. Participants were then asked to indicate their preference: whether they would prefer to purchase from the original owner, who previously owned a charity NFT and, depending on their assigned condition, may have relisted it for sale, or from the newly discovered, alternative owner. The dependent variable capturing a reduced willingness to engage in transactions with the focal owner is denoted as *PreferAlternativeSeller*.

---

[25] The SHAQ NFT collection, available at https://rarible.com/shaq-gives-back/items, is a charity NFT initiative unrelated to war. The collection description explicitly states that "100% of the proceeds from the SHAQ GIVES BACK NFT sale & ALL REVENUE from OpenSea creator fees will go to The Shaquille O'Neal Foundation." This foundation actively supports underprivileged youth across the nation by providing essential resources such as toys, clothing, and meals. Its mission emphasizes creating pathways for underserved youth to realize their full potential. As detailed in Appendix I, we will conduct an observational study based on this charity collection to demonstrate the generalizability of our findings.

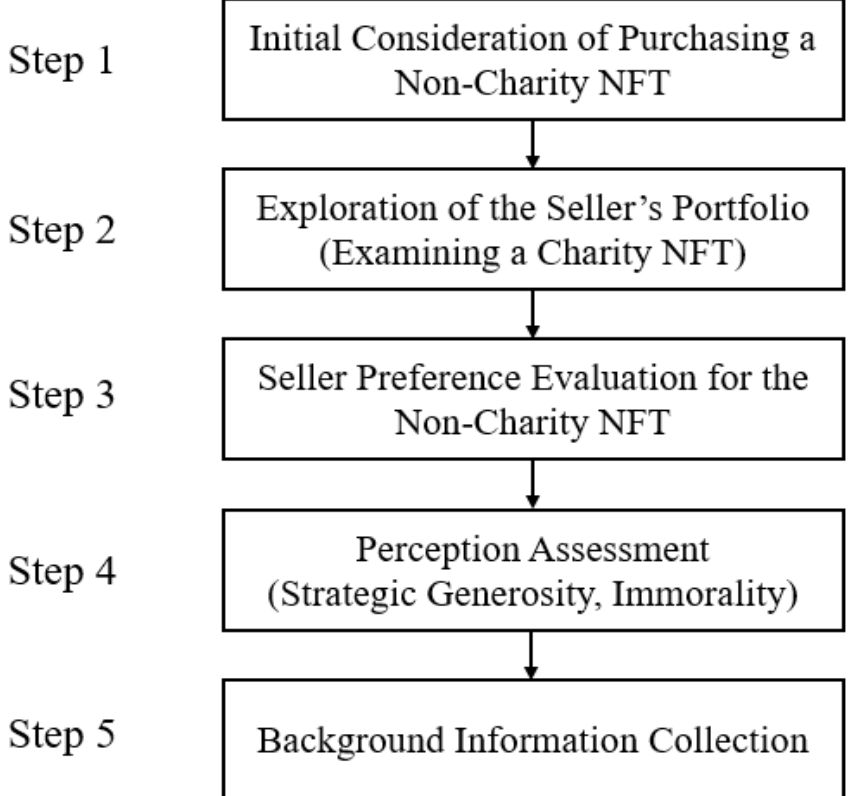

**Figure 5. Overview of Experimental Design**

In Step 4, participants were asked to evaluate their perceptions of how generous the owner appeared, based on their acquiring charity NFTs under the assigned condition. Participants assessed whether the acquisition was perceived as being motivated by strategic generosity (i.e., self-interest) versus altruism and evaluated the immorality of relisting charity NFTs (Newman and Cain 2014). Specifically, they rated the extent to which they perceived the act of relisting a charity NFT as unethical, as immoral, and whether they disapproved of such behavior (Cronbach's alpha = 0.858). Finally, in Step 5, participants completed demographic questions and responded to a hypothetical scenario about relisting a charity NFT for profit, assessing their behavioral intentions.

### 6.2.2. Experiment Results

We obtained complete responses from 265 individuals. A randomization check (Appendix U) confirms the validity of randomization. Based on our experimental data, we find that observing an owner relisting a charity NFT at a price higher than its original purchase price significantly increases participants' preference for purchasing the (non-charity) NFT from an alternative seller. Consistent with our earlier findings, our results also indicate that owners who relist a charity NFT at a higher price are more likely to be perceived as being strategic in their generosity, purchasing charity NFTs with opportunistic motives, compared to those who relist the charity NFT at a lower price or those who do not relist the charity NFT at all. These findings are visualized in Figure 5.

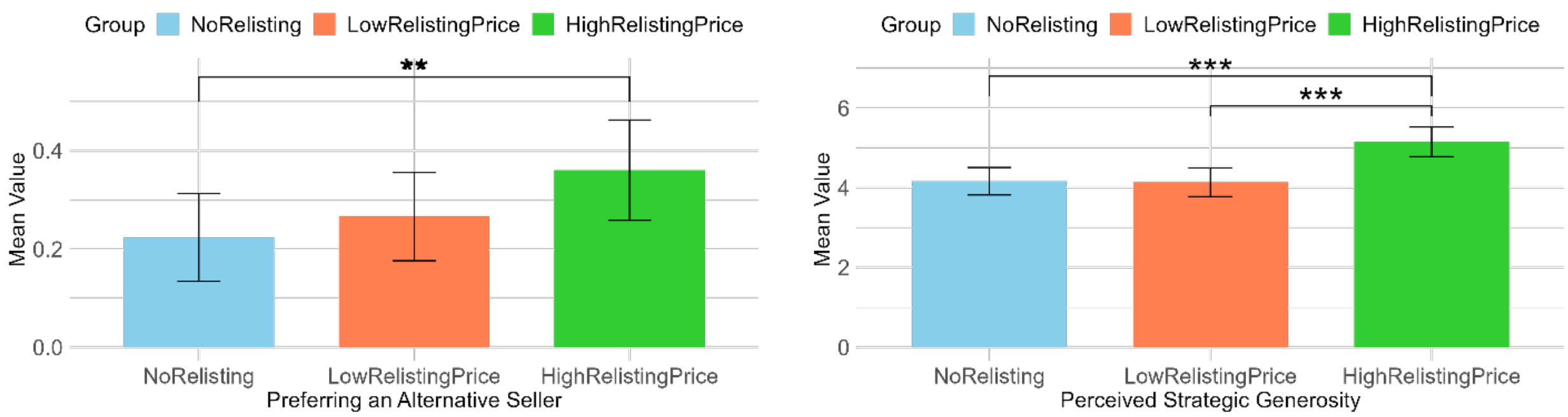

**Figure 5. Treatment Effects on Seller Preference and Perceived Strategic Generosity**
*Note: Error bars denote 95% confidence intervals. Significance stars are shown only for pairwise comparisons that are statistically significant.* According to the left side of Figure 5, we find that the probability of purchase decreases by 17.5% (calculated as $1 - (1 - 0.36)/(1 - 0.224)$) when donors relist charity NFTs above their original donation price, quantifying the market's response to perceived strategic generosity. * *p<0.1,* ** *p<0.05,* *** *p<0.01*

To further validate our proposed mechanism, we conduct a mediation analysis, with the results presented in Table 8. We find that, conditional on differences in perceived levels of strategic generosity, relisting a charity NFT at a higher price no longer has a significant impact on participants' preference for an alternative seller. This finding suggests that perceived strategic generosity is a key mediator for the negative effects of relisting a charity NFT at a high price on participants' reduced willingness to trade with the focal seller, supporting our claim of market punishment for such behavior.

**Table 8. Mediation Role of Perceived Strategic Generosity**

| | PreferAlternativeSeller | PerceivedStrategicGenerosity | PreferAlternativeSeller |
|---|---|---|---|
| | (1) | (2) | (3) |
| LowRelistingPrice | 0.042 | -0.026 | 0.044 |
| | (0.067) | (0.258) | (0.066) |
| HighRelistingPrice | $0.137^{**}$ | $0.986^{***}$ | 0.085 |
| | (0.069) | (0.264) | (0.069) |
| PerceivedStrategicGenerosity | | | $0.052^{***}$ |
| | | | (0.016) |
| Observations | 265 | 265 | 265 |
| R-squared | 0.016 | 0.069 | 0.055 |

*Note: The NoRelisting group is the baseline group. * p<0.1, ** p<0.05, *** p<0.01.*

Additionally, our data reveal that, relisting a charity NFT at a higher price, such that sale would result in a profit, significantly heightens participants' sense that the behavior is immoral ($p$-value $<0.05$), increases the likelihood they disapprove of the relisting behavior ($p$-value $<0.05$), and raises the proportion of participants who would intend to avoid purchasing NFTs from the owner ($p$-value $<0.10$). Detailed results can be found in Appendix V. Overall, this supplementary experiment corroborates our observational analyses, highlighting that the perceived strategic generosity of relisting a charity NFT for sale at a profit serves as a key mechanism underlying individuals' reluctance to purchase NFTs from the donor.

## 7. Discussion and Conclusion

### 7.1. Summary and Discussion

This paper investigates the emerging trend of crypto donations and NFT charity fundraisers, which have become a significant source of charitable donations worldwide. While the value of charity NFTs can increase rapidly, some purchasers may be seen as genuinely altruistic, while others may be viewed as motivated by the potential for quick financial gains. Our study examines how an individual's social image may affect their subsequent NFT market activity and lead to economic penalties. Leveraging a setting focused on a large NFT charity fundraiser, we find that charity NFT donors experience a significant 6.29% decline in the sale prices of other NFTs in their portfolio. This effect is driven by donors who eventually list their charity NFTs for resale, soon after purchase. By contrast, individuals who do not relist their charity NFTs are found to be positively impacted, consistent with the idea that market benefits and penalties are driven by motivational attributions made by onlookers.

### 7.2. Managerial Implications and Theoretical Contributions

Our work offers valuable implications for managers and platforms facilitating charity NFT fundraisers. NFTs offer a quick and scalable solution to charitable giving. The charity auction examined in our study provides a concrete example of this, raising more than $3 million within just five minutes. Moreover, the NFT charity campaign was launched within two days after the military conflict started. This rapid mobilization of funds through NFTs represents a new paradigm in charitable giving.

First, platform structures that highlight historical giving behavior and downplay short-term resale activity could discourage norms of strategic generosity. Our findings indicate that market punishment is more severe for sellers with greater social exposure, highlighting the crucial role platform design plays in shaping reputational outcomes. For instance, reputation badges or visible donor streaks, akin to GitHub contribution graphs or StackOverflow reputations, might reward consistency over time rather than one-off high-profile donations. Platforms could incorporate purposeful sources of friction, e.g., tiered resale lock-ins or time-based visibility bonuses, that make rapid flipping less socially desirable or less visible to peers.

Second, fundraisers operating in NFT ecosystems may consider co-designing campaigns with community leaders to ensure that philanthropic messaging resonates with local norms and expectations. Incorporating community governance features, e.g., Decentralized Autonomous Organization (DAO) voting on fund allocation, community moderation panels, and transparent proposal forums, can deepen engagement and elevate social accountability, particularly in communities where mutual monitoring and peer sanctioning are prevalent.

Finally, we suggest improving mechanisms that signal a donor's willingness to *"pay up for being a good person,"* as our interviewee put it. High-profile donors, in particular, might benefit from narrative signaling strategies that emphasize long-term commitment (e.g., time-locked NFTs, episodic giving narratives, or future donation pledges). Our findings suggest that in environments with visible transaction histories and active community discourse, reputational penalties for strategic flipping can propagate through informal sanctioning or withdrawal of status, creating longer-term economic consequences.

Our study offers significant theoretical contributions. First, we extend the literature on social image concerns and prosocial behavior, specifically within the domain of charitable giving and donations. The blockchain technology underlying NFTs allows for complete visibility of transactions and verifiable traceability of pseudonymous ownership history. The traceability and transparency, yet pseudonymity, create a unique context to examine social image and how it translates into market outcomes. Our research offers novel insights into the mechanisms of social penalties and ostracism that materialize when strategic generosity is perceived. Our findings illustrate how donors may encounter adverse social consequences, which may also lead to negative economic outcomes. They should be mindful of the potential market implications of quickly reselling charity NFTs, especially at higher prices, and consider the long-term value of maintaining ownership of these NFTs. Second, our study advances the emerging field of crypto philanthropy by highlighting the impact of the public visibility and traceability of transactions, an essential feature of NFTs and blockchain technology, on charitable giving. Our study sheds light on the significance of social image concerns in the context of charitable donations, highlighting the potential impact on both donors and the long-term effectiveness of NFT charity fundraisers.

Moreover, our approach to leveraging blockchain data for causal inference is novel. Empirical researchers often face challenges due to the lack of robust counterfactuals. We propose that failed transactions logged on the blockchain ledger could help identify individuals with similar intentions to successful transactors, intentions that would otherwise remain hidden in traditional datasets, who can serve as a valuable control group, particularly when successful and failed transactors are highly comparable in other aspects. This novel methodological approach offers significant implications for researchers studying other blockchain-based phenomena.

### 7.3. Limitations and Future Research

This study has several limitations. First, we estimate a local treatment effect, given that we focus on individuals who chose to participate in the Ukraine charity NFT fundraiser within a 5-minute window. Accordingly, these effects may or may not fully represent the behavior of the entire population of charity NFT donors. Moreover, in our study, we observe a persistent negative market penalty on the prices that donors can command for their other NFTs if they relist charity NFTs. This enduring price effect likely reflects either sustained disutility or hesitancy on the part of potential buyers, or a strategic adjustment by donors themselves to re-enter the market under less favorable terms. Additionally, charity NFTs could potentially be used for money laundering; future research could examine whether such actors are present and how their behavior differs from typical donors. We also acknowledge that individuals may create multiple accounts or wallets, though our findings remain meaningful since each wallet is evaluated independently by onlookers with no explicit linkage.

One might reasonably raise the concern that control users, those who failed to acquire the charity NFT, may have been prospective buyers of the treated users' subsequent NFTs. Disappointed by the competitive process, they might disengage or deliberately avoid treated users' sales. While our observational data cannot fully rule out such spillover effects, our experiments mitigate this concern by showing the sales decline reflects broader market-level punishment rather than targeted avoidance.

The larger-than-expected negative effect observed in the UKR charity NFT data likely reflects the high-profile nature and popularity of the RELI3F UKR collection. While punishment magnitude varies across collections, we consistently find robust negative treatment effects in both high-profile and low-profile charity NFTs, demonstrating our findings' generalizability. Nevertheless, the magnitude of market punishment may be influenced by specific features of these charity NFT fundraisers or broader cryptocurrency market characteristics.

Our work provides several potential avenues for future research. First, researchers could examine whether charity NFTs generate returns comparable to traditional financial assets, extending work by Whitaker and Kraussl (2020) on art market performance. Second, future work could explore how platforms might incentivize creators to participate in charitable NFT campaigns, connecting to broader research on creator economy incentives (Bhargava 2022). Third, while we focus on donors, researchers could investigate whether artists who contribute NFTs to charitable causes accrue social image benefits within NFT communities. Finally, scholars could examine whether similar dynamics of perceived strategic generosity extend to adjacent web3 philanthropy contexts, such as decentralized autonomous organizations (DAOs) or token-based social impact initiatives.

## References

Angrist JD, Pischke JS (2008) *Mostly harmless econometrics*. Princeton University Press.

Andreoni, J. (1989). Giving with impure altruism: Applications to charity and Ricardian equivalence. *Journal of Political Economy*, 97(6), 1447-1458

Athey, S., Bayati, M., Doudchenko, N., Imbens, G., & Khosravi, K. (2021). Matrix completion methods for causal panel data models. *Journal of the American Statistical Association*, 116(536), 1716-1730.

Autor, D. H. (2003). Outsourcing at will: The contribution of unjust dismissal doctrine to the growth of employment outsourcing. *Journal of Labor Economics*, 21(1), 1-42.

Bénabou, R., & Tirole, J. (2006). Incentives and prosocial behavior. *American Economic Review*, *96*(5), 1652-1678.

Berman, J. Z., Levine, E. E., Barasch, A., & Small, D. A. (2015). The Braggart's dilemma: On the social rewards and penalties of advertising prosocial behavior. *Journal of Marketing Research*, *52*(1), 90-104.

Berman, J. Z., & Silver, I. (2022). Prosocial behavior and reputation: When does doing good lead to looking good? *Current Opinion in Psychology*, *43*, 102-107.

Bhattacharjee, A., Dana, J., & Baron, J. (2017). Anti-profit beliefs: How people neglect the societal benefits of profit. *Journal of Personality and Social Psychology*, 113(5), 671-696.

Bhargava, H. K. (2022). The creator economy: Managing ecosystem supply, revenue sharing, and platform design. *Management Science*, 68(7), 5233-5251.

Blackwell, M., Iacus, S., King, G., & Porro, G. (2009). cem: Coarsened exact matching in Stata. The *Stata Journal*, 9(4), 524-546.

Bliege Bird, R., Ready, E., & Power, E. A. (2018). The social significance of subtle signals. *Nature Human Behaviour*, *2*(7), 452-457.

Brynjolfsson, E., Li, D., & Raymond, L. (2025). Generative AI at work. *The Quarterly Journal of Economics*, 140(2), 889-942.

Bursztyn, L., & Jensen, R. (2017). Social image and economic behavior in the field: Identifying, understanding, and shaping social pressure. *Annual Review of Economics*, *9*, 131-153.

Burtch, G., Ghose, A., & Wattal, S. (2015). The hidden cost of accommodating crowdfunder privacy preferences: A randomized field experiment. *Management Science*, 61(5), 949-962.

Burtch, G., Ghose, A., & Wattal, S. (2016). Secret admirers: An empirical examination of information hiding and contribution dynamics in online crowdfunding. *Information Systems Research*, 27(3), 478-496.

Burtch, G., Carnahan, S., & Greenwood, B. N. (2018). Can you gig it? An empirical examination of the gig economy and entrepreneurial activity. *Management Science, 64*(12), 5497-5520.

Carpenter, J., Holmes, J., & Matthews, P. H. (2008). Charity auctions: A field experiment. *The Economic Journal*, *118*(525), 92-113.

Carpenter, J., & Myers, C. K. (2010). Why volunteer? Evidence on the role of altruism, image, and incentives. *Journal of Public Economics*, *94*(11-12), 911-920.

Carpenter, J., Holmes, J., & Matthews, P. H. (2010). Endogenous participation in charity auctions. *Journal of Public Economics*, *94*(11-12), 921-935.

Catalini, C., de Gortari, A., & Shah, N. (2022). Some simple economics of stablecoins. *Annual Review of Financial Economics*, *14*.

Chao, M., & Fisher, G. (2022). Self-interested giving: The relationship between conditional gifts, charitable donations, and donor self-interestedness. *Management Science*, 68(6), 4537-4567.

Cong, L. W., Hui, X., Tucker, C., & Zhou, L. (2023). Scaling smart contracts via layer-2 technologies: Some empirical evidence. *Management Science,* 69(12), 7306-7316.

Cox, D. R. (1972). Regression models and life-tables. *Journal of the Royal Statistical Society: Series B (Methodological)*, 34(2), 187-202.

Critcher, C. R., & Dunning, D. (2011). No good deed goes unquestioned: Cynical reconstruals maintain belief in the power of self-interest. *Journal of Experimental Social Psychology*, 47(6), 1207-1213.

Elfenbein, D. W., & McManus, B. (2010). A greater price for a greater good? Evidence that consumers pay more for charity-linked products. *American Economic Journal: Economic Policy*, *2*(2), 28-60.

Elfenbein, D. W., Fisman, R., & McManus, B. (2012). Charity as a substitute for reputation: Evidence from an online marketplace. *Review of Economic Studies*, 79(4), 1441-1468.

Exley, C. (2018). Incentives for prosocial behavior: The role of reputations. *Management Science*, *64*(5), 2460-2471.

Forehand, M. R., & Grier, S. (2003). When is honesty the best policy? The effect of stated company intent on consumer skepticism. *Journal of Consumer Psychology*, 13(3), 349-356.

Gambetta, D., & Przepiorka, W. (2014). Natural and strategic generosity as signals of trustworthiness. *PloS one*, 9(5), e97533.

Halaburda, H., Sarvary, M., & Haeringer, G. (2022). Smart Contracts and Blockchain. In *Beyond Bitcoin: Economics of Digital Currencies and Blockchain Technologies* (pp. 135-178). Springer.

Hur, Y. Y., Jin, F., Li, X., Cheng, Y., & Hu, Y. J. (2023). Does social influence change with other information sources? A large-scale randomized experiment in medical crowdfunding. *Information Systems Research*, 34(4), 1476-1492.

Kafashan, S., Sparks, A., Griskevicius, V., & Barclay, P. (2014). Prosocial behavior and social status. In *The psychology of social status* (pp. 139-158). Springer, New York, NY.

Kanellopoulos, I. F., Gutt, D., & Li, T. (2025). NFT Disruption in Platform Competition: Evidence from Trading Card Collectibles. *Information Systems Research,* Forthcoming.

Kugler, L. (2021). Non-fungible tokens and the future of art. *Communications of the ACM*, *64*(9), 19-20.

Lacetera, N., & Macis, M. (2010). Social image concerns and prosocial behavior: Field evidence from a nonlinear incentive scheme. *Journal of Economic Behavior & Organization*, *76*(2), 225-237.

Lee, D., Gopal, A., Lee, D., & Shin, D. (2023). Nudging Private Ryan: Mobile Microgiving under Economic Incentives and Audience Effects. *MIS Quarterly*, 47(3), 1101-1146.

Leszczyc, P. T. P., & Rothkopf, M. H. (2010). Charitable motives and bidding in charity auctions. *Management Science*, *56*(3), 399-413.

Li, J., Wan, X. S., Cheng, H. K., & Zhao, X. (2024). Operation Dumbo Drop: To Airdrop or Not to Airdrop for Initial Coin Offering Success?, *Information Systems Research*, *35(4)*, 1928-1947.

Lin-Healy, F., & Small, D. A. (2012). Cheapened altruism: Discounting personally affected prosocial actors. *Organizational Behavior and Human Decision Processes*, 117(2), 269-274.

Lin-Healy, F., & Small, D. A. (2013). Nice guys finish last and guys in last are nice: The clash between doing well and doing good. *Social Psychological and Personality Science*, 4(6), 692-698.

Liu, L., Suh, A., & Wagner, C. (2018). Empathy or perceived credibility? An empirical study on individual donation behavior in charitable crowdfunding. *Internet Research*, 28(3), 623-651

Nadini, M., Alessandretti, L., Di Giacinto, F., Martino, M., Aiello, L. M., & Baronchelli, A. (2021). Mapping the NFT revolution: market trends, trade networks, and visual features. *Scientific Reports*, *11*(1), 20902.

Newman, G. E., & Cain, D. M. (2014). Tainted altruism: When doing some good is evaluated as worse than doing no good at all. Psychological Science, 25(3), 648-655.

Rosenbaum, P. R., & Rubin, D. B. (1983). The central role of the propensity score in observational studies for causal effects. *Biometrika*, *70*(1), 41–55.

Schamp, C., Heitmann, M., Bijmolt, T. H. A., & Katzenstein, R. (2022). The Effectiveness of Cause-Related Marketing: A Meta-Analysis on Consumer Responses. *Journal of Marketing Research*, 60(1), 189-215.

Simpson, B., & Willer, R. (2015). Beyond altruism: Sociological foundations of cooperation and prosocial behavior. *Annual Review of Sociology*, *41*(1), 43-63.

Stock, J. H., & Yogo, M. (2005). Testing for weak instruments in Linear Iv regression. In *Identification and Inference for Econometric Models: Essays in Honor of Thomas Rothenberg* (pp. 80-108). Cambridge University Press.

Tahmasbi, N., & Fuchsberger, A. (2022). Non-fungible Tokens - Exploring Suspicious Washtrader Communities in NFT Networks. *ICIS 2022 Proceedings*. 5.

Tan, X. J., & Tan, Y. (2022). Crypto rewards in fundraising: Evidence from crypto donations to Ukraine. *Available at SSRN 4161068*.

Weiss, J., & Obermeier, D. (2021). How Blockchain Can Enhance Trust and Transparency of Online Surveys. *ICIS 2021 Proceedings*. 7.

Whitaker, A., & Kräussl, R. (2020). Fractional equity, blockchain, and the future of creative work. *Management Science*, *66*(10), 4594-4611.

Wooldridge, J. M. (2010). Econometric analysis of cross section and panel data. MIT Press, 154.

Yoon, Y., Gürhan-Canli, Z., & Schwarz, N. (2006). The effect of corporate social responsibility (CSR) activities on companies with bad reputations. *Journal of Consumer Psychology*, 16(4), 377-390.

Zaucha, T., & Agur, C. (2022). Newly minted: Non-fungible tokens and the commodification of fandom. *New Media & Society*, 14614448221080481.

# SUPPLEMENTAL APPENDICES

## Appendix A. Semi-Structured Interview Background and Results

An online interview was conducted in March 2025 using a structured protocol. The interview lasted approximately 30 minutes and followed a predefined set of questions, with an open-ended question at the end of the interview. The interviewee was a verified donor to the Ukraine Charity NFT initiative. All identifying information was removed from the recording to ensure confidentiality.

### *Part 1. Background*

**Question: Could you share what motivated you to purchase the Ukraine charity NFT? What specific aspects of the cause or the NFT itself appealed to you?**

First, the cause is always obviously important. I wouldn't buy if I wouldn't support the cause, even though it was the most beautiful NFT, but second, the artwork is also important for me because it will show up in my portfolio of art collection and I'm also happy to share it with others. Others look into it and also it for me, it also helps show identity also a little bit in the sense of an NFT also closely links to identity. If people see my art collection they see what causes I support and therefore more or less who I am, and it also attracts other people to get in touch with other peoples that have similar views. For example Ukraine charity, once I bought one I posted it on X formerly Twitter. And then I got in touch with other like minded people and we talked about the cause and we talked about ideas etcetera. So it's definitely more than just a one time donation. And what you also see is that often lasting communities are built by sharing this identity on a specific topic or sharing the importance of a cause and often also that leads to other initiatives either new funding rounds, or the support of projects. What is also interesting in charity NFTs is that once they get results, a percentage of the royalties goes to the founders of the charity NFT. And they can also use that money and so some of these NFTs also hold value long term and that also depends on how active they are. For example, in World of Women, as active community and NFT is still their NFTs, their 10,000 of them. They still worth like 800 U.S. dollar somewhat even more almost 1000 U.S. dollars. So some people also buy it for profit. Or at least that they it doesn't cost them as much, and that also matters if I buy an NFT which is quite expensive, I want to the art to be good because I know I might be able to resell it if I want to later. So it's a store of value in that respect.

**Question: Before making your purchase, did you consider the potential financial value of the NFT? To what extent did investment potential factor into your decision?**

It matters because especially if the NFT is a bit more expensive and usually NFTs are relatively expensive compared to if somebody donates at the door. If somebody comes for charity at the door. So you pay more. But the good thing is that it doesn't feel like full financial loss. Because if the art is good, if the project is good, if the founders are good, usually they hold their value pretty well. So I'm more likely to spend more than just the amount I would normally donate.

**Question: When buying charity NFTs versus regular NFTs, how do your thought processes or considerations differ, if at all?**

Yeah, it does because normally if you buy NFTs, it's either for the art that you buy them often because it's a famous artist, but also because you expect that the value is going up. Otherwise if the value would go down, you would wait. Simple as that. You know, if you don't trust that the value goes up, then you wait, even though you find the art beautiful. Charity NFT is first, it's much more important that you really support the cause and want to signal that to others as well. So showing others, it's sometimes also maybe also a little bit of a flex in the sense that you show okay, I'm willing to support this and are willing to pay up for that. So people also will see, oh, that's a good guy. Whereas regular NFTs are a different kind of flex. It's more like showing how much money you have, or how much taste you have, or how good of an art person you are. So it's definitely different for me, yes.

### *Part 2. Motivations*

**Question: To what extent do you believe that the primary motivation for purchasing charity NFTs is altruism?**

I think the primary reason is altruism, but more than in other NFTs because other NFTs I also see as an investment, whereas in charity NFTs I'm willing to lose it all basically. So in the end I was always in the back of my mind it might go to zero and I would still be happy to have supported the cause.

**Question: Beyond altruistic intentions, what other factors might influence individuals' decisions to acquire charity NFTs?**

The other motivations are things like becoming part of a community, signaling your support of this group, little bit of some sort of personal flex in the community that you're willing to pay up for being a good person and being liked in the community. And like I said, if sometimes you really have a feeling that this is sometimes these NFTs are made by famous artists because they also want to support that cause. If that is the case, there is also an investment aspect to it so that you know okay, if we can buy this artist, that will provide me with probably it will be a good investment long term.

**Question: What potential indicators could be used to assess whether an individual's purchase of charity NFTs is driven purely by altruistic motives or influenced by other considerations?**

I would say whether that person resells or puts it up for sale. So somebody with purely financial motives, they might put it up for sale soon after they bought it or when price rises, whereas others that have pure altruistic motives, they might not sell it and be willing to hold on to it. Also, whether they are active in the community. If people are active in the community, it's probably more altruistic than just buying it and flexing it online.

**Question: If an individual purchases a charity NFT and subsequently resells it for profit within a short timeframe, how might this affect your perception of their initial motivation for the purchase?**

Yeah, basically. Like I said, in that case, it's more like somebody who is - that often quite often happens that people immediately sell it afterwards for even a small profit, and then obviously it signals that people are not really there for the cause, but basically for the financial gain.

**Question: Additionally, would this behavior influence your willingness to engage in future transactions with them?**

Yeah. In my view, their reputation goes down and I would be less willing to engage with them later on, yes.

### *Part 3. Decision to Hold (vs. Resell)*

**Question: Have you ever relisted or resold a charity NFT, and in particular the Ukraine charity NFT?**

No, I hold it. I still hold it and I will probably never sell it. That goes basically for my charity NFTs.

**Question: What factors influenced your decision to hold onto the Ukraine charity NFT rather than reselling it?**

Because on one end reselling, in my view in my book is a bit dirty. You know, in the sense that I think it will affect my identity in the community, how people view me. They might also think, okay, he's not supporting the cause anymore. And more practical, but it usually they don't hold their value much. So it's also sometimes no use as selling it anymore because you don't get much from them. But the most important issue is the signaling and the blockchain is transparent, so people see when I sell and for how much I sell. So I consider it a bit dirty.

**Question: Would you have considered selling it if it was not a charity NFT?**

This particular one you mean? I don't usually, I'm very - I want to limit my collection. I'm buying quite a lot, but I also want to limit my collection to special ones, to the things that are very particular, which artists I really love. And this charity NFT art-wise is not the most beautiful I have, so normally I would have sold it, yes.

**Question: Have you been approached by others interested in buying your charity NFT? If so, how did you respond?**

You get offers, but sometimes collection offers on OpenSea, so indirectly - you know how that works for OpenSea? So if there is an NFT and the lowest NFT is being sold for one ETH for example, I can put an offer on the whole collection and say, okay, I'm willing to pay .5 ETH, half an ETH. Anyone who wants to sell it to me can just dump it into my bid and I own it and they get the half ETH. So basically, there's almost always a bid on your NFTs, sometimes a relatively low bid. But there was almost always a bid, so in that sense, people have bid for it. But I haven't sold and to be honest, I see in general that there is much less circulation in charity NFTs than in others.

**Question: At any point, did you consider reselling the NFT? What would need to happen for you to consider selling it?**

Well, "never" is - you know, if somebody said I'll pay you €10,000 for your NFT, obviously there's always a price, right?

***Part 4. Social Image***

**Question: Do you think owning a charity NFT affects how others perceive you or interact with you in the NFT marketplace? If so, how?**

Maybe it's something to think about. Sometimes it affects you positively, but like you know in the case of Ukraine, it might also affect you negatively to some other people. So some people will become hostile and some people share your views. So it's more about building your identity and trying to be in touch with people who are likewise minded that you have.

**Question: Have you noticed any reactions from others in the NFT community regarding your ownership of the Ukraine charity NFT?**

Both. I got a lot of support and I got a lot of hate. So both actually. I got some very negative reactions on me holding World of Women NFT because especially in the US nowadays there are a lot of people who are against the "woke" ideas. And so even beyond war. And so you're usually safe if it is about children or cancer or these kind of things, then you are safe. Or like the elephants. But obviously Ukraine is something which might be controversial.

I'm mostly active on Twitter because there is also Discord, but those are all the like-minded people because they are token gated - you have to have an NFT to be able to access that environment. But on Twitter, you know, it's open and everybody can react and I get a lot of these reactions, especially on my

NFT Twitter. I have like 15,000 followers. And so there are always a lot of different people and then a lot of reactions on my posts.

**Question: Have you noticed any differences in your ability to sell other NFTs in your portfolio since acquiring the charity NFT?**

Yes, in the sense that I have become more connected to others. Other NFT enthusiasts that first approached me because of the Ukraine NFT, and later they became online friends. And there, because they're also into NFTs, you often get into "do you want to share this NFT?" Or you know they're interested in a certain style or in a certain artwork. And if I have to sell, I approach them. So in that sense, yes.

### *Part 5. Transparency and Community Response*

**Question: In your opinion, is it easy to discover the NFTs owned by someone and track their transaction history?**

Yeah, it's very easy. The whole chain of transactions are always visible, especially on marketplaces like OpenSea. You can immediately see who owns a certain NFT and who bought it from whom, who minted it first and then the various transactions. Sometimes people - you can see the name of people. They can have an OpenSea name or they could be anonymous, but still you can track their wallet numbers, so this is relatively easy, much easier than regular art.

**Question: When you're considering buying an NFT from someone else, do you typically look at what other NFTs they own or their transaction history?**

Yeah. For a number of reasons. One is for safety. You don't want to buy from a scammer that stole NFTs from somebody else. But also because of provenance - we always say NFT provenance matters, so that means who has owned it before you? If that's a well known person in the space, for example, you want to buy from them. Also in this context of charity, people that do well for the community and for the broader society, I would rather buy from them than from somebody else.

**Question: How do you think the NFT community generally views those who quickly resell charity NFTs versus those who hold them?**

Many of them are called out by the community, also publicly, as sellers and especially people who are dumping them. So the community generally views this negatively.

**Question: If you noticed someone had previously purchased and quickly resold a charity NFT for profit, would this affect your willingness to buy from them? Why or why not?**

This would affect my willingness in the sense of negatively, yes.

**Question: Does the inclusion of a charity NFT in an individual's portfolio or the subsequent relisting of such an NFT have the potential to dilute buyer interest in their future non-charity NFTs?**

I think in general, yes.

### *Part 6. Arbitrage and Strategic Behavior in NFT Markets*

**Question: In your experience, do NFTs generally present arbitrage opportunities that can be easily exploited?**

if there's a new mint of NFT, there's generally a white list, and people, for example, famous people in the community, or people that already have NFTs in that particular series, they are prioritized. So sometimes they have 500 NFTs minted, but there is demand for 5000 and they get sold for .1 ETH as mint price. And the people can only mint if they're on this white list. And generally what you see is that a lot of them immediately sell them for a 5X or 10X. So there are these arbitrage opportunities in general.

**Question: How would you personally view someone who takes advantage of arbitrage opportunities for regular NFTs? Is this behavior generally accepted in the NFT community?**

Well, it's done, but it's also looked at badly in the sense of - because for example, what you often see is, for example, I get approached to be on a white list because people either like me or they think okay, I'm doing well for the community and they want to reward me for that, for example, and to give me a white list spot. If I take that and I immediately resell it for five X, I would feel very bad about doing that myself. It's like you're handed a present on your birthday and you immediately sell it the same day. So you know, and this is generally looked down upon by the NFT community, yes. And some people are really known for being flippers.

**Question: Do you think charity NFTs in particular might create unique arbitrage opportunities due to inefficient pricing or sentiment-driven purchasing?**

I'd rather say there are two things. One is they're less liquid, so because people are selling them less, there's only less on the market. So that might increase arbitrage opportunities. On the other hand, usually because they want initially to earn as much money for the cause as possible, the mint price is generally relatively low and sometimes open. So they sell a lot of them. So either it's an open mint that everybody can mint as much as possible, so there could be 500, but could also be 10,000 or 20,000 or whatever the demand is. So everybody mints at the same price. So then you don't see opportunities, and sometimes they are usually less bought for reselling. So people that buy don't see profit. So I'm not sure in that sense whether there are arbitrage opportunities because, like I said, you'll rarely see that they are sold for much more than they mint for.

**Question: In your opinion, is there a difference between exploiting price inefficiencies in regular NFTs versus charity NFTs?**

Overall I think it's probably easier in regular NFTs to exploit because people are buying them on the expectation of profits.

**Question: Would you view someone who strategically purchases charity NFTs with the intention of profiting from resale less favorably than someone doing the same with regular NFTs?**

Very, very differently. Therefore, I would look much less favorably on people that make money from charity NFTs. So I don't mind people making money on regular NFTs, but I would mind it with charity NFTs, yes.

### *Part 7. Open Ended Question*

**Question: Any other factor that might play a role in how you interact with charity NFTs, which our questions did not address?**

There are two types of charity NFTs. One is the one-time donation and the other is an ongoing project with an active team behind it that builds a community around it, maybe like for example, World of Women, which has subsequent NFTs for sale to raise more money and actively supports projects etcetera. Maybe if you introduce me, you might indeed say that I'm over five years in the space. I'm - maybe I wouldn't call myself, but you could call me a prominent person in the space that has a good view on the dynamics of the space. So that you didn't just talk to anybody that has one or two NFTs, but somebody who has been consistently contributing to the space for over five years now and who is a very active community member in the NFT community.

## Appendix B. Unique Features of Blockchain-Based Charity.

Drawing on both the fresh qualitative insights from our interview, as well as the scholarly literature in crypto philanthropy, crowdfunding, and prosocial behavior, this section identifies three distinctive structural features that meaningfully shape user behavior and market dynamics. These features distinguish blockchain-enabled charity from both traditional charitable giving and conventional NFT markets.

### B.1. Potential for Personal Gain Under the Guise of Charitable Intent in Charity NFTs

Charity NFTs allow individuals to publicly signal generosity while retaining financial upside. As our interviewee remarked, *"You're willing to pay up for being a good person."* However, the ability to resell these NFTs, sometimes at a profit, creates a potential for personal gain. Hence, individuals can appear altruistic while pursuing financial gains. This can give rise to a perception of what Gambetta and Przepiorka (2014) refer to as strategic generosity, behavior that imitates prosocial intent but is ultimately self-interested. The NFT expert we interviewed summarized this dynamic clearly, stating: *"Obviously, it signals that people are not really there for the cause, but basically for the financial gain."* The ability to gain through charities, buying into a cause and cashing out for personal benefit, poses a reputational risk, especially in transparent, peer-monitored environments.

Prior work has conceptualized prosocial behavior driven by mixed motives through several theoretical frameworks. Andreoni (1989) introduced the concept of impure altruism in the context of charitable giving, where both genuine concern for others and personal benefits, such as a warm glow or social recognition, are acknowledged. Lin-Healy and Small (2012) coined the term cheapened altruism specifically to describe situations where prosocial actors are personally affected by the cause they support, showing how observers discount the altruistic value of charitable acts when the donor appears to have 'skin in the game' or stands to benefit personally from the outcome. Newman and Cain (2014) extended this work with their concept of tainted altruism to describe the finding that prosocial behavior with self-benefit can be evaluated more negatively than no action; that is, 'doing some good' while gaining personally is often viewed more negatively than 'doing no good at all.' This behavior has been referred to as strategic generosity by Gambetta and Przepiorka (2014), who distinguished between *natural generosity*, driven by genuine prosocial motives, and its *strategic alternative*, performed to cultivate a favorable reputation. Our study draws from and builds upon the concept of strategic generosity in the context of charity NFTs. The ability to monetize charity NFTs after donation might lead to a shift in attributed motives for what would otherwise be perceived as prosocial acts. Thus, relisting charity NFTs in the pursuit of personal gain may be interpreted as strategic generosity by onlookers. The erosion of perceived trustworthiness might then, in turn, spill over to affect a donor's broader market reputation and future transaction outcomes.

### B.2. Transparent Transaction History as A Reputational Signal in Pseudonymous Environments

A defining feature of blockchain-based NFT marketplaces is the transparency of transaction history. Every token's past ownership and trading activity is publicly recorded, enabling buyers to inspect a seller's behavior before making a purchase. As our interviewee noted, he routinely reviews seller wallets to avoid buying *"from a scammer"* and instead looks for *"people that do well for the community and for the broader society,"* noting that he *"would rather buy from them than from somebody else."* This behavior reflects how participants use observable transaction patterns as social cues to infer credibility and trustworthiness (Burtch et al. 2016; Liu et al., 2018). Burtch et al. (2016) demonstrate that participants in online crowdfunding platforms use the visibility of contributor information—including donation amounts and donor identities—as signals to assess campaign quality and credibility. Liu et al. (2018) further demonstrate that observers process such observable social actions through cognitive credibility assessments to evaluate participant motivations. These individual behaviors aggregate into systematic network structures in NFT markets. Vasan et al. (2022) provide evidence that successful artists receive repeated investment from collectors, showing that "the success for NFT artists rests on their ability to build relationships with collectors." Similarly, Nadini et al. (2021) demonstrate that traders typically form tight clusters with other traders, revealing specialized trading communities based on shared interests and familiarity. These network-level patterns illustrate how individual behaviors create systematic network structures where reputation and identity fundamentally shape trading relationships in NFT markets.

This on-chain visibility plays a critical reputational role, especially in charity contexts where altruism is expected. In pseudonymous markets, where conventional identity markers are missing, past transaction behavior serves as a primary signal of trust and social alignment. Participants actively interpret these signals to evaluate others' motivations. For instance, when discussing how altruism can be inferred from transactional behavior, the charity NFT donor we interviewed stated that he pays attention to *"whether that person resells or puts it [the NFT] up for sale. So somebody with purely financial motives, they might put it up for sale soon after they bought it or when price rises, whereas others that have pure altruistic motives, they might not sell it and be willing to hold on to it."* This comment demonstrates that market participants do indeed use observable behavior (relisting/reselling) to infer underlying motivations, regardless of the donor's actual motivation.

This qualitative anecdote is also supported by the literature (Burtch et al. 2015, Elfenbein et al. 2012, Lee et al 2023). Burtch et al. (2015) show that when crowdfunding platforms make the option to conceal donor identity more salient, it reduces donation amounts, suggesting that the default visibility of donor information serves as a motivator for giving behavior through social recognition. Lee et al. (2023) demonstrate that observable donation behavior in digital platforms creates social pressure that influences charitable decisions. Elfenbein et al. (2012) provide empirical evidence that charitable giving in pseudonymous contexts functions as a reputation-building tool, as sellers strategically increase charitable donations when their reputations are damaged. Moreover, Hur et al. (2023) demonstrate through online medical crowdfunding experiments that participants rely heavily on these social signals when making donation decisions. When charitable behavior is recorded on-chain, observers can readily examine these digital social cues to make inferences about whether prosocial behavior reflects genuine or strategic altruism. This makes transaction history a powerful mechanism for reputation signaling in blockchain-based NFT charity.

These reputational effects can also be persistent (Elfenbein et al. 2012, Schamp et al. 2022). Elfenbein et al. (2012) found in their longitudinal empirical investigation that charitable behavior serves as a lasting quality signal, with reputation effects continuing to influence seller success rates over time. Similarly, Schamp et al. (2022) provide meta-analytic evidence that cause-related marketing builds persistent and long-term brand reputation and image. As our interviewee further noted, confirming these ideas: *"This [re-listing charity NFTs] is generally looked down upon by the NFT community, yes. And some people are really known for being flippers."* Thus, the consequences of attributed motivations, whether perceived as genuine or strategic generosity, can be both salient and enduring.

**B.3. Attributed Motivations Lead to Economic Punishment**

A perception that one is acting strategically can carry real economic consequences. A considerable amount of prior research has explored social punishment, examining how onlookers respond when they perceive that ulterior motives underlie an individual's prosocial actions. These studies typically rely on lab experiments to infer motivations and social judgments, measuring behavioral responses rather than real-world economic consequences. Prior laboratory experiments have demonstrated that observers tend to maintain cynical interpretations of prosocial acts (Critcher & Dunning, 2011). Other work has found that prosocial actors face social penalties, including reduced trust and lower moral character evaluations, when observers detect mixed motives (Berman et al., 2015). Vignette experiments demonstrate that onlookers may withhold "charitable credit" from prosocial actors if they perceive self-interested motives (Lin-Healy & Small, 2012). These social penalties extend beyond individuals to corporations, where work has also shown that corporate social responsibility efforts backfire for companies when profit motives are salient (Yoon et al., 2006). Further, work has demonstrated systematic skepticism toward corporate prosocial efforts when company intent appears self-serving (Forehand & Grier, 2003) as observers view business profits and social goods as "necessarily in conflict" (Bhattacharjee et al., 2017).

Our study advances the literature, introducing the concept of economic punishment for perceived strategic generosity, documenting observable, downstream financial consequences in a real-world, market-based context. We show that individuals perceived as exploiting charitable causes through charity NFT resale are penalized across the broader market. As our interviewee explained, *"In my view, their reputation goes down and I would be less willing to engage [in future transactions] with them later on."* Buyers also make sharp ethical distinctions between profiting from regular NFTs and profiting from charity-linked assets. As the interviewee further explained, *"I would look much less favorably on people that make money from charity NFTs. So I don't mind people making money on regular NFTs, but I would mind it with charity NFTs."* Because of this, when charity NFTs are re-listed for sale, the act not only has social, intangible consequences; it has lasting economic consequences.

Together, these three structural features create a distinctive context in which deviations from expected prosocial behavior can trigger both reputational and market-level punishment. This makes blockchain-enabled charity a rich setting for understanding how strategic generosity may be evaluated and sanctioned in public digital markets.

## Appendix C. Balance Check and Relative Time Estimates

As Table C1 shows, all the covariates exhibit high comparability between the treatment and control groups after matching.

**Table C1. Balance Check for the CEM Sample**

| | Mean | | | t-test | | |
|---|---|---|---|---|---|---|
| Variable | Treated | Control | %bias | t | *p*-value | V(T)/V(C) |
| MinuteId | 3.357 | 3.357 | 0.000 | 0.000 | 1.000 | 1.000 |
| LowGas | 0.219 | 0.219 | 0.000 | 0.000 | 1.000 | . |
| ModeGas | 0.339 | 0.339 | 0.000 | 0.000 | 1.000 | . |
| HighGas | 0.443 | 0.443 | 0.000 | 0.000 | 1.000 | . |
| LnAvgMints | 0.700 | 0.675 | 3.100 | 0.610 | 0.539 | 0.980 |
| LnAvgPurchases | 0.098 | 0.132 | -8.000 | -1.600 | 0.110 | 0.970 |
| LnAvgSales | 0.363 | 0.354 | 1.200 | 0.240 | 0.809 | 1.020 |
| LnAvgListing | 0.402 | 0.406 | -0.500 | -0.100 | 0.922 | 0.970 |
| LnAvgTransfers | 0.364 | 0.368 | -0.800 | -0.150 | 0.880 | 0.980 |
| TenureMonth | 2.750 | 2.647 | 5.200 | 1.060 | 0.289 | 0.930 |

*Note: In the full sample, 43.10% of individuals have a gas limit of 122,839, which appears to be the prevailing default value. In the matching process, we categorize individuals into three distinct groups: LowGas, ModeGas, and HighGas. This classification is contingent upon whether their designated gas limits fall below, align with, or exceed the default benchmark.*

Dynamic DID estimates using the CEM sample (Table C2) confirm that treatment and control groups exhibit parallel trends in the pre-treatment period.

**Table 9. Relative Time Estimates of the Impact of Charity NFT Purchase on Other NFT Sales**

| | LnNumSales | LnTotRev | LnNumSales | LnTotRev |
|---|---|---|---|---|
| | (1) | (2) | (3) | (4) |
| Relative time -7 | 0.077 | -0.247 | 0.122 | -0.215 |
| | (0.340) | (0.264) | (0.343) | (0.266) |
| Relative time -6 | 0.003 | -0.038 | 0.064 | 0.006 |
| | (0.207) | (0.152) | (0.202) | (0.146) |
| Relative time -5 | -0.096 | -0.046 | -0.070 | -0.028 |
| | (0.115) | (0.072) | (0.116) | (0.073) |
| Relative time -4 | 0.013 | -0.011 | 0.046 | 0.013 |
| | (0.095) | (0.057) | (0.091) | (0.056) |
| Relative time -3 | -0.029 | -0.095* | 0.009 | -0.068 |
| | (0.079) | (0.050) | (0.077) | (0.050) |
| Relative time -2 | 0.003 | -0.003 | 0.014 | 0.005 |
| | (0.059) | (0.036) | (0.058) | (0.036) |
| Relative time -1 | Baseline (omitted) | | | |
| Relative time 0 | -0.232*** | -0.118*** | -0.223*** | -0.111*** |
| | (0.067) | (0.037) | (0.066) | (0.036) |
| Relative time 1 | -0.141** | -0.073* | -0.127* | -0.063 |
| | (0.069) | (0.040) | (0.069) | (0.040) |
| Relative time 2 | -0.113* | -0.102*** | -0.097 | -0.091** |
| | (0.060) | (0.039) | (0.059) | (0.038) |
| Relative time 3 | -0.116* | -0.101*** | -0.098 | -0.088** |
| | (0.065) | (0.038) | (0.065) | (0.037) |
| Relative time 4 | 0.038 | -0.070* | 0.057 | -0.056 |
| | (0.068) | (0.039) | (0.067) | (0.038) |
| Relative time 5 | 0.016 | -0.079** | 0.036 | -0.065* |
| | (0.066) | (0.040) | (0.066) | (0.039) |
| Relative time 6 | 0.015 | -0.077* | 0.036 | -0.063 |
| | (0.065) | (0.039) | (0.064) | (0.038) |
| Relative time 7 | -0.004 | -0.087** | 0.017 | -0.072* |
| | (0.069) | (0.040) | (0.069) | (0.039) |
| LnTenure | | | 0.482*** | 0.343*** |
| | | | (0.051) | (0.031) |
| Individual FE | Yes | Yes | Yes | Yes |
| Year-month FE | Yes | Yes | Yes | Yes |
| Observations | 16,780 | 16,780 | 16,780 | 16,780 |
| Num of Individuals | 1,574 | 1,574 | 1,574 | 1,574 |
| R-squared | 0.481 | 0.452 | 0.491 | 0.469 |

*Note: Robust standard errors clustered at the individual level. * p<0.1, ** p<0.05, *** p<0.01.*

## Appendix D. Additional Analysis on Relisted Prices of Charity NFTs

As a further investigation into whether the negative effect of relisting charity NFTs on the sales performance of other non-charity NFTs is due to market punishment for perceived strategic generosity, we also examine the potential role played by the relisted price. Specifically, we classify the scenarios based on whether the relisted price for the charity NFT is set higher than its original price or lower, and we estimate the average treatment effect on the treated (ATT) for these two scenarios separately. By distinguishing between these two pricing strategies, we aim to determine if the market's negative response is particularly pronounced when individuals relist charity NFTs at a higher price, which could be interpreted as a stronger signal of strategic generosity. Therefore, we expect that the market punishment would be stronger in the latter case.

Related ATT estimation results are in Table D1. We find that the negative effect of relisting a charity NFT is more pronounced when the relisted price is high, compared to when the price is low. Additionally, the dynamic treatment effects, illustrated in Figures D1-D2, support the parallel trends assumption and demonstrate a persistent negative impact on sales performance when charity NFTs are relisted for sale at a profit (i.e., flipping a charity NFT). Our results suggest that market punishment intensifies when charity NFTs are relisted at higher prices, a response likely driven by perceptions of opportunistic strategic generosity.

**Table D1. ATT Estimate Based on the MC Method**

| | LnNumSales<br>(1) | LnTotRev<br>(2) |
|---|---|---|
| ATT of relisting a charity NFT at a high price | -0.201*<br>(-0.438, 0.026) | -.273***<br>(-0.415, -0.142) |
| ATT of relisting a charity NFT at a low price | -0.171<br>(-0.571, 0.361) | -0.019<br>(-0.234, 0.245) |

*Note: Error bars denote bootstrapped 95% confidence intervals.*

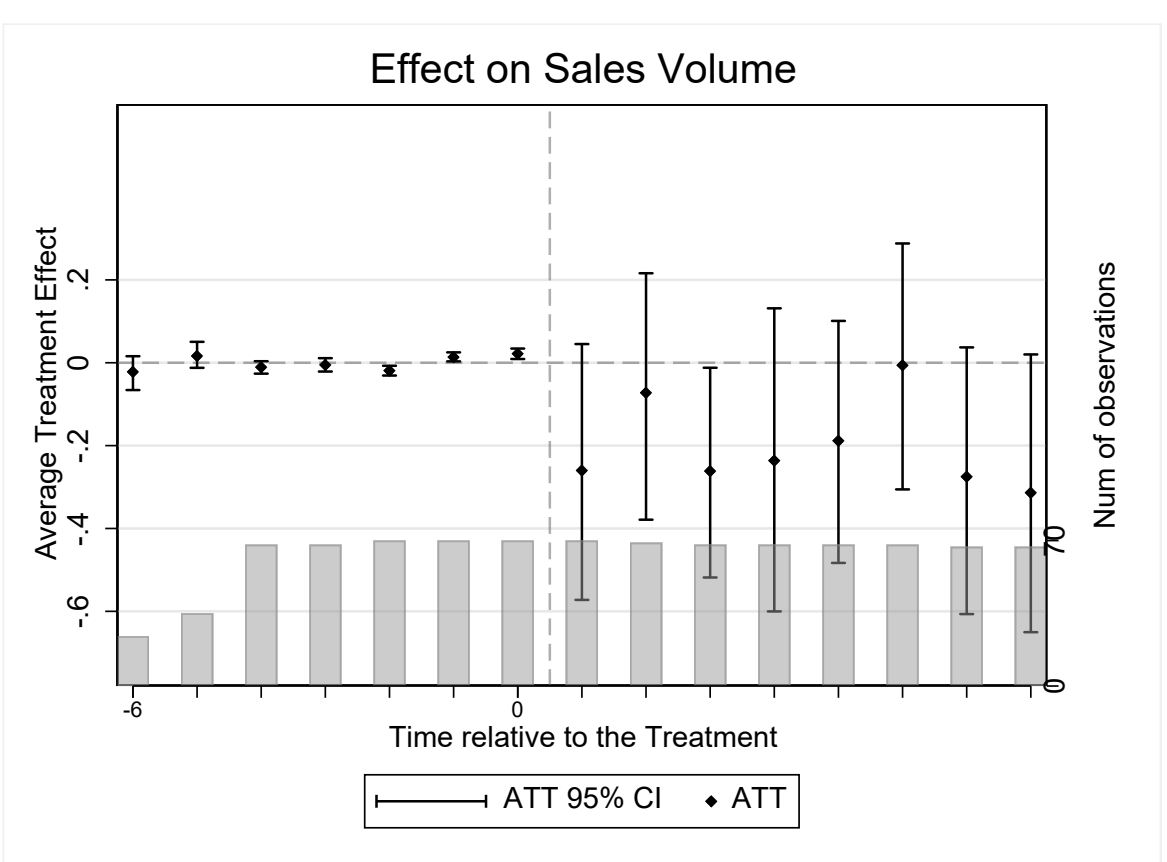

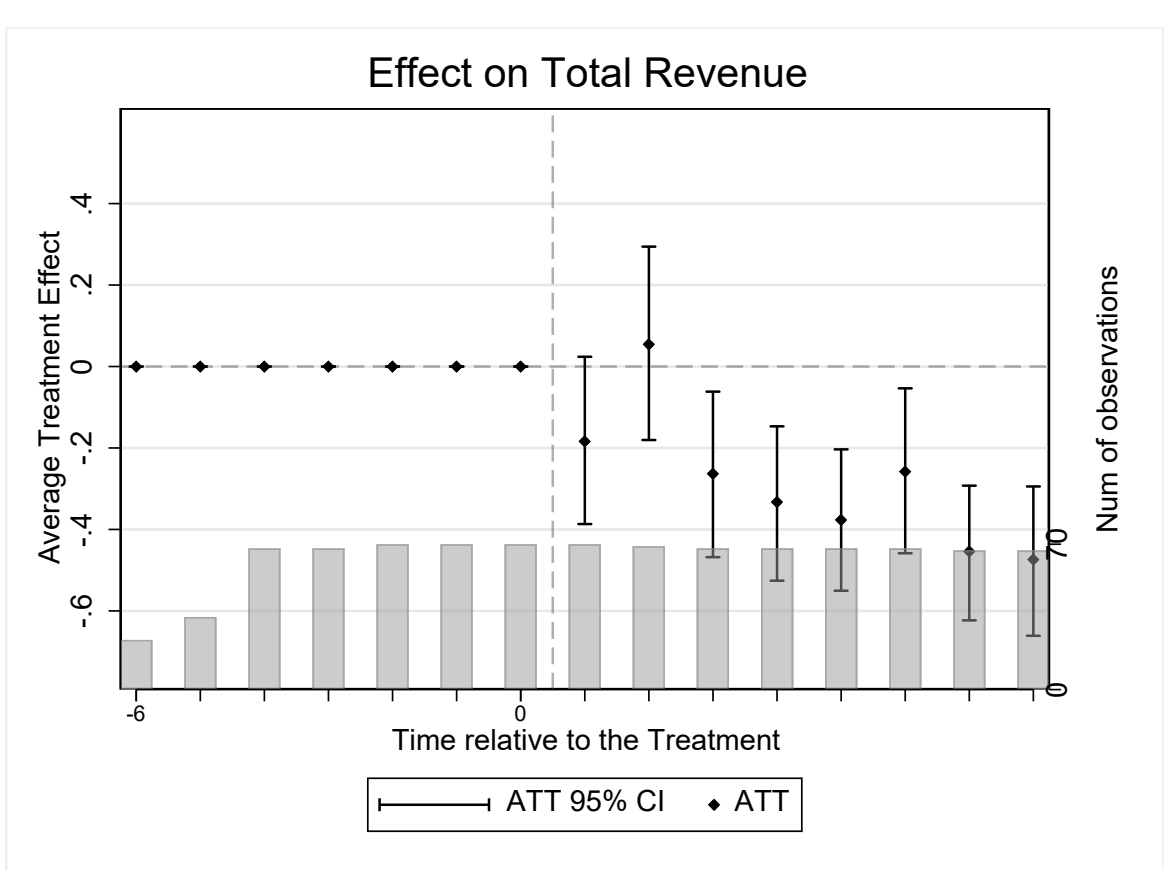

**Figure D1. Dynamic Treatment Effects of Relisting a Charity NFT at a High Price Based on the MC Method**

Notes: Both plots presented were produced utilizing the “fect” package in Stata. Error bars denote bootstrapped 95% confidence intervals.

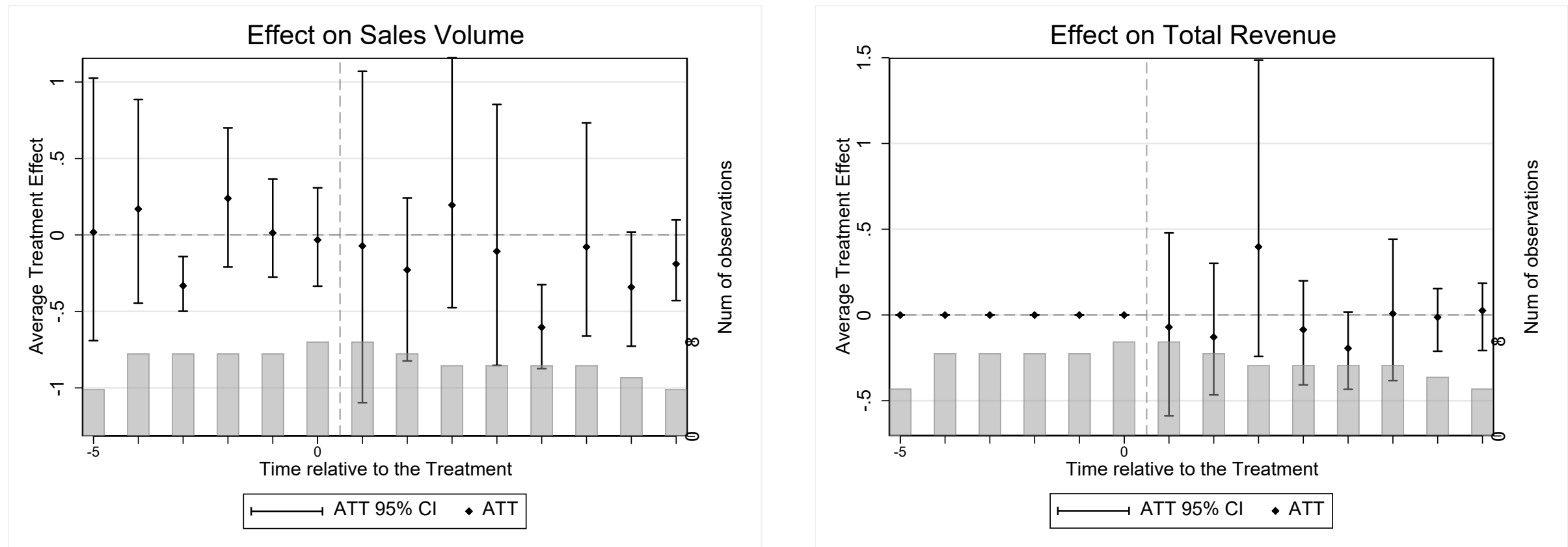

**Figure D2. Dynamic Treatment Effects of Relisting a Charity NFT at a Low Price Based on the MC Method**

Notes: Both plots presented were produced utilizing the “fect” package in Stata. Error bars denote bootstrapped 95% confidence intervals.

## Appendix E. Additional Analysis on Social Exposure as a Moderator

We evaluate whether the effects of the three treatment statuses discussed in Section 4.4.1 vary with an individual's social exposure. For clarity in interpretation, we demean the follower base measure ($FollowersCum$). Based on the results in Table E1, we find that the negative effect of perceived strategic generosity is more pronounced for individuals who have a larger follower base. Further, we observe a systematically larger effect from actual resale, versus mere relisting.

**Table E1. Moderating Effect of Social Exposure**

| | LnNumSales<br>(1) | LnTotRev<br>(2) | LnNumSales<br>(3) | LnTotRev<br>(4) | LnNumSales<br>(5) | LnTotRev<br>(6) |
|---|---|---|---|---|---|---|
| TreatedNoCharityListing | 0.090* | 0.047* | 0.081 | 0.041 | 0.203*** | 0.090*** |
| | (0.052) | (0.027) | (0.049) | (0.025) | (0.052) | (0.033) |
| TreatedCharityListingNoSold | -0.160** | -0.136*** | -0.159*** | -0.135*** | -0.183*** | -0.169*** |
| | (0.062) | (0.036) | (0.060) | (0.034) | (0.064) | (0.042) |
| TreatedCharitySold | -0.091 | -0.235*** | -0.031 | -0.193* | -0.141 | -0.247** |
| | (0.156) | (0.088) | (0.170) | (0.103) | (0.176) | (0.116) |
| FollowersCum | 0.036 | -0.001 | 0.067 | 0.021 | 0.021 | 0.000 |
| | (0.075) | (0.049) | (0.072) | (0.046) | (0.067) | (0.047) |
| TreatedNoCharityListing × FollowersCum | -0.082 | -0.038 | -0.084 | -0.039 | -0.050 | -0.019 |
| | (0.069) | (0.042) | (0.066) | (0.040) | (0.060) | (0.039) |
| TreatedCharityListingNoSold × FollowersCum | -0.007 | -0.014* | 0.000 | -0.009 | -0.004 | -0.012* |
| | (0.012) | (0.008) | (0.011) | (0.007) | (0.010) | (0.007) |
| TreatedCharitySold × FollowersCum | -1.211*** | -0.635*** | -1.082*** | -0.545*** | -0.901*** | -0.520*** |
| | (0.151) | (0.094) | (0.142) | (0.107) | (0.148) | (0.133) |
| LnTenure | | | 0.471*** | 0.329*** | 0.173** | 0.246*** |
| | | | (0.049) | (0.030) | (0.073) | (0.049) |
| LagLnNumSales | | | | | 0.190*** | 0.066*** |
| | | | | | (0.030) | (0.017) |
| LagLnNumListings | | | | | 0.137*** | 0.085*** |
| | | | | | (0.023) | (0.014) |
| Individual FE | Yes | Yes | Yes | Yes | Yes | Yes |
| Year-month FE | Yes | Yes | Yes | Yes | Yes | Yes |
| Observations | 16,780 | 16,780 | 16,780 | 16,780 | 15,206 | 15,206 |
| Num of Individuals | 1,574 | 1,574 | 1,574 | 1,574 | 1,574 | 1,574 |
| R-squared | 0.483 | 0.458 | 0.492 | 0.472 | 0.561 | 0.529 |

*Note: Robust standard errors clustered at the individual level. * p<0.1, ** p<0.05, *** p<0.01.*

Given that market punishment for perceived strategic generosity is likely influenced by the visibility of the charity NFT involved, we next explore the extent to which the negative effect of relisting a charity NFT depends on the popularity of the relisted charity NFTs. To measure popularity, we use the number of Likes received by the charity NFT on Rarible. Specifically, we estimate whether individuals who relist more popular charity NFTs are subject to greater market punishment. For ease of interpretation, we use a demeaned log-transformed number of Likes (denoted as *LnUKRNFTLikes*). Since we cannot identify the specific charity NFT that individuals in the control group attempted to purchase, *LnUKRNFTLikes* is set to zero for those individuals. The results in Table E2, reveal that relisting a charity

NFT negatively affects the sales performance of individuals' other NFTs. Additionally, while the popularity of the charity NFT (as indicated by Likes) generally has a positive impact on boosting the sales of individuals' other NFTs, relisting such popular NFTs leads to a significant negative effect. This finding suggests that the marginal effect of the charity NFT's popularity turns significantly negative when these popular NFTs are relisted, underscoring the potential backfire effect when generosity is perceived as strategic.

**Table E2. HTE by UKR NFT Popularity**

| | LnNumSales<br>(1) | LnTotRev<br>(2) | LnNumSales<br>(3) | LnTotRev<br>(4) | LnNumSales<br>(5) | LnTotRev<br>(6) |
|---|---|---|---|---|---|---|
| Treated | 0.075 | 0.040 | 0.066 | 0.033 | 0.188*** | 0.082** |
| | (0.052) | (0.027) | (0.050) | (0.025) | (0.053) | (0.033) |
| Treated × UKRListing | -0.246*** | -0.194*** | -0.227*** | -0.180*** | -0.376*** | -0.265*** |
| | (0.060) | (0.034) | (0.058) | (0.033) | (0.060) | (0.040) |
| Treated × LnUKRNFTLikes | 0.369*** | 0.183*** | 0.368*** | 0.182*** | 0.361*** | 0.190** |
| | (0.111) | (0.070) | (0.106) | (0.066) | (0.114) | (0.086) |
| Treated × UKRListing × LnUKRNFTLikes | -0.507** | -0.327** | -0.453** | -0.288** | -0.614*** | -0.429*** |
| | (0.228) | (0.139) | (0.215) | (0.129) | (0.219) | (0.158) |
| LnTenure | | | 0.472*** | 0.338*** | 0.189** | 0.267*** |
| | | | (0.051) | (0.031) | (0.074) | (0.049) |
| LagLnNumSales | | | | | 0.191*** | 0.066*** |
| | | | | | (0.030) | (0.017) |
| LagLnNumListings | | | | | 0.137*** | 0.085*** |
| | | | | | (0.023) | (0.014) |
| Individual FE | Yes | Yes | Yes | Yes | Yes | Yes |
| Year-month FE | Yes | Yes | Yes | Yes | Yes | Yes |
| Observations | 16,780 | 16,780 | 16,780 | 16,780 | 15,206 | 15,206 |
| Num of Individuals | 1,574 | 1,574 | 1,574 | 1,574 | 1,574 | 1,574 |
| R-squared | 0.481 | 0.455 | 0.491 | 0.471 | 0.560 | 0.528 |

*Note: Robust standard errors clustered at the individual level. * p<0.1, ** p<0.05, *** p<0.01. Since we can observe only a single instance of the number of likes received by each charity NFT, LnUKRNFTLikes is a time-invariant measure. As a result, its main effect cannot be identified.*

## Appendix F. Additional Analyses Ruling Out NFT Listing Behavior as an Alternative Explanation

To rule out listing behavior as a confound, we restrict the analysis to NFTs that individuals had listed before the charity fundraiser and did not subsequently modify. The results of this analysis are reported in Table F1. We consistently find that a negative treatment effect only arose for individuals who purchased a charity NFT and sought to resell it; that is, individuals who purchased yet refrained from relisting the charity NFT benefited from its purchase.

**Table F1. Effect on the Sales Performance of NFTs Listed Prior to the NFT Charity Fundraiser**

| | LnNumSales<br>(1) | LnTotRev<br>(2) | LnNumSales<br>(3) | LnTotRev<br>(4) | LnNumSales<br>(5) | LnTotRev<br>(6) |
|---|---|---|---|---|---|---|
| TreatedNoCharityListing | 0.201*** | 0.093*** | 0.193*** | 0.088*** | 0.252*** | 0.104*** |
| | (0.039) | (0.026) | (0.037) | (0.025) | (0.049) | (0.033) |
| TreatedCharityListingNoSold | -0.134*** | -0.112*** | -0.128*** | -0.108*** | -0.119** | -0.118*** |
| | (0.047) | (0.032) | (0.044) | (0.030) | (0.056) | (0.040) |
| TreatedCharitySold | -0.405*** | -0.318*** | -0.324*** | -0.266*** | -0.387** | -0.297** |
| | (0.121) | (0.086) | (0.123) | (0.092) | (0.153) | (0.122) |
| LnTenure | | | 0.431*** | 0.276*** | 0.330*** | 0.205*** |
| | | | (0.036) | (0.024) | (0.046) | (0.030) |
| LagLnNumSales | | | | | 0.291*** | 0.200*** |
| | | | | | (0.020) | (0.017) |
| LagLnNumListings | | | | | 0.000 | -0.005** |
| | | | | | (0.005) | (0.003) |
| Individual FE | Yes | Yes | Yes | Yes | Yes | Yes |
| Year-month FE | Yes | Yes | Yes | Yes | Yes | Yes |
| Observations | 16,780 | 16,780 | 16,780 | 16,780 | 15,206 | 15,206 |
| Num of Individuals | 1,574 | 1,574 | 1,574 | 1,574 | 1,574 | 1,574 |
| R-squared | 0.386 | 0.327 | 0.421 | 0.361 | 0.514 | 0.446 |

*Note: Robust standard errors clustered at the individual level. * $p<0.1$, ** $p<0.05$, *** $p<0.01$.*

Second, we re-estimate our regressions conditioning on the volume of NFTs an individual had listed in a given month, as well as their listing prices. Our findings, reported in Table F2, suggest that the negative treatment effect persists, ruling out individuals' listing behavior as an explanation for our findings.

**Table F2. Treatment Effect Conditional on Individuals' NFT Listing Behavior**

| | LnNumSales (1) | LnTotRev (2) | LnNumSales (3) | LnTotRev (4) | LnNumSales (5) | LnTotRev (6) |
|---|---|---|---|---|---|---|
| TreatedNoCharityListing | 0.072 | 0.039 | 0.039 | 0.022 | 0.039 | 0.016 |
| | (0.053) | (0.028) | (0.024) | (0.017) | (0.039) | (0.017) |
| TreatedCharityListingNoSold | -0.162*** | -0.143*** | -0.067** | -0.100*** | -0.082* | -0.088*** |
| | (0.063) | (0.036) | (0.031) | (0.024) | (0.047) | (0.024) |
| TreatedCharitySold | -0.379* | -0.384*** | -0.108 | -0.242*** | -0.128 | -0.209*** |
| | (0.213) | (0.114) | (0.080) | (0.070) | (0.147) | (0.062) |
| LnTenure | | | 0.141*** | 0.193*** | 0.257*** | 0.189*** |
| | | | (0.026) | (0.020) | (0.047) | (0.027) |
| LnNumListings | | | 0.689*** | 0.300*** | | |
| | | | (0.009) | (0.008) | | |
| LnListingPrices | | | | | 0.552*** | 0.379*** |
| | | | | | (0.068) | (0.047) |
| Individual FE | Yes | Yes | Yes | Yes | Yes | Yes |
| Year-month FE | Yes | Yes | Yes | Yes | Yes | Yes |
| Observations | 16,780 | 16,780 | 16,780 | 16,780 | 15,206 | 15,206 |
| Num of Individuals | 1,574 | 1,574 | 1,574 | 1,574 | 1,574 | 1,574 |
| R-squared | 0.481 | 0.455 | 0.797 | 0.663 | 0.619 | 0.672 |

*Note: Robust standard errors clustered at the individual level. * p<0.1, ** p<0.05, *** p<0.01*

## Appendix G. Survival Analysis

Furthermore, we consider an alternative estimation approach. Specifically, we employ a listing-level specification, estimating a Cox proportional hazard model (Cox model thereafter) (Cox 1972) that models the hazard of other NFTs being sold as a function of the individual engaging in behavior reflective of strategic generosity. We specify our hazard function, $\lambda_{ij}(t)$, at time $t$ as follows:

$$\lambda_{ij}(t) = \lambda_0(t) \cdot exp\left(\beta \cdot x_{ijt} + \mu \cdot TreatStatus_{it}\right) \tag{G1}$$

where $\lambda_0(t)$ is the nonparametric baseline hazard rate at a given point in time, $t$. In Equation (G1), $i$ indexes individuals and $j$ indexes a given (non-charity) NFT, listed by individual $i$. The time-varying covariate $x_{ijt}$ denotes the listing price of NFT $j$ associated with individual $i$ at time $t$, while $TreatStatus_{it}$ denotes the treatment status of individual $i$ at that time.

Table G1 presents the estimates from this Cox model. We find that the coefficients for $TreatedCharityListingNoSold$ and $TreatedCharitySold$ are significantly negative, indicating that perceived strategic generosity reduces the hazard of NFT sale, i.e., NFT listings take longer to sell for these individuals.

**Table G1. Survival Analysis of NFT Listings (Hazard of Sale)**

| | Cox proportional hazard estimates | |
|---|---|---|
| | (1) | (2) |
| TreatedNoCharityListing | 0.465*** | 0.469*** |
| | (0.019) | (0.019) |
| TreatedCharityListingNoSold | -0.071** | -0.071** |
| | (0.015) | (0.016) |
| TreatedCharitySold | -1.225*** | -1.225*** |
| | (0.075) | (0.075) |
| LnNFTListPrice | -0.442*** | -0.443*** |
| | (0.021) | (0.021) |
| LnTenure | | 0.017 |
| | | (0.011) |
| Observations | 101,403 | 101,403 |
| Log-Likelihood | -232,214.600 | -232,213.300 |

*Notes: Robust standard errors clustered by NFT-listing are reported in parentheses. * p<0.1, ** p<0.05, *** p<0.01.*

Conversely, the coefficient associated with $TreatedNoCharityListing$ is positive and significant, again suggesting that individuals who purchase a charity NFT and who refrain from relisting it begin to experience more rapid NFT sales. Bearing in mind that the estimations condition on the NFT listing price, our results are once again attributable only to changes in the demand for an individual's listed NFTs, not changes in an individual's listing behavior.

## Appendix H. Instrumental Variable Regression Results

We report the results of the IV regression in Table H1. In the first stage IV regression, we find strong evidence of the instrument's relevance. The interaction between the log-transformed processing time and the *After* dummy, i.e., the instrument, is significantly and negatively correlated with the *Treated* dummy (coefficient = -0.304, *p*-value = 0.000). This aligns with our expectation that the longer the processing time for a charity NFT purchase request, the lower the likelihood of an individual successfully acquiring an NFT (being treated). Further, the Kleibergen-Paap rk Wald F statistic exceeds the critical values of the Stock-Yogo weak instrument test (Stock and Yogo 2005), providing additional evidence against the presence of a weak instrument problem. The results of the second stage are consistent with our main results again; the treatment leads to a statistically significant negative impact on revenue from other NFTs.

**Table H1. Instrumental Variable Regression**

| | Treated<br>(1) | LnNumSales<br>(2) | LnTotRev<br>(3) | Treated<br>(4) | LnNumSales<br>(5) | LnTotRev<br>(6) |
|---|---|---|---|---|---|---|
| LnTransProcessDur | -0.304***<br>(0.008) | | | -0.304***<br>(0.008) | | |
| Treated | | -0.113**<br>(0.057) | -0.102***<br>(0.033) | | -0.098*<br>(0.055) | -0.091***<br>(0.031) |
| LnTenure | | | | 0.025**<br>(0.012) | 0.486***<br>(0.052) | 0.347***<br>(0.032) |
| Individual FE | Yes | Yes | Yes | Yes | Yes | Yes |
| Year-month FE | Yes | Yes | Yes | Yes | Yes | Yes |
| Observations | 16,397 | 16,397 | 16,397 | 16,397 | 16,397 | 16,397 |
| Num of Individuals | 1,535 | 1,535 | 1,535 | 1,535 | 1,535 | 1,535 |
| Kleibergen-Paap rk Wald F statistic | | 1450 | 1450 | | 1459 | 1459 |
| R-squared | 0.817 | 0.085 | 0.112 | 0.818 | 0.103 | 0.139 |

*Note: Results are highly consistent if the full sample is considered. Robust standard errors clustered at the individual level. * p<0.1, ** p<0.05, *** p<0.01.*

## Appendix I. Replication with Another Charity NFT Collection

To demonstrate the generalizability of our findings on the market punishment of perceived strategic generosity due to the transparent nature of the blockchain, we conducted a replication study using another charity NFT collection: the SHAQ Gives Back NFT collection (referred to as SHAQ NFT).[26]

There are several advantages to using the SHAQ NFT dataset for this replication test, as it shares key similarities with the UKR NFT collection used in our main analysis. First, the SHAQ NFT collection is a charity NFT series sold at a fixed price (0.05 Ether), with its initial sale also occurring within a short time window. Second, similar to the UKR NFT collection, some individuals attempted to purchase SHAQ NFTs but failed during the initial sale. This allows us to apply a comparable identification strategy, where the treatment group consists of individuals who successfully purchased the NFTs, and the control group consists of those who attempted but failed. However, the SHAQ dataset presents a challenge in that the number of individuals who failed to purchase is remarkably smaller than the number of successful buyers. As a result, traditional matching approaches struggle to find individuals in the control group closely matched to the treatment group.

Given this limitation, our analysis mainly focuses on results from the full sample. We further augment these results with the MC method to strengthen the causal inference. As shown in Table I1, we find that while the overall treatment effect of purchasing SHAQ NFTs or relisting them for sale (denoted as *SHAQListing*) does not have a significant negative effect on the sales performance of other NFTs in individuals' portfolios, there is a consistent a significant negative effect on sales performance when individuals list the charity NFT for sale at a price higher than the original sale price (denoted by *HighListPrice*) shown in Column 6 of Table I1. The interaction term *Treated* × *SHAQListing* is negative but statistically insignificant in Column 4 of Table H1, this is likely due to the relatively low proportion of individuals who listed SHAQ NFTs for sale at prices higher than the original sale price. This contrasts with the UKR NFT dataset, where a higher percentage of individuals listed their NFTs for sale at prices above the original sale price.

---

[26] The SHAQ NFT collection, which can be found at https://rarible.com/shaq-gives-back/items. The collection description clearly states that "100% of the proceeds of the SHAQ GIVES BACK NFT sale & **ALL REVENUE** from Opensea creator fee will go to The Shaquille O'Neal Foundation…the Foundation creates pathways for underserved youth, to help them achieve their full potential. The Foundation works to instill hope and bring about change in communities, collectively shaping a brighter future for our children."

**Table I1. Average Treatment Effect Based on an Alternative Charity NFT Collection**

| | LnNumSales (1) | LnTotRev (2) | LnNumSales (3) | LnTotRev (4) | LnNumSales (5) | LnTotRev (6) |
|---|---|---|---|---|---|---|
| Treated | -0.074 | 0.033 | 0.072 | 0.132* | -0.055 | 0.056 |
| | (0.089) | (0.063) | (0.116) | (0.076) | (0.095) | (0.066) |
| Treated × SHAQListing | | | -0.057 | -0.071 | -0.034 | -0.042 |
| | | | (0.084) | (0.060) | (0.089) | (0.063) |
| Treated × SHAQListing ×HighListPrice | | | | | -0.255* | -0.304*** |
| | | | | | (0.135) | (0.093) |
| Individual FE | Yes | Yes | Yes | Yes | Yes | Yes |
| Year-month FE | Yes | Yes | Yes | Yes | Yes | Yes |
| Observations | 14,357 | 14,357 | 14,357 | 14,357 | 14,357 | 14,357 |
| Num of Individuals | 1,136 | 1,136 | 1,136 | 1,136 | 1,136 | 1,136 |
| R-squared | 0.682 | 0.673 | 0.677 | 0.671 | 0.682 | 0.673 |

*Note: Robust standard errors clustered at the individual level. * p<0.1, ** p<0.05, *** p<0.01.*

We further assess the impact of relisting a charity NFT at a high versus low price using the MC method. As shown in Table I2, relisting a charity NFT leads to a stronger negative effect on individuals' NFT sales performance, both in magnitude and significance, when the relist price is high. Additionally, Figures I1 and I2 provide supporting evidence for the parallel trends assumption and demonstrate the sustained market punishment for perceived strategic generosity.

**Table I2. ATT Estimate Based on the MC Method**

| | LnNumSales (1) | LnTotRev (2) |
|---|---|---|
| ATT of relisting a charity NFT at a high price | -0.496*** | -0.375*** |
| | (-0.785, -0.188) | (-0.555, -0.186) |
| ATT of relisting a charity NFT at a low price | -0.247*** | -0.126* |
| | (-0.419, -0.084) | (-0.256, -0.001) |

*Note:* Error bars denote bootstrapped 95% confidence intervals.

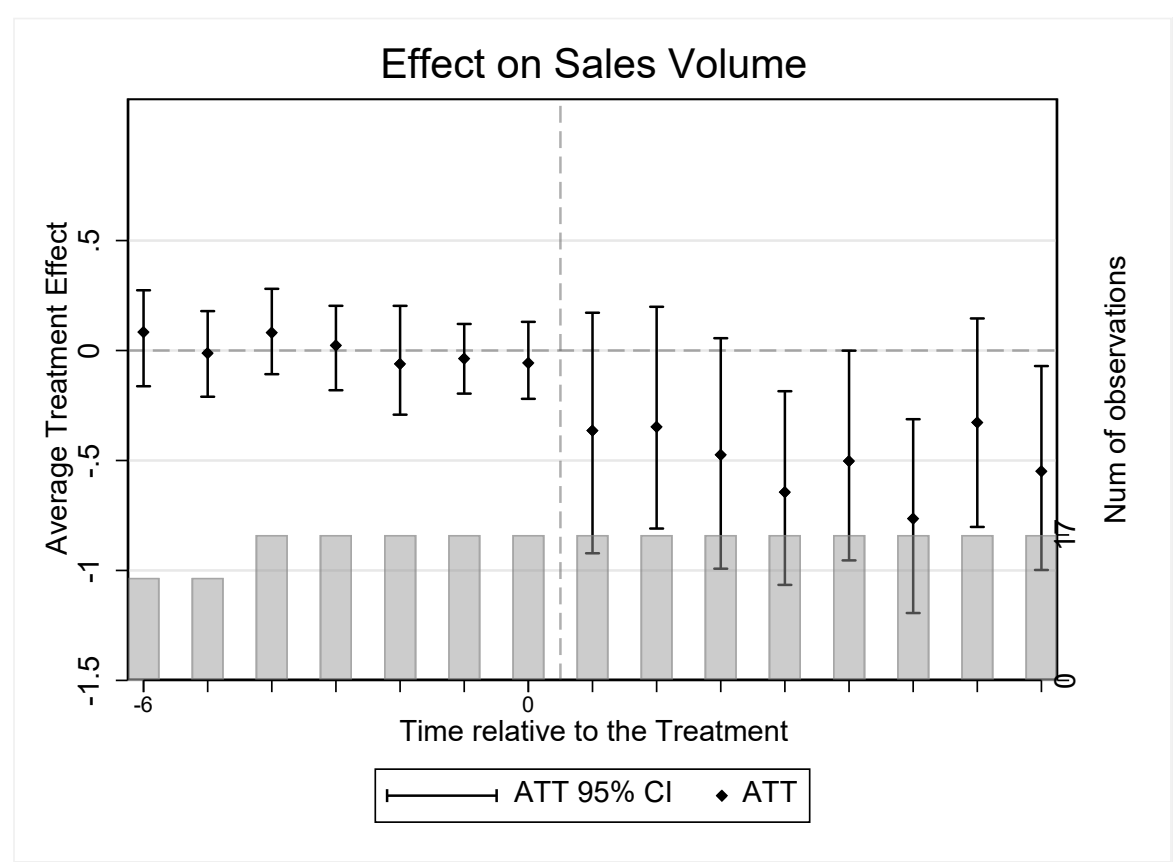

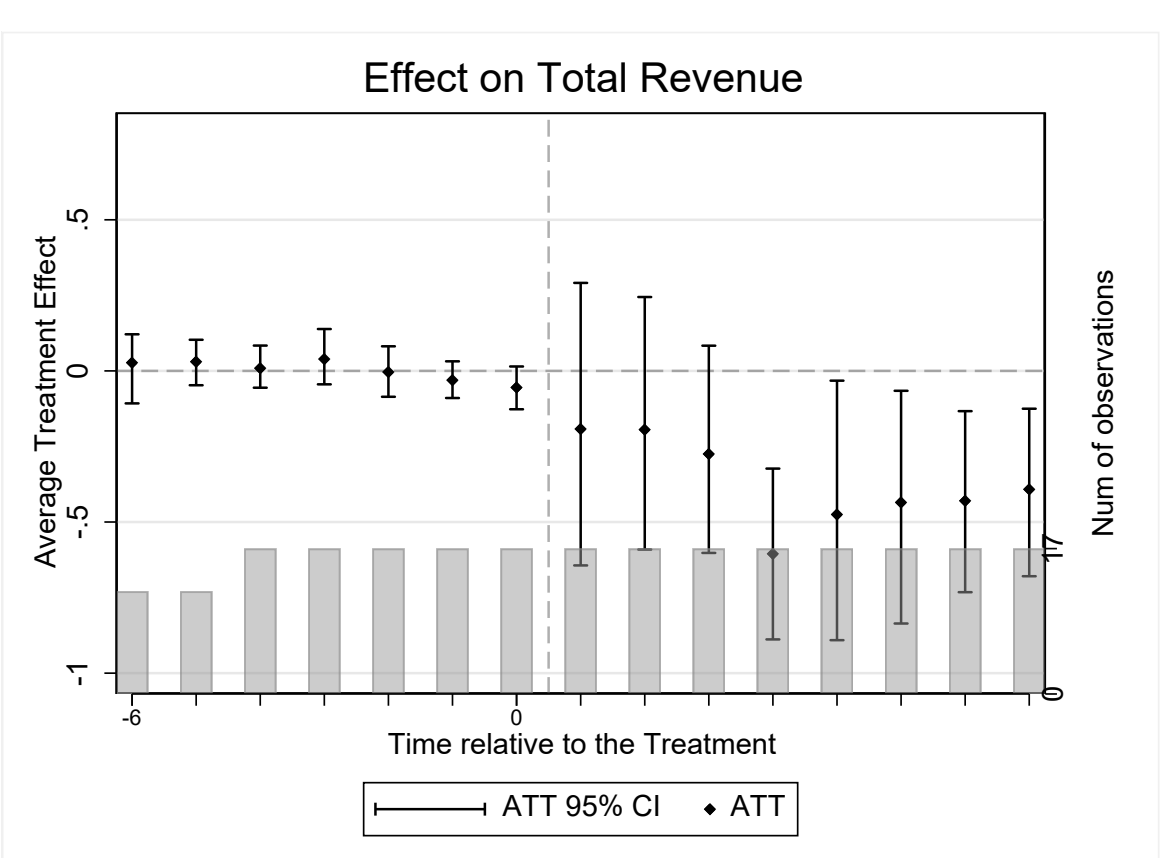

**Figure I1. Dynamic Treatment Effects of Relisting a Charity NFT at a High Price**

Notes: Both plots presented were produced utilizing the "fect" package in Stata. Error bars denote bootstrapped 95% confidence intervals.

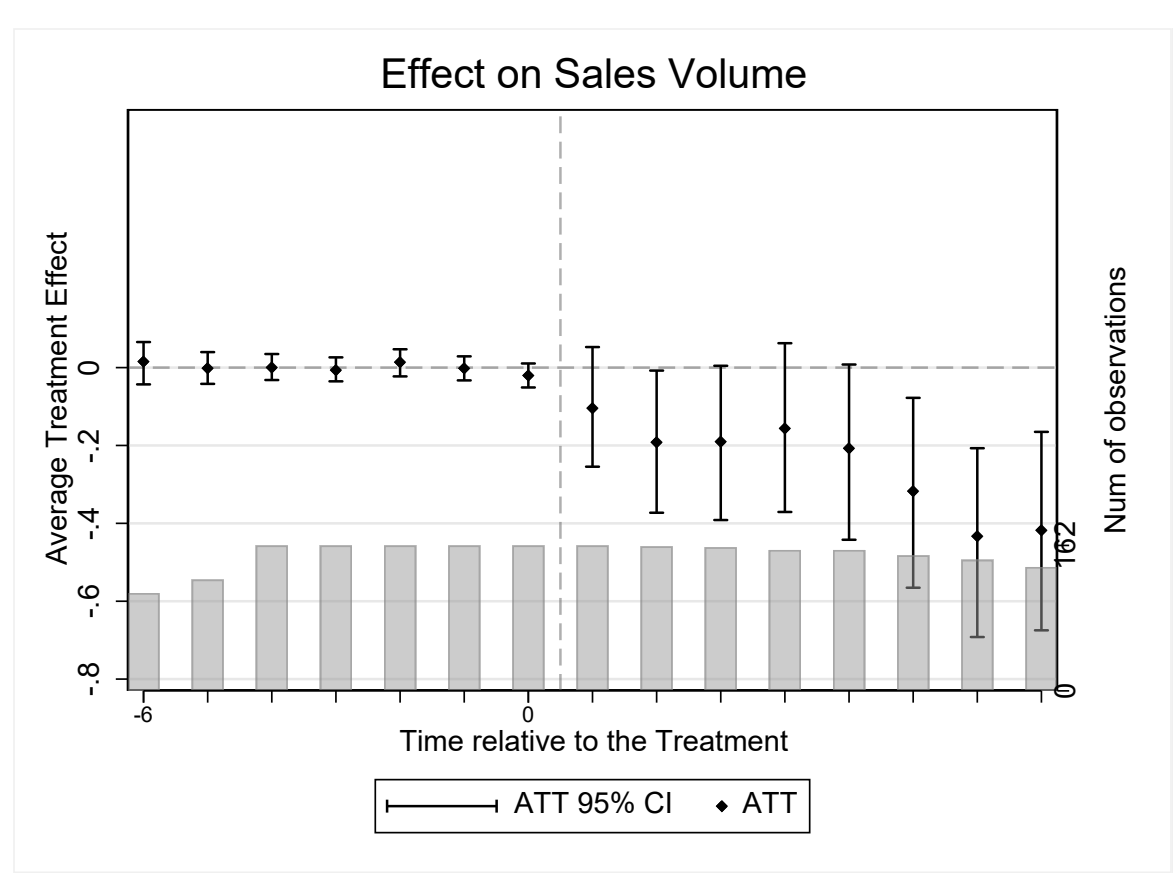

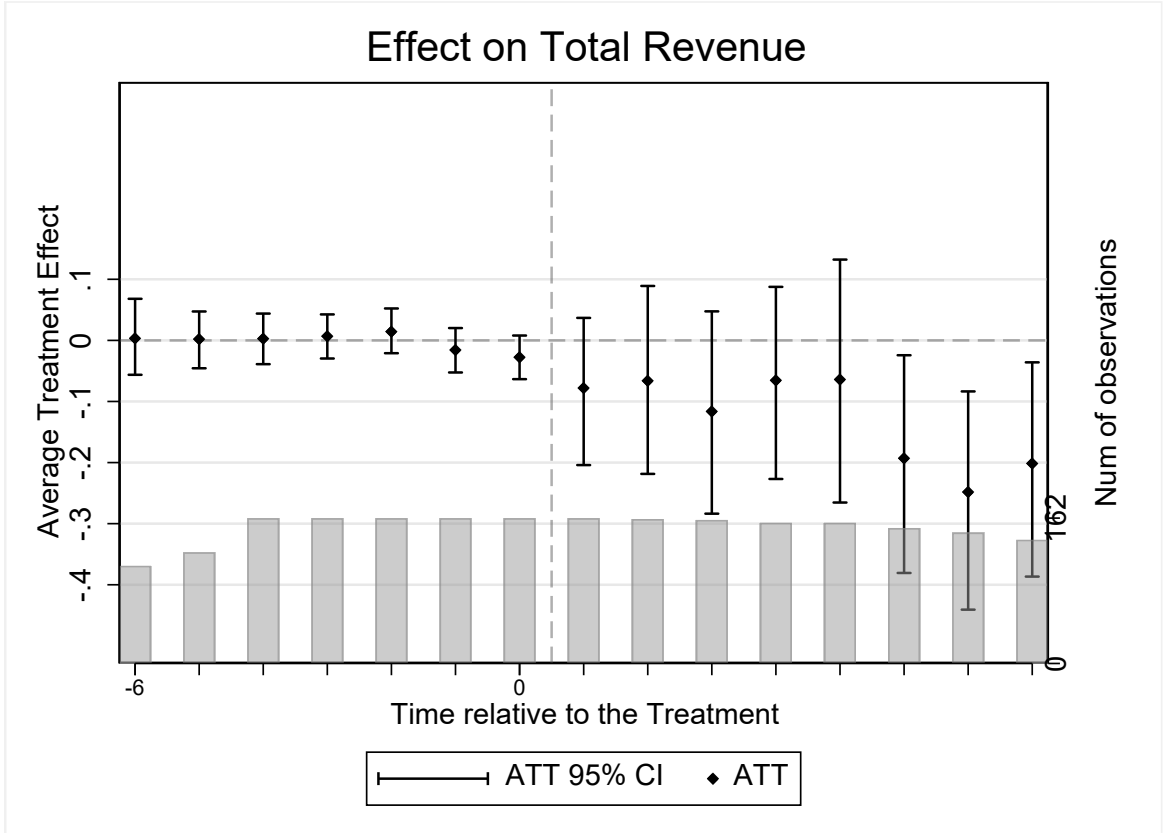

**Figure I2. Dynamic Treatment Effects of Relisting a Charity NFT at a Low Price**

Notes: Both plots presented were produced utilizing the "fect" package in Stata. Error bars denote bootstrapped 95% confidence intervals.

Taken in sum, results consistently suggest that the market penalizes individuals perceived as engaging in strategic generosity. This pattern is not unique to the UKR NFT context but represents a broader phenomenon that applies to other charity NFT collections as well, such as the SHAQ NFT collection. These findings highlight the unintended consequences of blockchain transparency, wherein individuals attempting to profit from charity-related NFTs may face market punishment due to the high visibility of their actions.

## Appendix J. Falsification Test with a Non-Charity NFT Collection

To further support our conclusion regarding market punishment for perceived strategic generosity, we conduct a falsification test using data from individuals who purchased a high-profile non-charity NFT. Specifically, we collected data on all individuals who purchased Porsche NFTs, totaling 1,405 individuals, of whom 502 have ever relisted these NFTs for sale. In this setting, we were unable to identify a control group of individuals who attempted to purchase Porsche NFTs but failed. Therefore, our analysis focuses on whether relisting a high-profile non-charity NFT, such as a Porsche NFT, would result in a similar negative impact on the sales performance of other NFTs in individuals' portfolios.

To this end, we introduce a time-varying binary variable, *PorscheListing*, to capture the listing status, where *PorscheListing* = 1 indicates that a user has relisted their Porsche NFT for sale, and *PorscheListing* = 0 otherwise. As shown in Tables J1 and J2, our results indicate that relisting Porsche NFTs for sale has a short-term positive effect on the sales of other NFTs in individuals' portfolios. This effect is likely driven by the high-profile nature of Porsche NFTs, which may attract more viewers to the individuals' portfolios, increasing exposure and demand for their other NFTs. These findings further emphasize the contrast between the market response to high-profile non-charity NFTs and charity NFTs, supporting the conclusion that the negative market penalties observed in our main analysis are associated with strategic generosity perceived in the context of charity NFTs.

**Table J1. Average Treatment Effect (Full Sample)**

| | LnNumSales | LnTotRev | LnNumSales | LnTotRev |
|---|---|---|---|---|
| | (1) | (2) | (5) | (6) |
| PorscheListing | 0.168*** | 0.080*** | 0.066*** | 0.028* |
| | (0.028) | (0.020) | (0.018) | (0.016) |
| LagLnNumSales | | | 0.240*** | 0.127*** |
| | | | (0.023) | (0.017) |
| LagLnNumListings | | | 0.172*** | 0.088*** |
| | | | (0.015) | (0.012) |
| Individual FE | Yes | Yes | Yes | Yes |
| Year-month FE | Yes | Yes | Yes | Yes |
| Observations | 21,075 | 21,075 | 19,670 | 19,670 |
| Num of Individuals | 1,405 | 1,405 | 1,405 | 1,405 |
| R-squared | 0.578 | 0.674 | 0.727 | 0.610 |

*Note: Robust standard errors clustered at the individual level. * $p<0.1$, ** $p<0.05$, *** $p<0.01$.*

**Table J2. Relative Time Estimates of the Impact of Relisting Non-Charity NFTs for Sales (Full Sample)**

| | LnNumSales<br>(1) | LnTotRev<br>(2) | LnNumSales<br>(3) | LnTotRev<br>(4) |
|---|---|---|---|---|
| Relative time -7 | -0.177*** | -0.085*** | | |
| | (0.045) | (0.031) | | |
| Relative time -6 | -0.138*** | -0.040 | -0.071* | -0.005 |
| | (0.045) | (0.033) | (0.040) | (0.031) |
| Relative time -5 | -0.144*** | -0.064* | -0.093** | -0.037 |
| | (0.044) | (0.034) | (0.040) | (0.032) |
| Relative time -4 | -0.116*** | -0.074** | -0.062 | -0.046 |
| | (0.044) | (0.032) | (0.040) | (0.031) |
| Relative time -3 | -0.107*** | -0.057* | -0.066 | -0.036 |
| | (0.041) | (0.031) | (0.042) | (0.031) |
| Relative time -2 | -0.047 | -0.029 | -0.008 | -0.009 |
| | (0.036) | (0.030) | (0.042) | (0.031) |
| Relative time -1 | Baseline (omitted) | | | |
| Relative time 0 | 0.115*** | 0.099** | 0.089** | 0.085** |
| | (0.040) | (0.039) | (0.045) | (0.040) |
| Relative time 1 | 0.164*** | 0.100** | 0.081 | 0.057 |
| | (0.051) | (0.043) | (0.051) | (0.042) |
| Relative time 2 | 0.110** | 0.045 | 0.028 | 0.002 |
| | (0.055) | (0.041) | (0.054) | (0.041) |
| Relative time 3 | 0.082 | -0.015 | 0.006 | -0.054 |
| | (0.050) | (0.038) | (0.048) | (0.036) |
| Relative time 4 | 0.010 | -0.018 | -0.029 | -0.039 |
| | (0.052) | (0.039) | (0.047) | (0.036) |
| Relative time 5 | -0.031 | -0.030 | -0.059 | -0.045 |
| | (0.053) | (0.036) | (0.044) | (0.033) |
| Relative time 6 | -0.042 | -0.059 | -0.043 | -0.059* |
| | (0.054) | (0.037) | (0.046) | (0.034) |
| Relative time 7 | 0.028 | 0.024 | 0.036 | 0.028 |
| | (0.058) | (0.045) | (0.048) | (0.042) |
| LagLnNumSales | | | 0.240*** | 0.128*** |
| | | | (0.023) | (0.017) |
| LagLnNumListed | | | 0.171*** | 0.088*** |
| | | | (0.015) | (0.012) |
| Individual FE | Yes | Yes | Yes | Yes |
| Year-month FE | Yes | Yes | Yes | Yes |
| Observations | 21,075 | 21,075 | 19,670 | 19,670 |
| Num of Individuals | 1,405 | 1,405 | 1,405 | 1,405 |
| R-squared | 0.675 | 0.579 | 0.727 | 0.611 |

*Note: Robust standard errors clustered at the individual level. * p<0.1, ** p<0.05, *** p<0.01.*

We have also employed a similar CEM method as introduced in Section 4.2.1, matching individuals based on their NFT-related activities (i.e., purchases, sales, mints, listings, and transfers). As shown in Table J3, the matching process ensures that both groups are balanced across these different aspects. The results, reported in Tables J4 and J5, are highly consistent with our previous findings, reinforcing the short-term positive effect of relisting Porsche NFTs on the sales performance of other NFTs in individuals' portfolios. Consistent results are observed using an alternative matching approach (PSM). Details are omitted for brevity and are available upon request.

**Table J3. Balance Check for the CEM Sample**

| | Mean | | | t-test | | |
|---|---|---|---|---|---|---|
| Variable | Treated | Control | %bias | t | *p*-value | V(T)/V(C) |
| LnAvgMints | 0.931 | 0.931 | 0.000 | 0.010 | 0.995 | 0.980 |
| LnAvgPurchases | 0.276 | 0.281 | -0.900 | -0.120 | 0.907 | 0.990 |
| LnAvgSales | 0.085 | 0.078 | 3.200 | 0.410 | 0.682 | 1.040 |
| LnAvgListing | 0.223 | 0.191 | 7.900 | 1.010 | 0.313 | 1.080 |
| LnAvgTransfers | 0.959 | 0.967 | -0.800 | -0.120 | 0.908 | 1.000 |

**Table J4. Average Treatment Effect on the CEM Sample**

| | LnNumSales | LnTotRev | LnNumSales | LnTotRev |
|---|---|---|---|---|
| | (1) | (2) | (5) | (6) |
| ListPorscheNFT | 0.118*** | 0.052*** | 0.072*** | 0.028** |
| | (0.021) | (0.014) | (0.017) | (0.013) |
| LagLnNumSales | | | 0.076* | 0.032 |
| | | | (0.039) | (0.024) |
| LagLnNumListings | | | 0.130*** | 0.065*** |
| | | | (0.023) | (0.016) |
| Individual FE | Yes | Yes | Yes | Yes |
| Year-month FE | Yes | Yes | Yes | Yes |
| Observations | 14,250 | 14,250 | 13,300 | 13,300 |
| Num of Individuals | 950 | 950 | 950 | 950 |
| R-squared | 0.308 | 0.247 | 0.351 | 0.279 |

*Note: Robust standard errors clustered at the individual level. * p<0.1, ** p<0.05, *** p<0.01.*

**Table J5. Relative Time Estimates of the Impact of Relisting Non-Charity NFTs for Sales (CEM Sample)**

| | LnNumSales | LnTotRev | LnNumSales | LnTotRev |
|---|---|---|---|---|
| | (1) | (2) | (3) | (4) |
| Relative time -7 | -0.089*** | -0.074*** | | |
| | (0.030) | (0.027) | | |
| Relative time -6 | -0.048 | -0.045 | -0.033 | -0.038 |
| | (0.032) | (0.027) | (0.030) | (0.026) |
| Relative time -5 | -0.067** | -0.056** | -0.056* | -0.050* |
| | (0.030) | (0.027) | (0.029) | (0.026) |
| Relative time -4 | -0.068** | -0.047 | -0.056** | -0.041 |
| | (0.028) | (0.030) | (0.027) | (0.030) |
| Relative time -3 | -0.041 | -0.051* | -0.032 | -0.046* |
| | (0.029) | (0.027) | (0.030) | (0.028) |
| Relative time -2 | -0.019 | -0.039 | -0.010 | -0.035 |
| | (0.032) | (0.030) | (0.034) | (0.030) |
| Relative time -1 | Baseline (omitted) | | | |
| Relative time 0 | 0.065* | 0.018 | 0.060 | 0.016 |
| | (0.037) | (0.032) | (0.039) | (0.032) |
| Relative time 1 | 0.145*** | 0.075* | 0.113** | 0.060 |
| | (0.054) | (0.042) | (0.052) | (0.041) |
| Relative time 2 | 0.103** | 0.019 | 0.063 | -0.000 |
| | (0.052) | (0.036) | (0.050) | (0.035) |
| Relative time 3 | 0.037 | -0.008 | -0.002 | -0.026 |
| | (0.040) | (0.032) | (0.038) | (0.031) |
| Relative time 4 | 0.052 | -0.030 | 0.029 | -0.041 |
| | (0.038) | (0.028) | (0.037) | (0.028) |
| Relative time 5 | 0.028 | -0.024 | -0.002 | -0.038 |
| | (0.039) | (0.027) | (0.037) | (0.026) |
| Relative time 6 | 0.004 | -0.023 | -0.023 | -0.036 |
| | (0.033) | (0.030) | (0.031) | (0.030) |
| Relative time 7 | 0.015 | -0.012 | -0.002 | -0.020 |
| | (0.038) | (0.030) | (0.037) | (0.030) |
| LagLnNumSales | | | 0.076* | 0.031 |
| | | | (0.039) | (0.024) |
| LagLnNumListed | | | 0.132*** | 0.065*** |
| | | | (0.023) | (0.016) |
| Individual FE | Yes | Yes | Yes | Yes |
| Year-month FE | Yes | Yes | Yes | Yes |
| Observations | 14,250 | 14,250 | 13,300 | 13,300 |
| Num of Individuals | 950 | 950 | 950 | 950 |
| R-squared | 0.309 | 0.249 | 0.352 | 0.281 |

*Note: Robust standard errors clustered at the individual level. * p<0.1, ** p<0.05, *** p<0.01.*

In the analysis of both the full sample and the CEM sample, we consistently find that relisting Porsche NFTs for sale has a significant *positive* impact on the sales performance of other NFTs in individuals' portfolios. Additionally, we find supporting evidence for the parallel trends assumption in this DID setting, and the results remain consistent when applying the MC method as an alternative causal inference approach (see Figure J1). This contrast reinforces our conclusion that the negative market response observed in our main analysis is specifically driven by the perception of strategic generosity associated with charity NFTs, rather than being a general reaction to the relisting of high-profile NFTs.

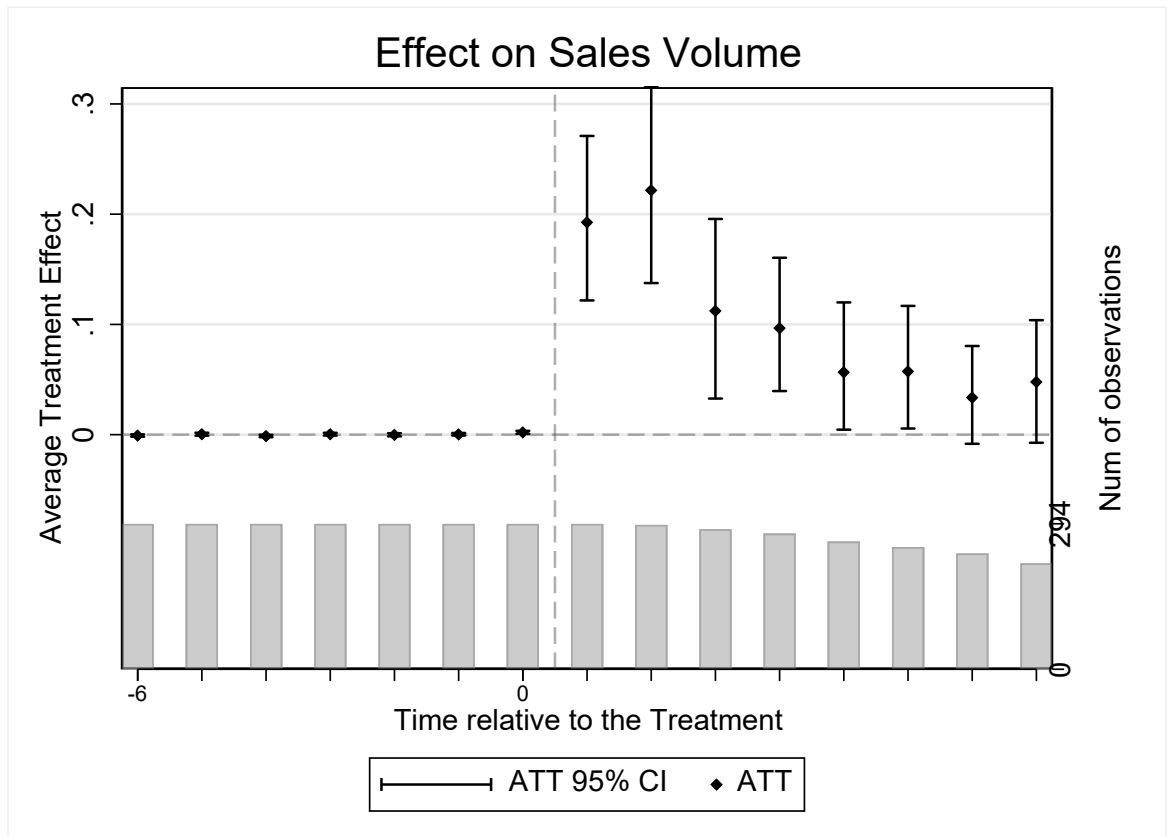

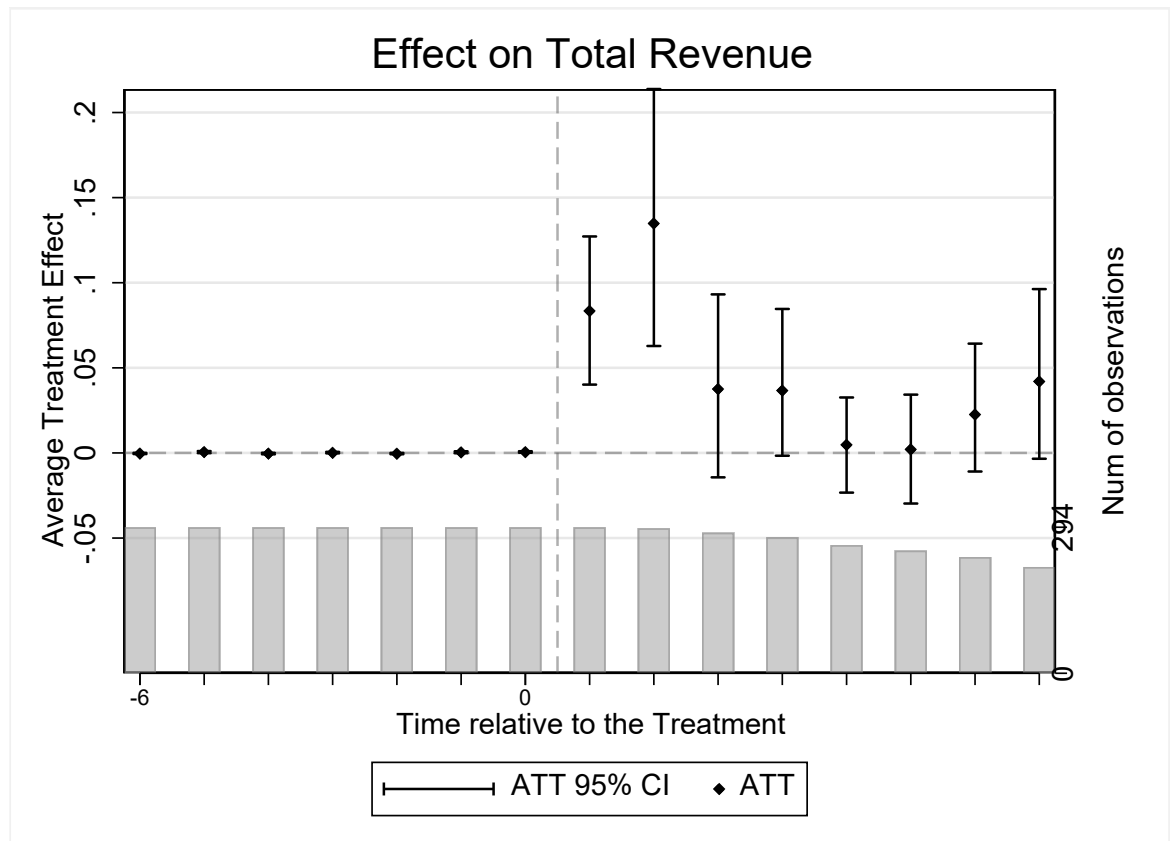

**Figure J1. Dynamic Treatment Effects Based on the MC Method**

Notes: Both plots presented were produced utilizing the "fect" package in Stata. Error bars denote bootstrapped 95% confidence intervals. Results are highly consistent when we include various time-varying controls or when the full sample is considered.

## Appendix K. Placebo Test

Following prior literature (Abadie et al. 2015, Burtch et al. 2018), we next conduct a randomization inference test in which the treatment indicator is repeatedly shuffled within the sample, and the regression is re-estimated, simulating the distribution of treatment effects that would be observed under the null. We repeat this process 1,000 times and recover the distribution of placebo treatment effects, comparing the estimate from the true data against that distribution. Our findings are reported in Table K1. The results indicate that there is no significant relationship between the shuffled treatment and NFT sales outcomes. Calculating an empirical *p*-value, we find that the observed negative impact of charity NFT purchases remains highly significant.

**Table K1. Results of the Placebo Test**

| | *Dependent variable:* | | | |
|---|---|---|---|---|
| | Without Controls | | With Controls | |
| | LnNumSales | LnTotRev | LnNumSales | LnTotRev |
| | (1) | (2) | (3) | (4) |
| Placebo effect (mean) | -0.0002 | -0.0002 | -0.0002 | -0.0002 |
| Placebo effect (st.d.) | 0.0139 | 0.0077 | 0.0134 | 0.0075 |
| Actual treatment effect (coeff) | -0.060 | -0.066** | -0.058 | -0.065** |
| Actual treatment effect (s.e.) | (0.050) | (0.027) | (0.047) | (0.025) |
| Replication | 1,000 | 1,000 | 1,000 | 1,000 |
| *z*-score | -4.3140 | -8.5642 | -1.3139 | -8.5387 |
| Empirical *p*-value | 0.0000 | 0.0000 | 0.1889 | 0.0000 |

*Note: Results are consistent when employing the full sample.*

## Appendix L. Robustness Check with Alternative Dependent Variable

In an ideal world, we would test directly whether the treatment has a significant impact on the average sales price of NFTs sold by individuals. In our main analysis, we focus on total NFT revenue as the key dependent variable, given that the average sales price becomes undefined for individuals with no sales in certain months. Here, we consider various estimators that focus on the average sales price as the outcome of interest, while accounting for missingness. First, we estimate the model introducing a binary variable, *NoSales*, which takes a value of 1 when a user has no NFT sales in a given month. For these observations, we also set the outcome to 0. This adjustment allows us to estimate the treatment effect of charity NFT purchases on the average sales price of non-charity NFTs, conditional on individuals having at least one sale. The results, presented in Table L1, confirm a significant negative treatment effect on the average sales price of NFTs sold. Further, as shown in Table L2, we find that this negative effect is predominantly observed among individuals who relist charity NFTs, indicating that these individuals face significant market penalties, as reflected in the lower average prices they can command for their other NFT listings.

**Table L1. Average Treatment Effect on Average Sale Price**

| | LnAvgSalePrice | | |
|---|---|---|---|
| Sample | CEM | CEM | CEM |
| | (1) | (3) | (5) |
| Treated | -0.016** | -0.016** | -0.026*** |
| | (0.007) | (0.007) | (0.009) |
| NoSales | -0.215*** | -0.213*** | -0.207*** |
| | (0.007) | (0.006) | (0.007) |
| LnTenure | | 0.033*** | 0.053*** |
| | | (0.007) | (0.012) |
| LagLnNumSales | | | -0.008** |
| | | | (0.003) |
| LagLnNumListings | | | 0.006** |
| | | | (0.003) |
| Individual FE | Yes | Yes | Yes |
| Year-month FE | Yes | Yes | Yes |
| Observations | 16,780 | 16,780 | 15,206 |
| Num of Individuals | 1,574 | 1,574 | 1,574 |
| R-squared | 0.473 | 0.474 | 0.480 |

*Note: Robust standard errors clustered at the individual level. * p<0.1, ** p<0.05, *** p<0.01.*

**Table L2. HTE by Intention to Resell the Charity NFTs**

| | LnAvgSalePrice | | |
|---|---|---|---|
| Sample | CEM | CEM | CEM |
| | (1) | (3) | (5) |
| Treated | 0.004 | 0.003 | -0.003 |
| | (0.007) | (0.007) | (0.009) |
| Treated × ListCharityNFT3Days | -0.040*** | -0.039*** | -0.045*** |
| | (0.009) | (0.009) | (0.012) |
| NoSales | -0.214*** | -0.212*** | -0.205*** |
| | (0.007) | (0.006) | (0.007) |
| LnTenure | | 0.032*** | 0.051*** |
| | | (0.007) | (0.013) |
| LagLnNumSales | | | -0.007** |
| | | | (0.003) |
| LagLnNumListings | | | 0.006** |
| | | | (0.003) |
| Individual FE | Yes | Yes | Yes |
| Year-month FE | Yes | Yes | Yes |
| Observations | 16,780 | 16,780 | 15,206 |
| Num of Individuals | 1,574 | 1,574 | 1,574 |
| R-squared | 0.474 | 0.476 | 0.481 |

*Note: Robust standard errors clustered at the individual level. * $p<0.1$, ** $p<0.05$, *** $p<0.01$.*

Next, we consider a Heckman correction (Heckman 1976, Shaver 1998). The Heckman model we use consists of two stages. In Stage 1, we estimate the probability of an individual having any NFT sales. In Stage 2, we examine the average sales price, accounting for selection into sales. We initially focus on model-based identification, exploiting assumed joint normality of the error terms at each stage (later, we also consider a plausibly exogenous instrument). We use a Probit regression in the first stage to estimate the probability of making a sale (*HasSales*), based on the individual's prior listing behavior, sales behavior, and Ethereum wallet tenure. In the second stage, we estimate the treatment effect on the average NFT sale price (*LnAvgSalePrice*) using a control function approach, conditioning on the Inverse Mills Ratio derived from the first stage.

Leveraging a joint log-likelihood function derived from the joint distribution of the selection indicator (*HasSales*) and the outcome variable (*LnAvgSalePrice*), we estimate the Heckman selection model via Maximum Likelihood Estimation (MLE). However, because the first stage of the model is specified as a Probit, the inclusion of high-dimensional fixed effects (e.g., user-level fixed effects) would lead to the Incidental Parameters Problem (Neyman and Scott 1948). Consequently, such fixed effects are excluded from the estimation to preserve consistency of the parameter estimates. Note that this is an inherent limitation of the estimator.

As shown in Column (1) of Table L3, the results from the second stage are consistent with our main analysis, indicating that the treatment effect on the average NFT sale price is still significantly negative after accounting for potential selection bias with the Heckman model.

Next, to further strengthen identification, we reevaluate the treatment effect on average sale price (*LnAvgSalePrice*), considering alternative specifications that incorporate a plausibly exogenous instrument: the log-transformed number of NFTs transferred out by the collector in the last month that were not a result of a sale (*LagLnTransOut*). Non-sale transfers refer to instances in which an NFT was sent to another wallet without monetary compensation, such as gifts, or direct peer-to-peer exchanges, rather than through sales. This variable is likely to satisfy the criteria for a valid instrument, as we discuss next.

**Table L3. Treatment Effect on Average Sale Price Based on the Heckman Model**

| Sample | CEM | CEM | CEM | CEM |
|---|---|---|---|---|
| Model | (1) | (2) | (3) | (4) |
| | | Stage 2: *LnAvgSalePrice* | | |
| Treated | -0.176*** | -0.177*** | -0.144*** | -0.145*** |
| | (0.019) | (0.024) | (0.018) | (0.021) |
| TreatmentGroup | 0.184*** | 0.185*** | 0.159*** | 0.159*** |
| | (0.019) | (0.023) | (0.019) | (0.022) |
| LagLnNumListings | 0.022 | 0.024 | 0.021* | 0.021 |
| | (0.014) | (0.020) | (0.012) | (0.017) |
| LagLnNumSales | -0.021 | -0.017 | -0.028 | -0.026 |
| | (0.048) | (0.071) | (0.039) | (0.057) |
| LnTenure | -0.043 | -0.045 | -0.064*** | -0.066** |
| | (0.027) | (0.040) | (0.022) | (0.032) |
| InitLagLnNumListings | | | -0.016 | -0.016 |
| | | | (0.011) | (0.012) |
| InitLagLnNumSales | | | 0.020 | 0.020 |
| | | | (0.014) | (0.015) |
| InitLnTenure | | | 0.098 | 0.099 |
| | | | (0.077) | (0.076) |
| | | Stage 1: *HasSales* | | |
| Treated | -0.197** | -0.227** | -0.156*** | -0.187*** |
| | (0.071) | (0.078) | (0.060) | (0.064) |
| LagLnTransOut | | 0.134* | | 0.143** |
| | | (0.066) | | (0.061) |
| TreatmentGroup | 0.203** | 0.209* | 0.174*** | 0.179** |
| | (0.074) | (0.093) | (0.064) | (0.078) |
| LagLnNumListings | 0.241*** | 0.231*** | 0.235*** | 0.224*** |
| | (0.028) | (0.031) | (0.029) | (0.031) |
| LagLnNumSales | 0.606*** | 0.594*** | 0.575*** | 0.560*** |
| | (0.086) | (0.122) | (0.073) | (0.100) |
| LnTenure | -0.326*** | -0.319*** | -0.367*** | -0.360*** |
| | (0.026) | (0.026) | (0.027) | (0.030) |
| InitLagLnTransOut | | | | 0.003 |
| | | | | (0.029) |
| InitLagLnNumListings | | | 0.005 | 0.013 |
| | | | (0.035) | (0.036) |
| InitLagLnNumSales | | | 0.170*** | 0.174*** |
| | | | (0.036) | (0.041) |
| InitLnTenure | | | 0.171 | 0.171 |
| | | | (0.138) | (0.149) |
| rho | -0.037 | 0.007 | -0.076 | -0.050 |
| | (0.395) | (0.503) | (0.302) | (0.441) |
| Observations | 15,206 | 15,206 | 15,206 | 15,206 |
| Log-Likelihood | -7061.835 | -7030.486 | -6972.47 | -6937.08 |

Note: Bootstrapped standard errors with 1,000 repetitions in parentheses. * $p$<0.1, ** $p$<0.05, *** $p$<0.01. Results are qualitatively consistent when we include the mean values of time-varying covariates (i.e., *MeanLagLnTransOut*, *MeanLagLnNumListings*, *MeanLagLnNumSales*, and *MeanLnTenure*) instead of initial values.

*Relevance (Stage 1: Probability of having sales):* NFT collectors frequently exchange or transfer NFTs through non-market transactions (e.g., peer-to-peer trades, gifts, or platform transfers) that do not involve posted sale prices (Huang et al. 2024). Collectors who transferred more NFTs out via trades in the previous period tend to exhibit greater engagement in NFT disposition activities in the current period and are therefore more likely to list NFTs for sale. Accordingly, *LagLnTransOut* is positively correlated with the probability of observing sales (*HasSales* = 1).

*Exclusion (Stage 2: Average sale price conditional on selling):* Critically, non-sale NFT transfers do not reveal any price information and occur through different mechanisms than market sales. Once we condition on a collector's decision to sell an NFT through the marketplace, the number of non-sale transfers they have conducted should not directly affect the price received, since activity pages on NFT platforms display prices only for completed sales, not for non-sale transfers. Consequently, prior non-sale transactions do not directly affect buyers' bids or willingness to pay. Instead, market prices are primarily determined by buyers' willingness to pay based on the NFT's characteristics and the seller's reputation—factors that are orthogonal to the seller's historical volume of non-sale transactions. Put differently, while active traders may be more likely to sell, this trading activity does not systematically increase or decrease the prices they command when they do sell. Results including *LagLnTransOut* as an exclusion restriction in the selection equation are only reported in Column (2) of Table L3. The first-stage results confirm that *LagLnTransOut* is a strong predictor of *HasSales* (coefficient = 0.134, $p < 0.1$), supporting its relevance for identification. Further, our main substantive findings in the second stage remain consistent.

To further account for individual-level heterogeneity within the Heckman framework, we considered approaches from the literature. The traditional Mundlak (1978) correction, including user-level means of time-varying covariates, is inappropriate in our setting because our specification includes lagged transaction measures, which violates the strict exogeneity assumption required by this approach.

As Wooldridge (2005) notes, when models include lagged dependent variables, one should condition on the initial value rather than on within-group means to model the distribution of the unobserved effect. Through Monte Carlo experiments, Rabe-Hesketh and Skrondal (2013) show that including initial-period explanatory variables as additional regressors can substantially reduce bias in constrained models when panels are not sufficiently long.

Based on these results in the literature, we condition on initial values of time-varying covariates (i.e., *InitLagLnTransOut*, *InitLagLnNumListings*, *InitLagLnNumSales*, *InitLnTenure*) rather than within-group means, which requires only sequential exogeneity rather than strict exogeneity. As reported in Table L3, results are highly consistent whether or not these initial-period controls are included. Results are qualitatively consistent when we use the mean values of time-varying covariates rather than initial-period values.

However, due to limitations in incorporating multiple fixed effects into the Heckman framework (noted above) and the challenge of including the lagged value of *LnAvgSalePrice* (which is only partially observed when sales occur), we continue to use *LnTotRev* (the log of total revenue) as the primary outcome variable in our main analysis. Moreover, when controlling for *LnNumSales* directly (Table L4), we still observe a significant negative treatment effect on *LnTotRev*, suggesting that the treatment effect operates at the level of average NFT sale prices rather than being driven solely by variation in sales volume.

**Table L4. Treatment Effect on Sales Price After Controlling for Sales Volume**

| | LnTotRev | | |
|---|---|---|---|
| Sample | CEM | CEM | CEM |
| | (1) | (3) | (5) |
| Treated | -0.066** | -0.065** | -0.041*** |
| | (0.027) | (0.025) | (0.014) |
| LnTenure | | 0.343*** | 0.144*** |
| | | (0.031) | (0.015) |
| LnNumSales | | | 0.416*** |
| | | | (0.009) |
| Individual FE | Yes | Yes | Yes |
| Year-month FE | Yes | Yes | Yes |
| Observations | 16,780 | 16,780 | 15,206 |
| Num of Individuals | 1,574 | 1,574 | 1,574 |
| R-squared | 0.451 | 0.468 | 0.762 |

*Note: Robust standard errors clustered at the individual level. * p<0.1, ** p<0.05, *** p<0.01.*

## Appendix M. Additional Results from Alternative Samples

To validate the robustness of our findings, we also conduct a similar DID analysis on the full sample. As shown in Table M1, the results reveal a significant negative treatment effect on individuals' NFT sales performance.

**Table M1. ATT Based on the Full (Unmatched) Sample**

| | LnNumSales<br>(1) | LnTotRev<br>(2) | LnNumSales<br>(3) | LnTotRev<br>(4) | LnNumSales<br>(5) | LnTotRev<br>(6) |
|---|---|---|---|---|---|---|
| Treated | -0.190***<br>(0.035) | -0.202***<br>(0.024) | -0.256***<br>(0.035) | -0.256***<br>(0.024) | -0.129***<br>(0.022) | -0.193***<br>(0.018) |
| LnTenure | | | 0.608***<br>(0.027) | 0.496***<br>(0.019) | 0.016<br>(0.027) | 0.197***<br>(0.021) |
| LagLnNumSales | | | | | 0.461***<br>(0.010) | 0.273***<br>(0.007) |
| LagLnNumListings | | | | | 0.042***<br>(0.007) | 0.039***<br>(0.005) |
| Individual FE | Yes | Yes | Yes | Yes | Yes | Yes |
| Year-month FE | Yes | Yes | Yes | Yes | Yes | Yes |
| Observations | 72,475 | 72,475 | 72,475 | 72,475 | 66,783 | 66,783 |
| Num of Individuals | 5,692 | 5,692 | 5,692 | 5,692 | 5,692 | 5,692 |
| R-squared | 0.537 | 0.549 | 0.549 | 0.564 | 0.678 | 0.661 |

*Note: Robust standard errors clustered at the individual level. * p<0.1, ** p<0.05, *** p<0.01.*

To bolster the credibility of our results, we also re-estimate our models on a refined sample, limited to individuals who placed their donation (i.e., charity NFT purchase) requests within a tight three-minute bandwidth surrounding the stock-out timestamp. Subsequently, we match individuals who successfully procured charity NFTs with their counterparts who were unsuccessful, conducting our match on the same set of covariates elaborated in Section 4.2.1.

**Table M2. Balance Check for the CEM Sample Within the Three-minute Bandwidth**

| | Mean | | | t-test | | |
|---|---|---|---|---|---|---|
| Variable | Treated | Control | %bias | t | *p*-value | V(T)/V(C) |
| MinuteId | 3.736 | 3.736 | 0.000 | 0.000 | 1.000 | 1.000 |
| LowGas | 0.181 | 0.181 | 0.000 | 0.000 | 1.000 | . |
| ModeGas | 0.395 | 0.395 | 0.000 | 0.000 | 1.000 | . |
| HighGas | 0.424 | 0.424 | 0.000 | 0.000 | 1.000 | . |
| LnAvgMints | 0.633 | 0.603 | 3.900 | 0.640 | 0.523 | 0.990 |
| LnAvgPurchases | 0.105 | 0.140 | -7.400 | -1.250 | 0.210 | 0.970 |
| LnAvgSales | 0.322 | 0.319 | 0.400 | 0.070 | 0.944 | 1.000 |
| LnAvgListing | 0.364 | 0.378 | -1.800 | -0.290 | 0.768 | 0.970 |
| LnAvgTransfers | 0.351 | 0.351 | -0.100 | -0.010 | 0.990 | 1.000 |
| TenureMonth | 2.753 | 2.688 | 3.200 | 0.530 | 0.599 | 0.890 |

*Note: In the full sample, 43.10% of individuals have a gas limit of 122,839, which appears to be the prevailing default value. In the matching process, we categorize individuals into three distinct groups: LowGas, ModeGas, and HighGas. This classification is contingent upon whether their designated gas limits fall below, align with, or exceed the default benchmark.*

Table M2, after applying the CEM method to the refined sample (consisting solely of individuals who made procurement bids for charity NFTs within a narrow three-minute window around the stock-out timestamp), indicates a notable similarity between the treatment and control groups across multiple dimensions. The DID estimations using this more stringent matched sample are provided in Table M3. These results mirror our main findings.

**Table M3. ATT Based on CEM Sample Using a Three-minute Bandwidth**

| | LnNumSales | LnTotRev | LnNumSales | LnTotRev | LnNumSales | LnTotRev |
|---|---|---|---|---|---|---|
| | (1) | (2) | (3) | (4) | (5) | (6) |
| Treated | -0.089 | -0.071** | -0.092 | -0.073** | -0.043 | -0.073* |
| | (0.059) | (0.033) | (0.057) | (0.032) | (0.060) | (0.038) |
| LnTenure | | | 0.433*** | 0.296*** | 0.223** | 0.260*** |
| | | | (0.060) | (0.037) | (0.091) | (0.058) |
| LagLnNumSales | | | | | 0.193*** | 0.076*** |
| | | | | | (0.039) | (0.022) |
| LagLnNumListings | | | | | 0.134*** | 0.077*** |
| | | | | | (0.030) | (0.017) |
| Individual FE | Yes | Yes | Yes | Yes | Yes | Yes |
| Year-month FE | Yes | Yes | Yes | Yes | Yes | Yes |
| Observations | 11,251 | 11,251 | 11,251 | 11,251 | 10,202 | 10,202 |
| Num of Individuals | 1,049 | 1,049 | 1,049 | 1,049 | 1,049 | 1,049 |
| R-squared | 0.465 | 0.449 | 0.474 | 0.462 | 0.538 | 0.511 |

*Note: Robust standard errors clustered at the individual level. * p<0.1, ** p<0.05, *** p<0.01.*

To bolster the robustness of our findings, we perform the parallel check test on the two alternative samples that is, the full sample and a narrowly defined matched sample concentrating on individuals who attempted to purchase charity NFTs within a tight three-minute bandwidth surrounding the stock-out timestamp. The results, as presented in Table M4, consistently support the parallel trend assumption for both samples.

**Table M4. Dynamic Treatment Effect Based on Two Alternative Samples**

| Sample | Full Sample | | CEM (3 Mins) | |
|---|---|---|---|---|
| | LnNumSales | LnTotRev | LnNumSales | LnTotRev |
| | (1) | (2) | (3) | (4) |
| Relative time -7 | 0.103 | 0.035 | 0.203 | -0.109 |
| | (0.087) | (0.069) | (0.340) | (0.280) |
| Relative time -6 | 0.132* | 0.056 | 0.102 | 0.041 |
| | (0.068) | (0.052) | (0.220) | (0.178) |
| Relative time -5 | 0.018 | -0.020 | -0.123 | -0.096 |
| | (0.053) | (0.040) | (0.129) | (0.076) |
| Relative time -4 | 0.080* | -0.008 | 0.096 | 0.015 |
| | (0.045) | (0.034) | (0.113) | (0.065) |
| Relative time -3 | 0.039 | -0.034 | 0.064 | -0.036 |
| | (0.039) | (0.029) | (0.095) | (0.060) |
| Relative time -2 | 0.007 | 0.010 | 0.030 | -0.006 |
| | (0.031) | (0.024) | (0.074) | (0.045) |
| Relative time -1 | Baseline (omitted) | | | |
| Relative time 0 | -0.177*** | -0.180*** | -0.243*** | -0.125*** |
| | (0.030) | (0.023) | (0.087) | (0.046) |
| Relative time 1 | -0.054* | -0.069*** | -0.132* | -0.059 |
| | (0.032) | (0.026) | (0.077) | (0.047) |
| Relative time 2 | -0.216*** | -0.126*** | -0.114* | -0.082* |
| | (0.034) | (0.029) | (0.069) | (0.046) |
| Relative time 3 | -0.181*** | -0.261*** | -0.079 | -0.084* |
| | (0.041) | (0.030) | (0.079) | (0.045) |
| Relative time 4 | -0.115*** | -0.239*** | -0.019 | -0.073 |
| | (0.044) | (0.032) | (0.083) | (0.046) |
| Relative time 5 | -0.172*** | -0.238*** | 0.010 | -0.073 |
| | (0.043) | (0.032) | (0.079) | (0.047) |
| Relative time 6 | -0.151*** | -0.257*** | 0.044 | -0.071 |
| | (0.045) | (0.033) | (0.077) | (0.046) |
| Relative time 7 | -0.140*** | -0.240*** | 0.011 | -0.093** |
| | (0.046) | (0.033) | (0.085) | (0.047) |
| Individual FE | Yes | Yes | Yes | Yes |
| Year-month FE | Yes | Yes | Yes | Yes |
| Observations | 72,475 | 72,475 | 11,251 | 11,251 |
| Num of Individuals | 5,692 | 5,692 | 1,049 | 1,049 |
| R-squared | 0.537 | 0.550 | 0.467 | 0.449 |

*Note: Robust standard errors clustered at the individual level; * p<0.1, ** p<0.05, *** p<0.01.*

## Appendix N. Alternative Matching Method

We also explored the use of an alternative matching method, namely, PSM (Rosenbaum and Rubin 1983; Liu et al. 2024). Both Figure N1 and Table N1 reveal that individuals in the treatment and control groups are highly comparable across various dimensions. The findings based on this matched sample, presented in Table M2, indicate that all results are consistent with our main findings.

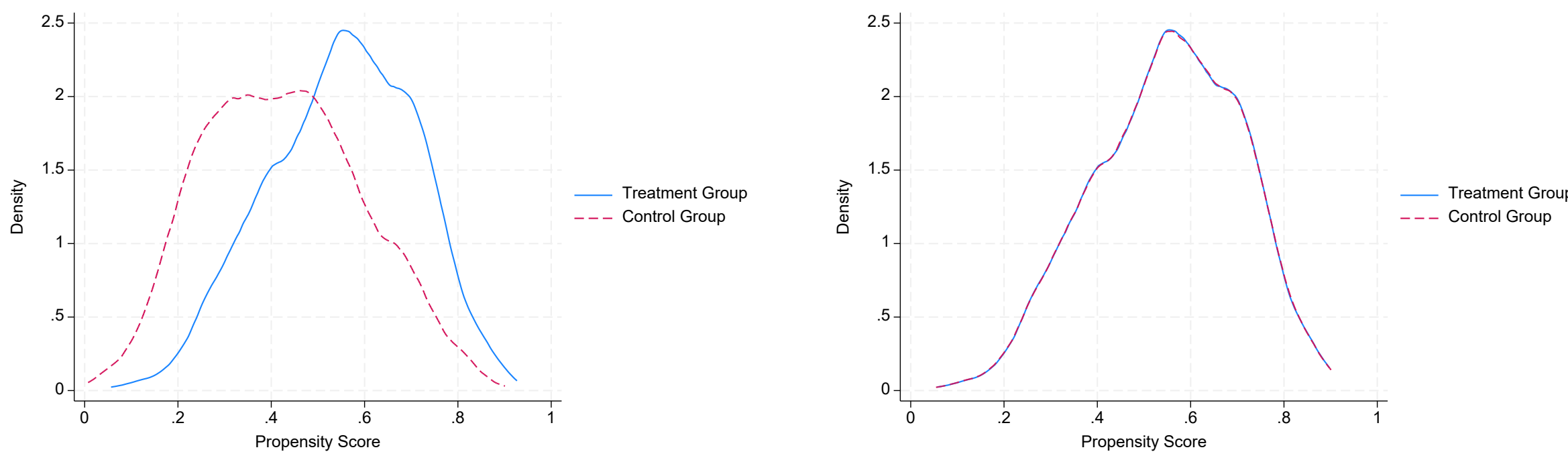

(1) Before Matching (2) After Matching

**Figure N1. Density Distribution of Propensity Scores Before and After Matching**

**Table N1. Balance Check for the PSM Sample**

| Variable | Mean Treated | Mean Control | %bias | t-test t | t-test *p*-value | V(T)/V(C) |
|---|---|---|---|---|---|---|
| MinuteId | 3.558 | 3.541 | 1.400 | 0.550 | 0.582 | 0.92* |
| LnGas | 11.816 | 11.820 | -1.300 | -0.480 | 0.631 | 0.41* |
| LnAvgMints | 2.042 | 2.020 | 1.600 | 0.570 | 0.570 | 1.19* |
| LnAvgPurchases | 0.900 | 0.931 | -2.800 | -1.020 | 0.309 | 1.24* |
| LnAvgSales | 1.678 | 1.674 | 0.300 | 0.100 | 0.917 | 1.18* |
| LnAvgListing | 1.875 | 1.953 | -5.000 | -1.830 | 0.067 | 1.16* |
| LnAvgTransfers | 1.414 | 1.405 | 0.700 | 0.270 | 0.786 | 1.15* |
| TenureMonth | 1.454 | 1.426 | 3.600 | 1.340 | 0.181 | 0.940 |

**Table N2. ATT Based on the PSM Sample**

| | LnNumSales (1) | LnTotRev (2) | LnNumSales (3) | LnTotRev (4) | LnNumSales (5) | LnTotRev (6) |
|---|---|---|---|---|---|---|
| Treated | -0.032 | -0.081*** | -0.039 | -0.086*** | 0.047* | -0.040* |
| | (0.046) | (0.031) | (0.045) | (0.030) | (0.028) | (0.023) |
| LnTenure | | | 0.725*** | 0.588*** | 0.052 | 0.250*** |
| | | | (0.034) | (0.024) | (0.036) | (0.028) |
| LagLnNumSales | | | | | 0.449*** | 0.261*** |
| | | | | | (0.011) | (0.008) |
| LagLnNumListings | | | | | 0.043*** | 0.043*** |
| | | | | | (0.008) | (0.006) |
| Individual FE | Yes | Yes | Yes | Yes | Yes | Yes |
| Year-month FE | Yes | Yes | Yes | Yes | Yes | Yes |
| Observations | 61,841 | 61,841 | 61,841 | 61,841 | 56,965 | 56,965 |
| Num of Individuals | 4,876 | 4,876 | 4,876 | 4,876 | 4,876 | 4,876 |
| R-squared | 0.537 | 0.557 | 0.552 | 0.577 | 0.677 | 0.670 |

*Note: Robust standard errors clustered at the individual level. * p<0.1, ** p<0.05, *** p<0.01.*

## Appendix O. Results from Alternative Time Windows

We next consider whether the negative treatment effect extends beyond the seven-month post-treatment period. To evaluate this, we expanded our data collection, tracing individuals' NFT sales performance for up to 12 months subsequent to the charity NFT fundraiser. Note that we did not consider expanding the period of observation earlier in time because only 22.9% of individuals had established their Ethereum wallets more than 7 months prior to the NFT charity auction.

The DID analysis based on the extended observational window is presented in Table O1 and demonstrates strong consistency with our main findings. To further elucidate how the dynamic treatment effect evolves over an extended post-treatment period, we revisit the relative time model, employing a dataset with an expanded observation window. As delineated in Figure O1 and Table O2, the negative impact of acquiring charity NFTs persists, enduring beyond a 12-month span. This protracted influence can be attributed to a market punishment mechanism, culminating in individuals' heightened detachment from NFT transactions. As demand-side punishment prompts individuals to withdraw, this results in a continued deterioration of their performance in NFT sales, notably constraining their ability to command higher prices in the market. Additionally, as shown in Table O3, we consistently find that individuals are negatively affected by the charity NFT purchase only if they also sought to resell it.

**Table O1. ATT Within the Extended Observation Period**

| | LnNumSales<br>(1) | LnTotRev<br>(2) | LnNumSales<br>(3) | LnTotRev<br>(4) | LnNumSales<br>(5) | LnTotRev<br>(6) |
|---|---|---|---|---|---|---|
| Treated | -0.044 | -0.069** | -0.042 | -0.068** | -0.010 | -0.075** |
| | (0.049) | (0.030) | (0.047) | (0.028) | (0.048) | (0.034) |
| LnTenure | | | 0.458*** | 0.348*** | 0.164** | 0.271*** |
| | | | (0.052) | (0.034) | (0.065) | (0.049) |
| LagLnNumSales | | | | | 0.290*** | 0.126*** |
| | | | | | (0.030) | (0.018) |
| LagLnNumListings | | | | | 0.103*** | 0.056*** |
| | | | | | (0.022) | (0.013) |
| Individual FE | Yes | Yes | Yes | Yes | Yes | Yes |
| Year-month FE | Yes | Yes | Yes | Yes | Yes | Yes |
| Observations | 24,650 | 24,650 | 24,650 | 24,650 | 23,076 | 23,076 |
| Num of Individuals | 1,574 | 1,574 | 1,574 | 1,574 | 1,574 | 1,574 |
| R-squared | 0.437 | 0.390 | 0.447 | 0.410 | 0.543 | 0.487 |

Note: Results from the CEM sample are reported. Results are highly consistent if the full sample is considered. Robust standard errors clustered at the individual level. * $p$<0.1, ** $p$<0.05, *** $p$<0.01.

**Table O2. Relative Time Models on the Impact of Charity NFT Acquisition Within the Extended Observation Period**

| | LnNumSales<br>(1) | LnTotRev<br>(2) | LnNumSales<br>(3) | LnTotRev<br>(4) |
|---|---|---|---|---|
| Relative time -7 | 0.151 | -0.157 | 0.195 | -0.124 |
| | (0.347) | (0.269) | (0.346) | (0.269) |
| Relative time -6 | -0.014 | -0.015 | 0.049 | 0.032 |
| | (0.218) | (0.157) | (0.210) | (0.149) |
| Relative time -5 | -0.086 | -0.033 | -0.058 | -0.012 |
| | (0.115) | (0.070) | (0.115) | (0.070) |
| Relative time -4 | 0.008 | -0.006 | 0.044 | 0.021 |
| | (0.092) | (0.055) | (0.089) | (0.054) |
| Relative time -3 | -0.028 | -0.084* | 0.014 | -0.053 |
| | (0.076) | (0.049) | (0.074) | (0.049) |
| Relative time -2 | 0.004 | 0.001 | 0.018 | 0.011 |
| | (0.058) | (0.036) | (0.057) | (0.036) |
| Relative time -1 | | Baseline (omitted) | | |
| Relative time 0 | -0.232*** | -0.118*** | -0.223*** | -0.111*** |
| | (0.067) | (0.037) | (0.066) | (0.036) |
| Relative time 1 | -0.141** | -0.073* | -0.128* | -0.063 |
| | (0.069) | (0.040) | (0.069) | (0.040) |
| Relative time 2 | -0.113* | -0.102*** | -0.097* | -0.090** |
| | (0.060) | (0.039) | (0.059) | (0.038) |
| Relative time 3 | -0.116* | -0.101*** | -0.099 | -0.088** |
| | (0.065) | (0.038) | (0.064) | (0.037) |
| Relative time 4 | 0.038 | -0.070* | 0.057 | -0.056 |
| | (0.068) | (0.039) | (0.067) | (0.038) |
| Relative time 5 | 0.016 | -0.079** | 0.035 | -0.065* |
| | (0.066) | (0.040) | (0.066) | (0.039) |
| Relative time 6 | 0.015 | -0.077* | 0.035 | -0.062 |
| | (0.065) | (0.039) | (0.064) | (0.038) |
| Relative time 7 | -0.004 | -0.087** | 0.016 | -0.072* |
| | (0.069) | (0.040) | (0.069) | (0.039) |
| Relative time 8 | -0.024 | -0.082** | -0.003 | -0.067* |
| | (0.063) | (0.040) | (0.062) | (0.039) |
| Relative time 9 | -0.009 | -0.072* | 0.011 | -0.056 |
| | (0.060) | (0.040) | (0.059) | (0.039) |
| Relative time 10 | -0.046 | -0.090** | -0.025 | -0.074* |
| | (0.064) | (0.039) | (0.064) | (0.038) |
| Relative time 11 | 0.048 | -0.069* | 0.069 | -0.053 |
| | (0.065) | (0.040) | (0.065) | (0.039) |
| Relative time 12 | -0.076 | -0.101** | -0.055 | -0.084** |
| | (0.066) | (0.041) | (0.065) | (0.040) |
| LnTenure | | | 0.461*** | 0.348*** |
| | | | (0.052) | (0.034) |
| Individual FE | Yes | Yes | Yes | Yes |
| Year-month FE | Yes | Yes | Yes | Yes |
| Observations | 24,650 | 24,650 | 24,650 | 24,650 |
| Num of Individuals | 1,574 | 1,574 | 1,574 | 1,574 |
| R-squared | 0.439 | 0.390 | 0.449 | 0.411 |

*Note: This analysis utilizes the CEM sample. Results demonstrate high consistency when employing the full sample. Robust standard errors clustered at the individual level. * p<0.1, ** p<0.05, *** p<0.01.*

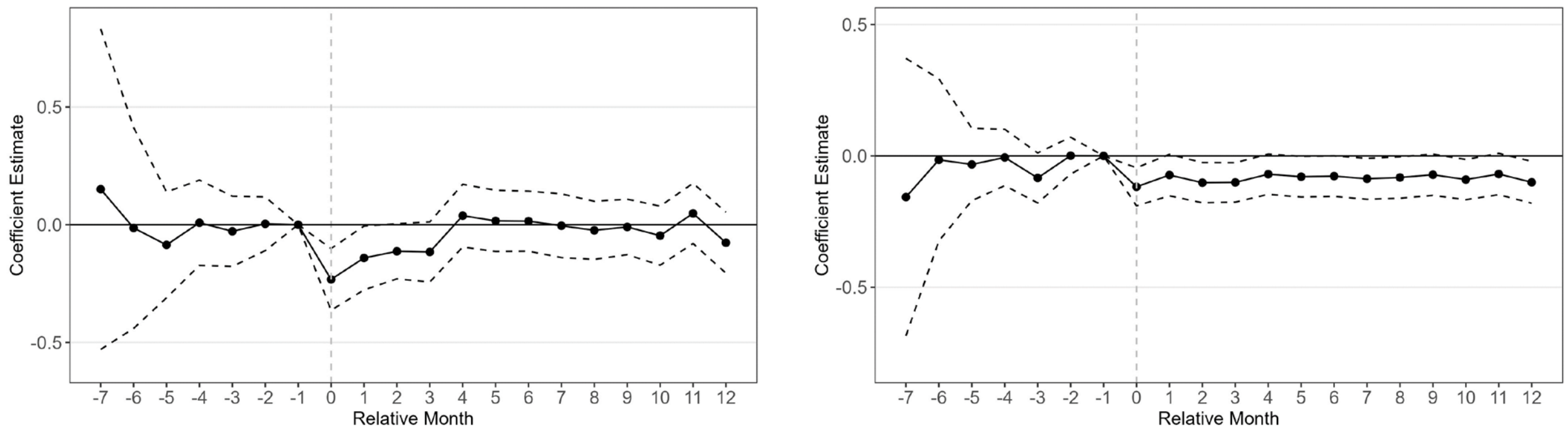

**Figure O1. Coefficients of Relative Time Estimates Within the Extended Observation Period (Left = Sales Volume Effect; Right = Total Revenue Effect)**

*Notes: The dashed vertical line denotes the treatment time. Error bars represent the 95% confidence intervals using the standard errors clustered at the individual level. This analysis utilizes the CEM sample. Results demonstrate high consistency when employing the full sample.*

**Table O3. HTE by Intention to Resell the Charity NFTs Within the Extended Observation Period**

| | LnNumSales | LnTotRev | LnNumSales | LnTotRev | LnNumSales | LnTotRev |
|---|---|---|---|---|---|---|
| | (1) | (2) | (3) | (4) | (5) | (6) |
| Treated | 0.122** | 0.061** | 0.112** | 0.054* | 0.197*** | 0.084** |
| | (0.054) | (0.031) | (0.051) | (0.029) | (0.051) | (0.035) |
| Treated × ListCharityNFT3Days | -0.322*** | -0.253*** | -0.300*** | -0.236*** | -0.400*** | -0.306*** |
| | (0.062) | (0.039) | (0.060) | (0.037) | (0.058) | (0.043) |
| LnTenure | | | 0.452*** | 0.343*** | 0.150** | 0.261*** |
| | | | (0.052) | (0.034) | (0.064) | (0.048) |
| LagLnNumSales | | | | | 0.289*** | 0.125*** |
| | | | | | (0.030) | (0.018) |
| LagLnNumListings | | | | | 0.103*** | 0.056*** |
| | | | | | (0.022) | (0.014) |
| Individual FE | Yes | Yes | Yes | Yes | Yes | Yes |
| Year-month FE | Yes | Yes | Yes | Yes | Yes | Yes |
| Observations | 24,650 | 24,650 | 24,650 | 24,650 | 23,076 | 23,076 |
| Num of Individuals | 1,574 | 1,574 | 1,574 | 1,574 | 1,574 | 1,574 |
| R-squared | 0.439 | 0.395 | 0.449 | 0.415 | 0.545 | 0.492 |

*Note: Robust standard errors clustered at the individual level. * $p<0.1$, ** $p<0.05$, *** $p<0.01$.*

We also consider a more constrained alternative observation window, focusing on the 3 months before and after the treatment time. The results within this alternative window, presented in Tables O4-O6 and Figure O2, are highly consistent with our main findings.

**Table O4. ATT Within the Constrained Observation Period**

| | LnNumSales<br>(1) | LnTotRev<br>(2) | LnNumSales<br>(3) | LnTotRev<br>(4) | LnNumSales<br>(5) | LnTotRev<br>(6) |
|---|---|---|---|---|---|---|
| Treated | -0.138*** | -0.075*** | -0.132*** | -0.071*** | -0.096 | -0.058* |
| | (0.051) | (0.028) | (0.049) | (0.027) | (0.061) | (0.035) |
| LnTenure | | | 0.522*** | 0.317*** | 0.486*** | 0.311*** |
| | | | (0.052) | (0.033) | (0.103) | (0.061) |
| LagLnNumSales | | | | | -0.116*** | -0.070*** |
| | | | | | (0.033) | (0.017) |
| LagLnNumListings | | | | | 0.186*** | 0.122*** |
| | | | | | (0.027) | (0.016) |
| Individual FE | Yes | Yes | Yes | Yes | Yes | Yes |
| Year-month FE | Yes | Yes | Yes | Yes | Yes | Yes |
| Observations | 9,604 | 9,604 | 9,604 | 9,604 | 8,444 | 8,444 |
| Num of Individuals | 1,574 | 1,574 | 1,574 | 1,574 | 1,574 | 1,574 |
| R-squared | 0.578 | 0.565 | 0.586 | 0.574 | 0.619 | 0.610 |

Note: Results from the CEM sample are reported. Results are highly consistent if the full sample is considered. Robust standard errors clustered at the individual level. * $p$<0.1, ** $p$<0.05, *** $p$<0.01.

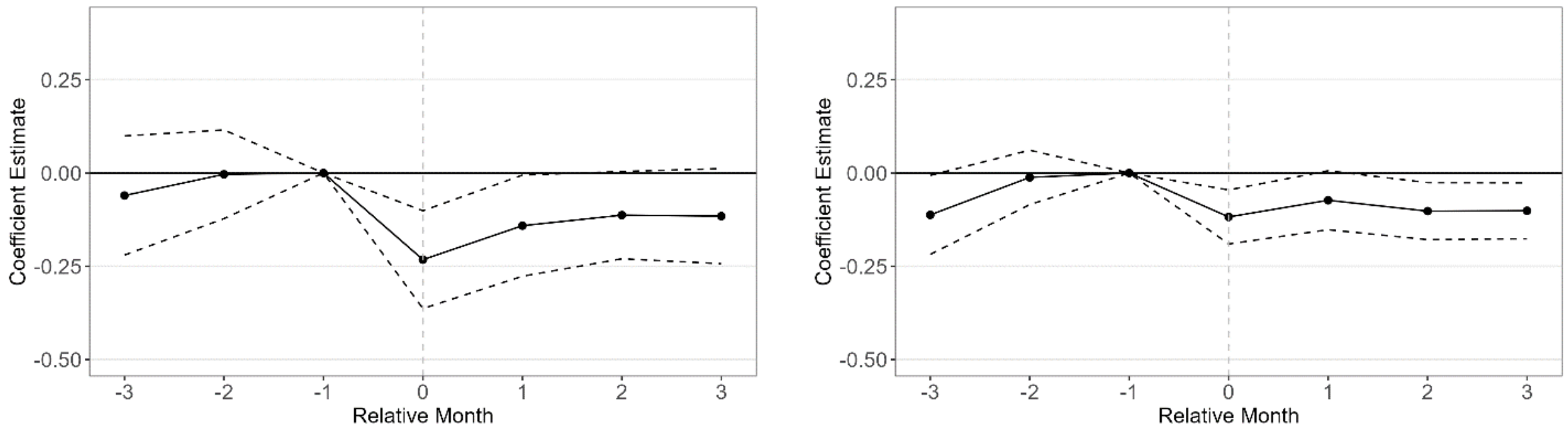

**Figure O2. Coefficients of Relative Time Estimates Within the Constrained Observation Period (Left = Sales Volume Effect; Right = Total Revenue Effect)**

*Notes: The dashed vertical line denotes the treatment time. Error bars represent the 95% confidence intervals using the standard errors clustered at the individual level. This analysis utilizes the CEM sample. Results demonstrate high consistency when employing the full sample.*

**Table O5. Relative Time Models on the Impact of Charity NFT Acquisition Within the Constrained Observation Period**

| | LnNumSales<br>(1) | LnTotRev<br>(2) | LnNumSales<br>(3) | LnTotRev<br>(4) |
|---|---|---|---|---|
| Relative time -3 | -0.060 | -0.112** | -0.026 | -0.092* |
| | (0.081) | (0.054) | (0.081) | (0.055) |
| Relative time -2 | -0.003 | -0.012 | 0.005 | -0.006 |
| | (0.061) | (0.037) | (0.060) | (0.038) |
| Relative time -1 | Baseline (omitted) | | | |
| Relative time 0 | -0.232*** | -0.118*** | -0.222*** | -0.112*** |
| | (0.067) | (0.037) | (0.066) | (0.036) |
| Relative time 1 | -0.141** | -0.073* | -0.126* | -0.064 |
| | (0.069) | (0.040) | (0.069) | (0.040) |
| Relative time 2 | -0.113* | -0.102*** | -0.095 | -0.092** |
| | (0.060) | (0.039) | (0.059) | (0.038) |
| Relative time 3 | -0.116* | -0.101*** | -0.096 | -0.089** |
| | (0.065) | (0.038) | (0.065) | (0.037) |
| LnTenure | | | 0.523*** | 0.315*** |
| | | | (0.052) | (0.033) |
| Individual FE | Yes | Yes | Yes | Yes |
| Year-month FE | Yes | Yes | Yes | Yes |
| Observations | 9,604 | 9,604 | 9,604 | 9,604 |
| Num of Individuals | 1,574 | 1,574 | 1,574 | 1,574 |
| R-squared | 0.578 | 0.565 | 0.587 | 0.574 |

*Note: This analysis utilizes the CEM sample. Results demonstrate high consistency when employing the full sample. Robust standard errors clustered at the individual level. * p<0.1, ** p<0.05, *** p<0.01.*

**Table O6. HTE by Intention to Resell the Charity NFTs Within the Constrained Observation Period**

| | LnNumSales | LnTotRev | LnNumSales | LnTotRev | LnNumSales | LnTotRev |
|---|---|---|---|---|---|---|
| | (1) | (2) | (3) | (4) | (5) | (6) |
| Treated | -0.070 | -0.011 | -0.076 | -0.014 | 0.054 | 0.045 |
| | (0.057) | (0.029) | (0.055) | (0.028) | (0.065) | (0.037) |
| Treated × ListCharityNFT3Days | -0.144** | -0.122*** | -0.128** | -0.112*** | -0.305*** | -0.208*** |
| | (0.061) | (0.036) | (0.059) | (0.035) | (0.072) | (0.046) |
| LnTenure | | | 0.458*** | 0.271*** | 0.402*** | 0.238*** |
| | | | (0.050) | (0.033) | (0.097) | (0.064) |
| LagLnNumSales | | | | | -0.040 | -0.021 |
| | | | | | (0.032) | (0.019) |
| LagLnNumListings | | | | | 0.147*** | 0.100*** |
| | | | | | (0.025) | (0.016) |
| Individual FE | Yes | Yes | Yes | Yes | Yes | Yes |
| Year-month FE | Yes | Yes | Yes | Yes | Yes | Yes |
| Observations | 10,484 | 10,484 | 10,484 | 10,484 | 8,910 | 8,910 |
| Num of Individuals | 1,574 | 1,574 | 1,574 | 1,574 | 1,574 | 1,574 |
| R-squared | 0.571 | 0.549 | 0.579 | 0.557 | 0.621 | 0.605 |

*Note: Robust standard errors clustered at the individual level. * p<0.1, ** p<0.05, *** p<0.01.*

## Appendix P. Robustness Check after Excluding Potential Suspicious Traders

One potential concern is that some individuals may employ bots to trade NFTs, which could make it easier for them to successfully purchase charity NFTs. However, we do not expect this behavior to substantially affect our main results regarding the sales performance of individuals in the treatment group, as onlookers are unlikely to recognize whether those individuals are utilizing bots, and thus, such behavior is unlikely to result in penalization of those individuals' NFT sales.

Nonetheless, to address this concern, we conduct a robustness check by excluding potentially suspicious NFT traders who exhibit unusually high volumes of NFT transfers prior to the treatment (specifically, those in the top 90$^{th}$ percentile of NFT transfers). We then reanalyze the data, as presented in Tables P1 and P2. The results remain highly consistent with our main findings, suggesting that our conclusions are unlikely to be driven by negative responses to bot users.

**Table P1. Average Treatment Effect After Excluding Potentially Suspicious Traders**

| | LnNumSales (1) | LnTotRev (2) | LnNumSales (3) | LnTotRev (4) | LnNumSales (5) | LnTotRev (6) |
|---|---|---|---|---|---|---|
| Treated | -0.064 | -0.066** | -0.063 | -0.066*** | -0.018 | -0.061* |
| | (0.049) | (0.026) | (0.047) | (0.025) | (0.051) | (0.032) |
| LnTenure | | | 0.448*** | 0.319*** | 0.177** | 0.252*** |
| | | | (0.050) | (0.030) | (0.076) | (0.050) |
| LagLnNumSales | | | | | 0.191*** | 0.065*** |
| | | | | | (0.030) | (0.017) |
| LagLnNumListings | | | | | 0.135*** | 0.083*** |
| | | | | | (0.023) | (0.014) |
| Individual FE | Yes | Yes | Yes | Yes | Yes | Yes |
| Year-month FE | Yes | Yes | Yes | Yes | Yes | Yes |
| Observations | 16,705 | 16,705 | 16,705 | 16,705 | 15,136 | 15,136 |
| Num of Individuals | 1,569 | 1,569 | 1,569 | 1,569 | 1,569 | 1,569 |
| R-squared | 0.473 | 0.444 | 0.482 | 0.459 | 0.550 | 0.514 |

*Note: Robust standard errors clustered at the individual level. * p<0.1, ** p<0.05, *** p<0.01.*

**Table P2. HTE by Intention to Resell the Charity NFTs**

| | LnNumSales (1) | LnTotRev (2) | LnNumSales (3) | LnTotRev (4) | LnNumSales (5) | LnTotRev (6) |
|---|---|---|---|---|---|---|
| Treated | 0.070 | 0.037 | 0.063 | 0.032 | 0.185*** | 0.082** |
| | (0.052) | (0.027) | (0.050) | (0.025) | (0.053) | (0.033) |
| Treated × ListCharityNFT3Days | -0.262*** | -0.202*** | -0.244*** | -0.190*** | -0.394*** | -0.275*** |
| | (0.061) | (0.035) | (0.059) | (0.033) | (0.061) | (0.041) |
| LnTenure | | | 0.442*** | 0.315*** | 0.161** | 0.241*** |
| | | | (0.049) | (0.030) | (0.075) | (0.050) |
| LagLnNumSales | | | | | 0.190*** | 0.065*** |
| | | | | | (0.030) | (0.017) |
| LagLnNumListings | | | | | 0.135*** | 0.084*** |
| | | | | | (0.023) | (0.014) |
| Individual FE | Yes | Yes | Yes | Yes | Yes | Yes |
| Year-month FE | Yes | Yes | Yes | Yes | Yes | Yes |
| Observations | 16,705 | 16,705 | 16,705 | 16,705 | 15,136 | 15,136 |
| Num of Individuals | 1,569 | 1,569 | 1,569 | 1,569 | 1,569 | 1,569 |
| R-squared | 0.475 | 0.447 | 0.483 | 0.462 | 0.553 | 0.518 |

*Note: Robust standard errors clustered at the individual level. * p<0.1, ** p<0.05, *** p<0.01.*

## Appendix Q. Robustness Check after Accounting for the Potential Confounding Effects of Other Charity NFTs

Concerns may arise regarding the potential confounding effects of other charity NFTs on our findings. To address this issue, we invested significant effort in collecting the collection descriptions of all NFTs in the portfolios of the individuals in both our treatment and control groups. Among all NFTs ever owned by the individuals within our sample, we identify only 29 charity fundraisers. To ensure a more conservative analysis that excludes the effects of other charity NFTs, we classify any NFT collection as a charity fundraiser if any portion (ranging from 1% to 100%) of its primary sales or royalties from secondary sales were donated to charities.

With this additional information, we conduct two robustness checks. First, we rerun all our analyses focusing solely on the sales performance of non-charity NFTs in individuals' portfolios. The results, presented in Tables Q1-Q2, are consistent with our original findings. Second, we limit our analysis to individuals who have never owned other charity NFTs in the treatment and control groups. Similarly, in Tables Q3-Q4, we observe consistent market penalties for individuals in the treatment group who are perceived as exhibiting strategic generosity, as evidenced by lower total revenue from their other non-charity NFT listings when they relist charity NFTs for sale.

These additional analyses further reinforce the robustness of our findings, suggesting that the observed market penalties are unlikely to be driven by the confounding effects of other charity NFTs.

**Table Q1. Average Treatment Effect on the Sales of Non-Charity NFTs**

| | LnNumSales | LnTotRev | LnNumSales | LnTotRev | LnNumSales | LnTotRev |
|---|---|---|---|---|---|---|
| | (1) | (2) | (3) | (4) | (5) | (6) |
| Treated | -0.070 | -0.071*** | -0.068 | -0.070*** | -0.031 | -0.069** |
| | (0.051) | (0.027) | (0.049) | (0.026) | (0.051) | (0.032) |
| LnTenure | | | 0.482*** | 0.350*** | 0.252*** | 0.315*** |
| | | | (0.051) | (0.031) | (0.072) | (0.050) |
| LagLnNumSales | | | | | 0.190*** | 0.063*** |
| | | | | | (0.023) | (0.014) |
| LagLnNumListings | | | | | 0.131*** | 0.089*** |
| | | | | | (0.016) | (0.011) |
| Individual FE | Yes | Yes | Yes | Yes | Yes | Yes |
| Year-month FE | Yes | Yes | Yes | Yes | Yes | Yes |
| Observations | 16,780 | 16,780 | 16,780 | 16,780 | 15,206 | 15,206 |
| Num of Individuals | 1,574 | 1,574 | 1,574 | 1,574 | 1,574 | 1,574 |
| R-squared | 0.510 | 0.487 | 0.520 | 0.503 | 0.588 | 0.560 |

*Note: Robust standard errors clustered at the individual level. * p<0.1, ** p<0.05, *** p<0.01.*

**Table Q2. HTE on the Sales of Non-Charity NFTs by Intention to Resell the Charity NFTs**

| | LnNumSales (1) | LnTotRev (2) | LnNumSales (3) | LnTotRev (4) | LnNumSales (5) | LnTotRev (6) |
|---|---|---|---|---|---|---|
| Treated | 0.056 | 0.032 | 0.048 | 0.026 | 0.163*** | 0.071** |
| | (0.055) | (0.028) | (0.052) | (0.026) | (0.055) | (0.034) |
| Treated × ListCharityNFT3Days | -0.246*** | -0.200*** | -0.226*** | -0.186*** | -0.375*** | -0.270*** |
| | (0.063) | (0.036) | (0.060) | (0.034) | (0.062) | (0.041) |
| LnTenure | | | 0.476*** | 0.345*** | 0.235*** | 0.303*** |
| | | | (0.050) | (0.031) | (0.072) | (0.050) |
| LagLnNumSales | | | | | 0.190*** | 0.063*** |
| | | | | | (0.023) | (0.014) |
| LagLnNumListings | | | | | 0.131*** | 0.088*** |
| | | | | | (0.016) | (0.011) |
| Individual FE | Yes | Yes | Yes | Yes | Yes | Yes |
| Year-month FE | Yes | Yes | Yes | Yes | Yes | Yes |
| Observations | 16,780 | 16,780 | 16,780 | 16,780 | 15,206 | 15,206 |
| Num of Individuals | 1,574 | 1,574 | 1,574 | 1,574 | 1,574 | 1,574 |
| R-squared | 0.512 | 0.490 | 0.521 | 0.505 | 0.590 | 0.564 |

*Note: Robust standard errors clustered at the individual level. * p<0.1, ** p<0.05, *** p<0.01.*

**Table Q3. Average Treatment Effect After Excluding Individuals Who Owned Other Non-Charity NFTs**

| | LnNumSales (1) | LnTotRev (2) | LnNumSales (3) | LnTotRev (4) | LnNumSales (5) | LnTotRev (6) |
|---|---|---|---|---|---|---|
| Treated | -0.011 | -0.012 | -0.019 | -0.017 | 0.026 | -0.006 |
| | (0.045) | (0.022) | (0.043) | (0.021) | (0.050) | (0.030) |
| LnTenure | | | 0.324*** | 0.222*** | 0.136* | 0.181*** |
| | | | (0.045) | (0.026) | (0.071) | (0.043) |
| LagLnNumSales | | | | | 0.107*** | 0.013 |
| | | | | | (0.024) | (0.011) |
| LagLnNumListings | | | | | 0.147*** | 0.080*** |
| | | | | | (0.021) | (0.011) |
| Individual FE | Yes | Yes | Yes | Yes | Yes | Yes |
| Year-month FE | Yes | Yes | Yes | Yes | Yes | Yes |
| Observations | 14,954 | 14,954 | 14,954 | 14,954 | 13,540 | 13,540 |
| Num of Individuals | 1,414 | 1,414 | 1,414 | 1,414 | 1,414 | 1,414 |
| R-squared | 0.438 | 0.381 | 0.443 | 0.391 | 0.500 | 0.436 |

*Note: Robust standard errors clustered at the individual level. * p<0.1, ** p<0.05, *** p<0.01.*

**Table Q4. HTE by Intention to Resell the Charity NFTs After Excluding Individuals Who Owned Other Non-Charity NFTs**

| | LnNumSales (1) | LnTotRev (2) | LnNumSales (3) | LnTotRev (4) | LnNumSales (5) | LnTotRev (6) |
|---|---|---|---|---|---|---|
| Treated | 0.088* | 0.064*** | 0.083* | 0.060*** | 0.184*** | 0.101*** |
| | (0.048) | (0.022) | (0.046) | (0.021) | (0.053) | (0.030) |
| Treated × ListCharityNFT3Days | -0.211*** | -0.162*** | -0.216*** | -0.165*** | -0.342*** | -0.231*** |
| | (0.053) | (0.028) | (0.052) | (0.028) | (0.060) | (0.038) |
| LnTenure | | | 0.326*** | 0.224*** | 0.134* | 0.181*** |
| | | | (0.045) | (0.026) | (0.070) | (0.043) |
| LagLnNumSales | | | | | 0.107*** | 0.013 |
| | | | | | (0.024) | (0.011) |
| LagLnNumListings | | | | | 0.147*** | 0.080*** |
| | | | | | (0.021) | (0.011) |
| Individual FE | Yes | Yes | Yes | Yes | Yes | Yes |
| Year-month FE | Yes | Yes | Yes | Yes | Yes | Yes |
| Observations | 14,954 | 14,954 | 14,954 | 14,954 | 13,540 | 13,540 |
| Num of Individuals | 1,414 | 1,414 | 1,414 | 1,414 | 1,414 | 1,414 |
| R-squared | 0.439 | 0.384 | 0.445 | 0.394 | 0.503 | 0.440 |

*Note: Robust standard errors clustered at the individual level. * p<0.1, ** p<0.05, *** p<0.01.*

## Appendix R. Robustness Checks Without Excluding Individuals with Insufficient Balance for Purchase

In our main analysis, we previously excluded individuals who had insufficient balance in their Ether wallets to cover the transaction fee and the price of the UKR NFT at one second prior to the launch of the UKR NFT collection. As a sanity check, we confirmed that all individuals with insufficient balance were placed in the control group.

To ensure the robustness of our findings, we conducted an additional analysis retaining these individuals who had an insufficient balance. We applied a similar CEM method, ensuring that both the treatment and control groups were balanced across different aspects, as shown in Table R1. The results, presented in Tables R2 and R3, consistently indicate a significant negative treatment effect. Notably, this negative effect is only observed among individuals who list charity NFTs for sale, reinforcing the notion of market punishment towards those perceived as displaying strategic generosity. These findings validate the robustness of our main findings.

**Table R1. Balance Check for the CEM Sample**

| | Mean | | | t-test | | |
|---|---|---|---|---|---|---|
| Variable | Treated | Control | %bias | t | *p*-value | V(T)/V(C) |
| MinuteId | 3.357 | 3.357 | 0.000 | 0.000 | 1.000 | 1.000 |
| LowGas | 0.219 | 0.219 | 0.000 | 0.000 | 1.000 | . |
| ModeGas | 0.339 | 0.339 | 0.000 | 0.000 | 1.000 | . |
| HighGas | 0.443 | 0.443 | 0.000 | 0.000 | 1.000 | . |
| LnAvgMints | 0.700 | 0.676 | 3.000 | 0.590 | 0.554 | 0.980 |
| LnAvgPurchases | 0.098 | 0.132 | -7.900 | -1.580 | 0.113 | 0.970 |
| LnAvgSales | 0.363 | 0.355 | 1.200 | 0.230 | 0.819 | 1.020 |
| LnAvgListing | 0.402 | 0.406 | -0.500 | -0.110 | 0.915 | 0.970 |
| LnAvgTransfers | 0.364 | 0.368 | -0.700 | -0.140 | 0.891 | 0.980 |
| TenureMonth | 2.750 | 2.640 | 5.600 | 1.140 | 0.255 | 0.930 |

*Note: In the full sample, 43.10% of individuals have a gas limit of 122,839, which appears to be the prevailing default value. In the matching process, we categorize individuals into three distinct groups: LowGas, ModeGas, and HighGas. This classification is contingent upon whether their designated gas limits fall below, align with, or exceed the default benchmark.*

**Table R2. Average Treatment Effect**

| | LnNumSales (1) | LnTotRev (2) | LnNumSales (3) | LnTotRev (4) | LnNumSales (5) | LnTotRev (6) |
|---|---|---|---|---|---|---|
| Treated | -0.060 | -0.066** | -0.057 | -0.064** | -0.016 | -0.063* |
| | (0.050) | (0.027) | (0.047) | (0.025) | (0.051) | (0.032) |
| LnTenure | | | 0.478*** | 0.344*** | 0.213*** | 0.289*** |
| | | | (0.051) | (0.031) | (0.075) | (0.050) |
| LagLnNumSales | | | | | 0.193*** | 0.067*** |
| | | | | | (0.030) | (0.017) |
| LagLnNumListings | | | | | 0.136*** | 0.083*** |
| | | | | | (0.023) | (0.014) |
| Individual FE | Yes | Yes | Yes | Yes | Yes | Yes |
| Year-month FE | Yes | Yes | Yes | Yes | Yes | Yes |
| Observations | 16,869 | 16,869 | 16,869 | 16,869 | 15,286 | 15,286 |
| Num of Individuals | 1,583 | 1,583 | 1,583 | 1,583 | 1,583 | 1,583 |
| R-squared | 0.479 | 0.451 | 0.489 | 0.468 | 0.557 | 0.523 |

*Note: Robust standard errors clustered at the individual level. * p<0.1, ** p<0.05, *** p<0.01.*

**Table R3. HTE by Intention to Resell the Charity NFTs**

| | LnNumSales<br>(1) | LnTotRev<br>(2) | LnNumSales<br>(3) | LnTotRev<br>(4) | LnNumSales<br>(5) | LnTotRev<br>(6) |
|---|---|---|---|---|---|---|
| Treated | 0.071 | 0.039 | 0.064 | 0.034 | 0.182*** | 0.080** |
| | (0.054) | (0.028) | (0.051) | (0.026) | (0.054) | (0.034) |
| Treated × ListCharityNFT3Days | -0.255*** | -0.205*** | -0.235*** | -0.191*** | -0.383*** | -0.276*** |
| | (0.061) | (0.035) | (0.059) | (0.034) | (0.062) | (0.042) |
| LnTenure | | | 0.472*** | 0.339*** | 0.196*** | 0.277*** |
| | | | (0.051) | (0.031) | (0.075) | (0.050) |
| LagLnNumSales | | | | | 0.193*** | 0.067*** |
| | | | | | (0.030) | (0.017) |
| LagLnNumListings | | | | | 0.136*** | 0.083*** |
| | | | | | (0.023) | (0.014) |
| Individual FE | Yes | Yes | Yes | Yes | Yes | Yes |
| Year-month FE | Yes | Yes | Yes | Yes | Yes | Yes |
| Observations | 16,869 | 16,869 | 16,869 | 16,869 | 15,286 | 15,286 |
| Num of Individuals | 1,583 | 1,583 | 1,583 | 1,583 | 1,583 | 1,583 |
| R-squared | 0.481 | 0.454 | 0.490 | 0.471 | 0.560 | 0.527 |

*Note: Robust standard errors clustered at the individual level. * p<0.1, ** p<0.05, *** p<0.01.*

## Appendix S. Additional Details and Results from the Incentive-Compatible Experiment

We conducted an incentive-compatible online experiment on Prolific. Participants were required to have prior experience minting or owning NFTs and to have not participated in our previous experiment. This experiment serves for two purposes: 1) to explicitly test whether onlookers (i.e., potential NFT buyers) tend to inspect a seller's broader portfolio and trading history; and 2) to examine whether and how onlookers interpret strategic motivations behind a donor's charity NFT listing behavior, and whether such perceptions lead to penalties in the form of reduced willingness to buy. In doing so, this study supplemented the closed-ended perception measures used in the scenario-based experiment (Section 5.2) with open-ended responses. These qualitative responses were analyzed using large language model (LLM) labeling to systematically extract common themes in onlookers' interpretations.

To avoid repeating similar descriptions of the experimental design across both experiments, this appendix provides additional details and analysis results specific to the incentive-compatible experiment. Since this preliminary experiment provides initial insights for our main study, we describe only elements unique to this design for brevity.

### *S.1. Experiment Design and Data Collection*

We designed a supplemental incentive-compatible online experiment simulating a real NFT purchase. Participants were informed they must select one NFT for purchase at the end of the experiment. They were informed that, due to budget constraints, only 20% of them were randomly selected to receive their chosen NFT, purchased by the research team with real funds and transferred to them. Any unused portion of their assigned $5 budget was returned in cash. Participants were instructed to treat their choice as a real purchase, given the possibility of an actual transaction. At the end of the experiment, selected participants were informed that, due to institutional restrictions on research spending, we could not make the purchase directly. Instead, they were invited to buy the NFT themselves and were offered a $5 Amazon gift card via email to offset the cost.

Our experiment consisted of four main steps. In Step 1, participants were presented with a (non-charity) NFT and given the option to either view the seller's full portfolio and trading history or proceed directly to purchase. Notably, 72.2% opted to examine the seller's portfolio before making a decision.

In Step 2, participants who opted to view the seller's portfolio were randomly assigned to one of three conditions based on the seller's behavior with a previously donated charity NFT: (a) the *HighRelistingPrice* group, where the seller relists the charity NFT at a price higher than its original purchase price; (b) the *LowRelistingPrice* group, where the seller relists the charity NFT at a price lower than its original purchase price; and (c) the *NoRelisting* group, where the seller does not relist the charity NFT for sale. Participants who chose to proceed directly to purchase were not exposed to this step.

In Step 3, all participants were given a $5 budget and asked to choose between: (1) purchasing the original NFT shown in Step 1 from the previous seller; or (2) purchasing a randomly selected NFT from a different seller at the same price, chosen by the research team. Participants were told they must select one NFT for purchase and that 20% would be randomly selected to have their choice fulfilled using real funds.

In Step 4, participants responded to demographic questions. More importantly, to gather additional qualitative evidence on onlookers' perceptions of individuals attempting to resell charity NFTs, participants were asked the following open-ended question: "Please share your thoughts on upon noticing someone reselling a charity NFT for profit within 3 days of purchase. What do you think motivates this behavior, and how might it be perceived by others in the NFT marketplace?" This question was designed to elicit participants' views on both the perceived motivations behind such behavior and its social interpretation, particularly how actors in the NFT community may consciously frame their actions as either strategic or altruistic.

At the end of the experiment, the selected 20% of participants were informed that, due to institutional restrictions on research spending, we could not make the purchase directly. Instead, they were invited to buy the NFT themselves and were offered a $5 Amazon gift card via email to offset the cost.

***S.2. Experiment Results***

***Do onlookers observe and respond to sellers' transaction histories?*** We collected 144 complete responses from participants. To examine whether donors tend to inspect the seller's full portfolio and trading history, we find that *72.2%* of participants chose to view this information before making an NFT purchase decision.

***Does the treatment influence onlookers' purchase decisions?*** Consistent with our market punishment argument, Table S1 reveals that onlookers are less willing to purchase from sellers who relist the charity NFT, compared to those who do not. This suggests that relisting a charity NFT is perceived negatively by potential buyers and may lead to reduced market interest in the seller's listings.

**Table S1. Treatment Effect on Seller Preference**

| | PreferAlternativeSeller<br>(1) |
|---|---|
| LowRelistingPrice | $0.326^{***}$ |
| | (0.109) |
| HighRelistingPrice | $0.281^{**}$ |
| | (0.109) |
| Controls | Yes |
| Observations | 104 |
| R-squared | 0.217 |

*Note: The NoRelisting group is the baseline group. We include participants' demographics (i.e., gender, race, and education level) as control variables. * p<0.1, ** p<0.05, *** p<0.01.*

***How do onlookers perceive the act of reselling charity NFTs by sellers?*** To obtain more qualitative evidence regarding onlookers' perceptions of sellers who resell charity NFTs, we conducted a text analysis of participants' responses to the open-ended question presented in Step 4. Following the approach outlined in Brynjolfsson et al. (2025), we undertook the following steps:

*Step 1: Response summarization.* We used the OpenAI GPT-4o-mini model to summarize each response using a small set of concise topics (1-5 words per summary). This allowed us to identify common themes and sentiments across participants' narratives.

*Step 2: Topic clustering.* We clustered the unique perception topic phrases extracted in Step 1 into semantically coherent groups using GPT-4o. After removing duplicates and non-informative entries, we prompted the model to organize the topic phrases into 5-7 distinct clusters based on underlying sentiment and meaning, each accompanied by a concise label, a brief thematic definition, and representative example topic phrases. The resulting clusters are summarized in Table E2 below. This approach allowed us to systematically extract themes without imposing a predefined framework. Five primary topic clusters emerged: *greed and self-interest, ethical concerns, opportunism, disappointment and distrust, and market dynamics and acceptance*.

*Step 3: Topic classification.* Since each participant's response may relate to more than one topic cluster, we classify responses using a multi-label approach. Based on the topic cluster labels and definitions developed in Step 2, we use the OpenAI GPT-4o-mini model to assign a Yes/No label to each topic cluster for every response. This allows us to systematically capture the presence or absence of each topic in participants' open-ended answers.

As shown in Figure S1, topics related to ethical concerns (60%), self-interest (56%), and opportunism (54%) were especially prevalent, providing strong qualitative support for our proposed mechanism of strategic generosity. Moreover, we find that discovering a seller's intent to resell a charity NFT negatively affected participants' willingness to purchase a separate, non-charity NFT from the same seller. Taken together, these findings further reinforce the mechanism outlined in earlier sections.

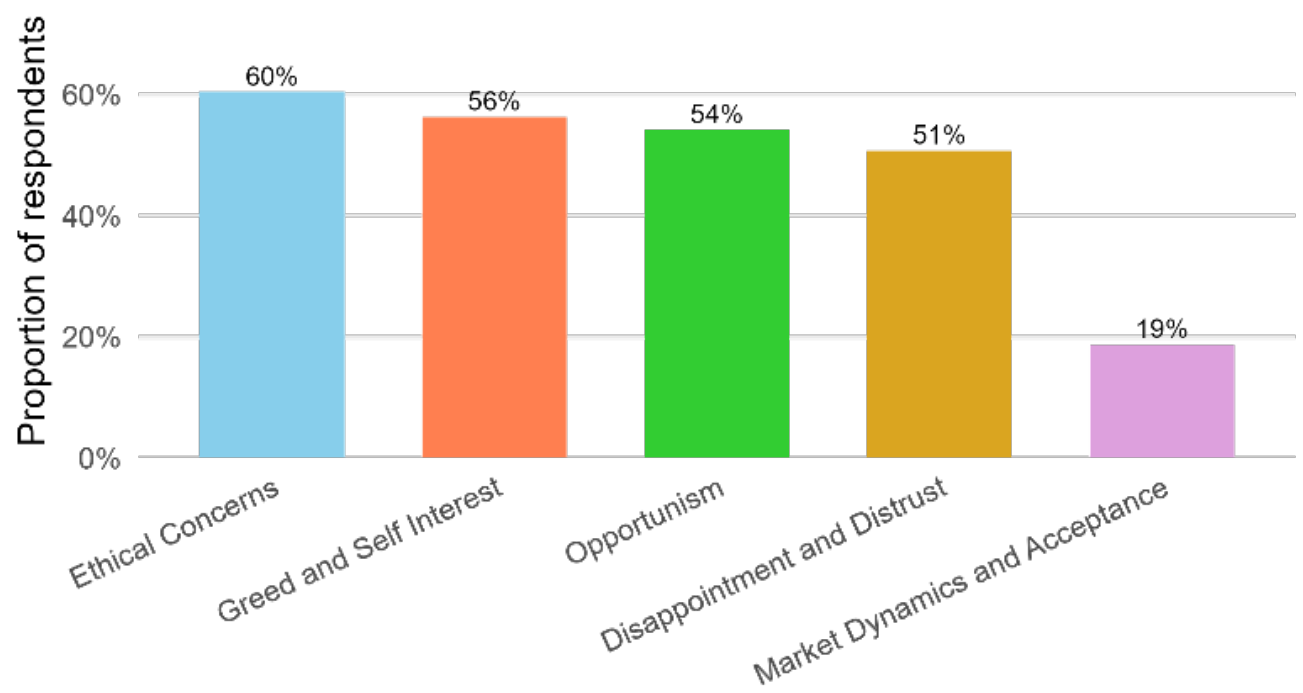

**Figure S1. Distribution of Topic Clustering Labels**

## Appendix T. Procedure and Design of the Scenario-based Experiment

We are conducting a study to understand individuals' willingness to pay for NFTs. In the survey below, you will be guided through a simulated decision-making process in an NFT marketplace. At the end of the survey, you will have an opportunity to place a bid on an NFT based on your level of interest.
Please review the NFT marketplace scenario carefully and complete the related tasks.

### *Step 1. Initial Consideration of Purchasing a Non-Charity NFT*

You have come across the following NFT, possessed by the owner (Oxc5efd..2ad2) at **0.05 ETH (approximately $134)**. Please examine the NFT's image and description closely.

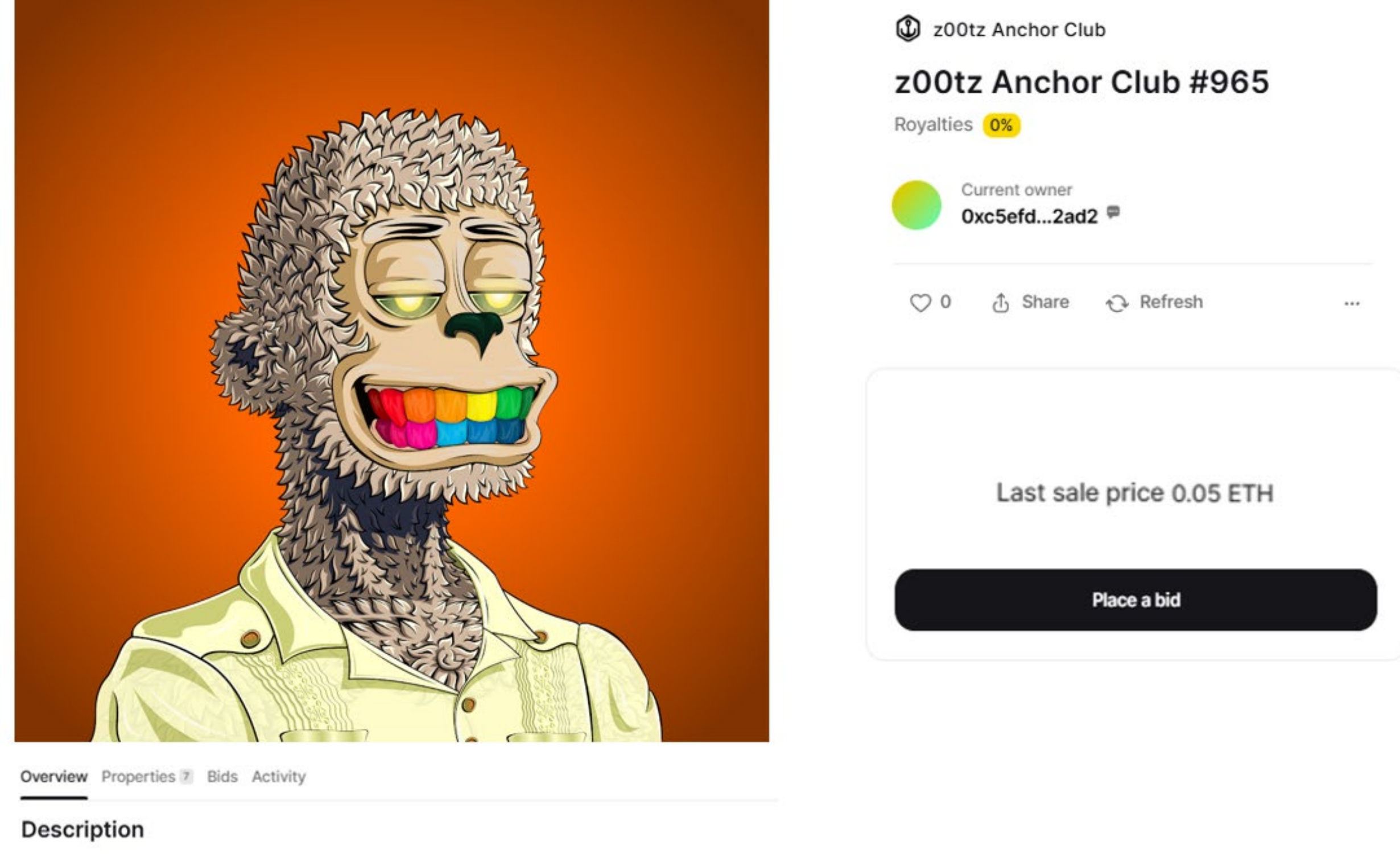

### *Step 2A. Treatment – Exploration of the Seller's Portfolio (Examining a Charity NFT)*

**[Information shown to participants in the HighRelistingPrice group]**

Out of curiosity, you explore the other NFTs in the seller's portfolio (Oxc5efd..2ad2) and discover another NFT in their collection, a SHAQ GIVES BACK charity NFT that the seller has recently purchased as part of the charity offering, and then listed for resale. Please examine the image and description of this charity NFT carefully. Note that when the seller purchased this charity NFT from the initial charity fundraiser, 100% of those proceeds went to the Shaquille O'Neal Foundation, which actively supports underprivileged youth nationwide by providing essential resources such as toys, clothing, and meals. The Foundation's mission emphasizes creating pathways for underserved youth, enabling them to achieve their full potential.

However, it is important to clarify that any proceeds from a successful resale will go to the current owner (the seller whose portfolio you are inspecting). That is, the initial minting price that this individual paid (Oxc5efd..2ad2) was **0.05 ETH (approximately $134)** and that amount was contributed to the Shaquille

O'Neil foundation. However, **any sales money from resale of the NFT will go directly and only to this individual**.

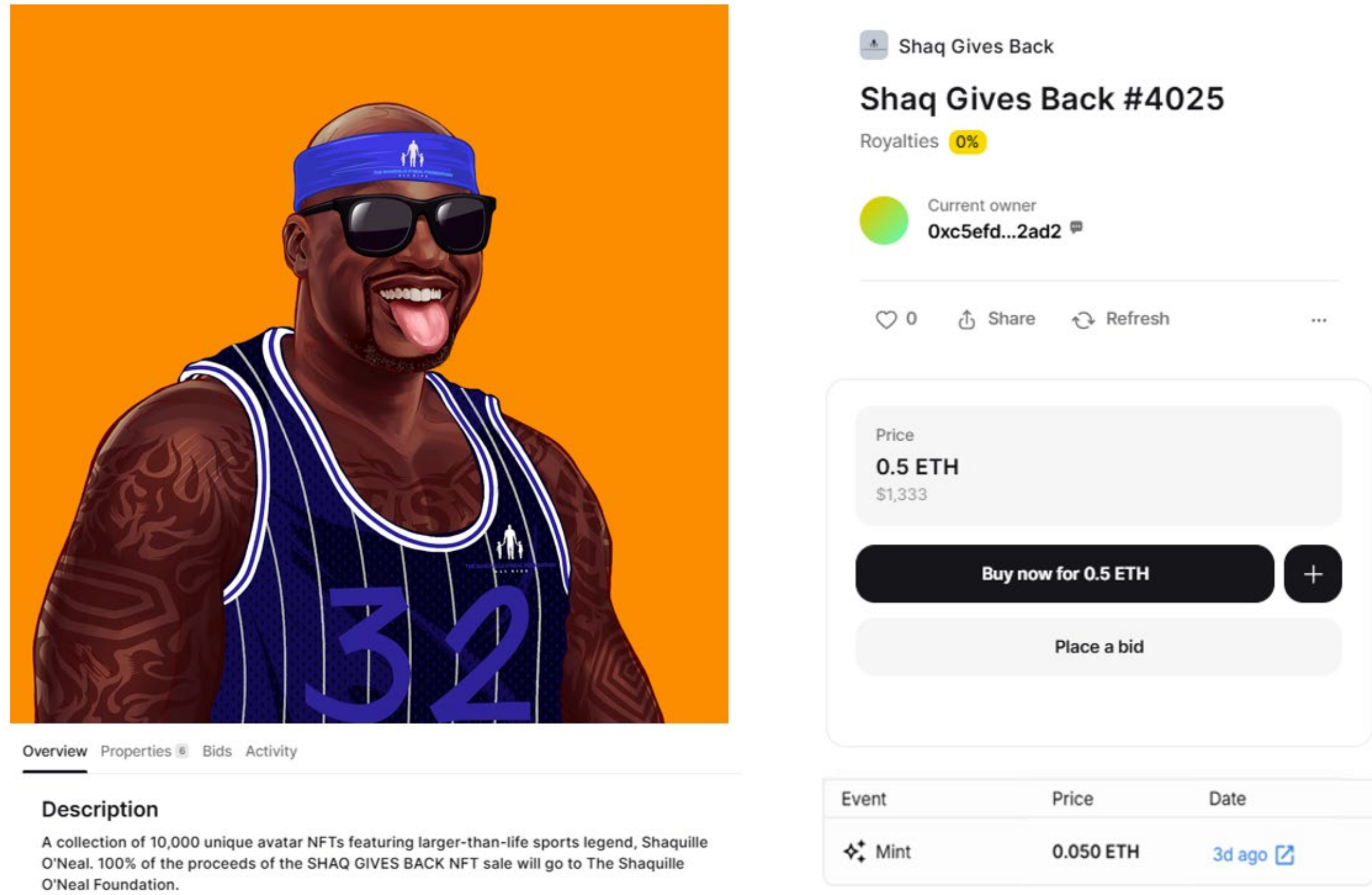

**[Information shown to participants in the LowRelistingPrice group]**

Out of curiosity, you explore the other NFTs in the seller's portfolio (Oxc5efd..2ad2) and discover another NFT in their collection, a SHAQ GIVES BACK charity NFT that the seller has recently purchased as part of the charity offering, and then listed for resale. Please examine the image and description of this charity NFT carefully. Note that when the seller purchased this charity NFT from the initial charity fundraiser, 100% of those proceeds went to the Shaquille O'Neal Foundation, which actively supports underprivileged youth nationwide by providing essential resources such as toys, clothing, and meals. The Foundation's mission emphasizes creating pathways for underserved youth, enabling them to achieve their full potential.

However, it is important to clarify that any proceeds from a successful resale will go to the current owner (the seller whose portfolio you are inspecting). That is, the initial minting price that this individual paid (Oxc5efd..2ad2) was **0.05 ETH (approximately $134)** and that amount was contributed to the Shaquille O'Neil foundation. However, **any sales money from the resale of the NFT will go directly and only to this individual**.

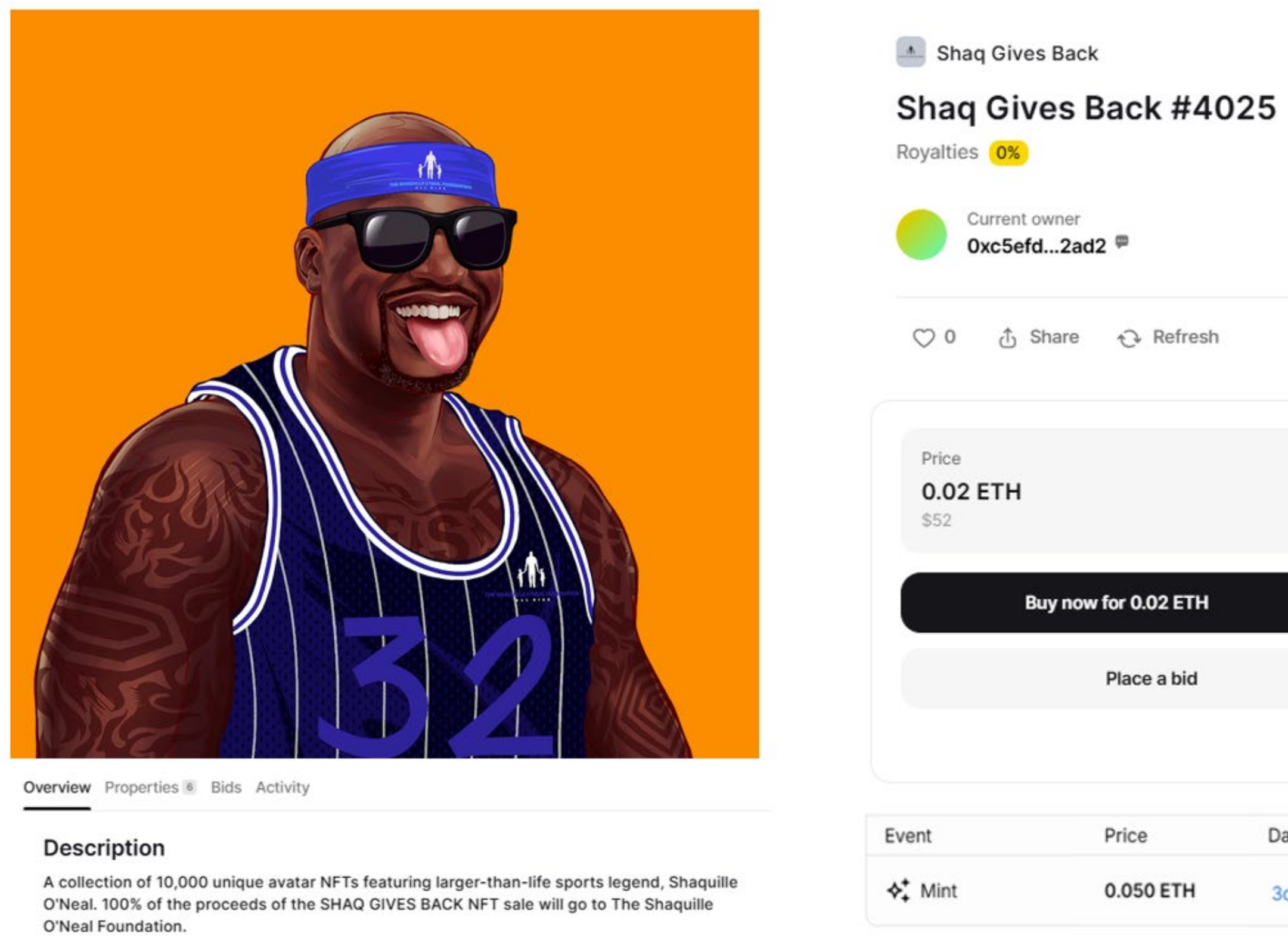

**[Information shown to participants in the NoRelisting group]**

Out of curiosity, you explore the other NFTs in the seller's portfolio (Oxc5efd..2ad2) and discover another NFT in their collection, a SHAQ GIVES BACK charity NFT that the seller has recently purchased as part of the charity offering. Please examine the image and description of this charity NFT carefully. Note that when the seller purchased this charity NFT from the initial charity fundraiser, 100% of those proceeds went to the Shaquille O'Neal Foundation, which actively supports underprivileged youth nationwide by providing essential resources such as toys, clothing, and meals. The Foundation's mission emphasizes creating pathways for underserved youth, enabling them to achieve their full potential.

However, it is important to clarify that any proceeds from a successful resale will go to the current owner (the seller whose portfolio you are inspecting). That is, the initial minting price that this individual paid (Oxc5efd..2ad2) was **0.05 ETH (approximately $134)** and that amount was contributed to the Shaquille O'Neil foundation. However, **any sales money from resale of the NFT will go directly and only to this individual**.

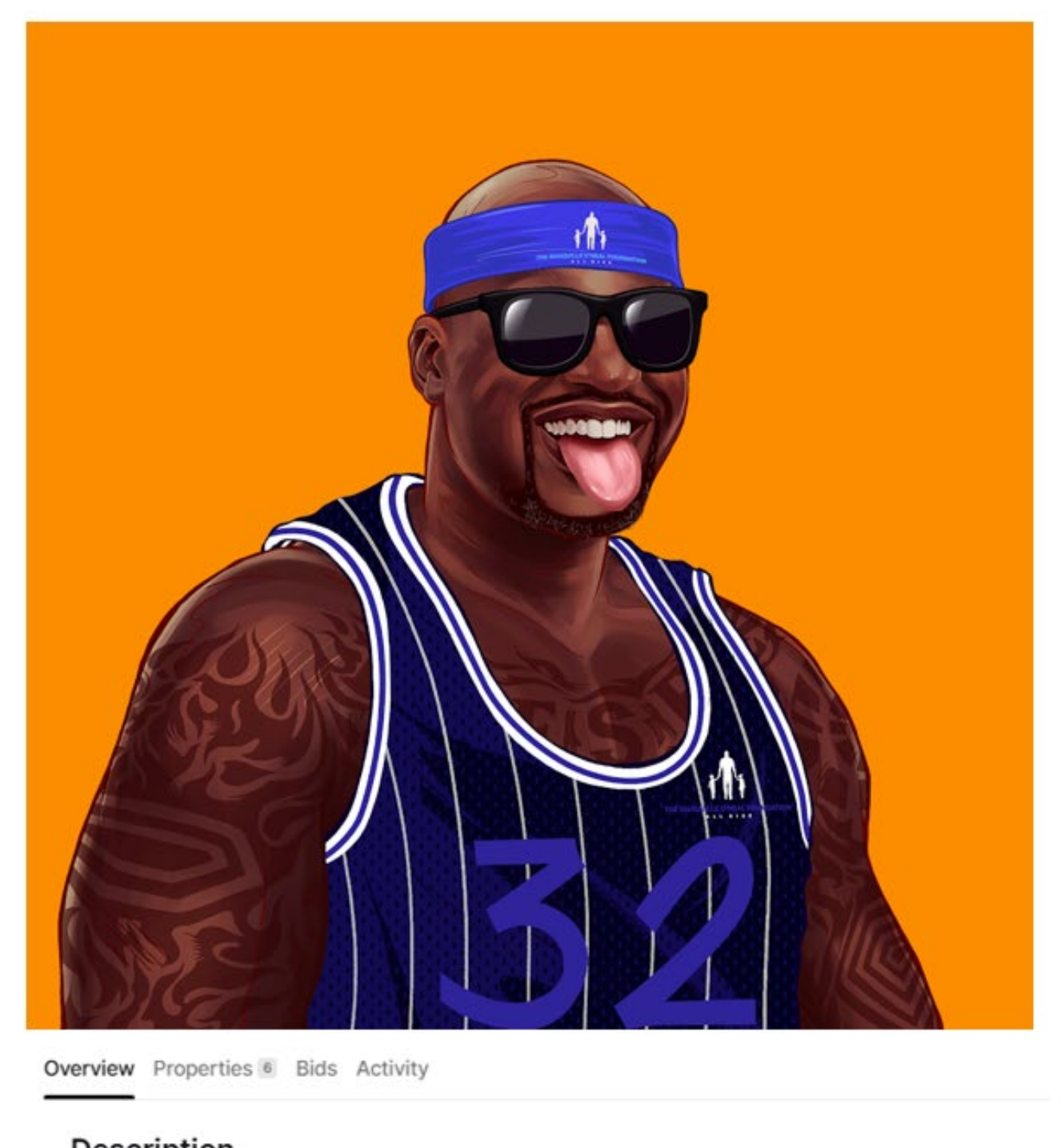

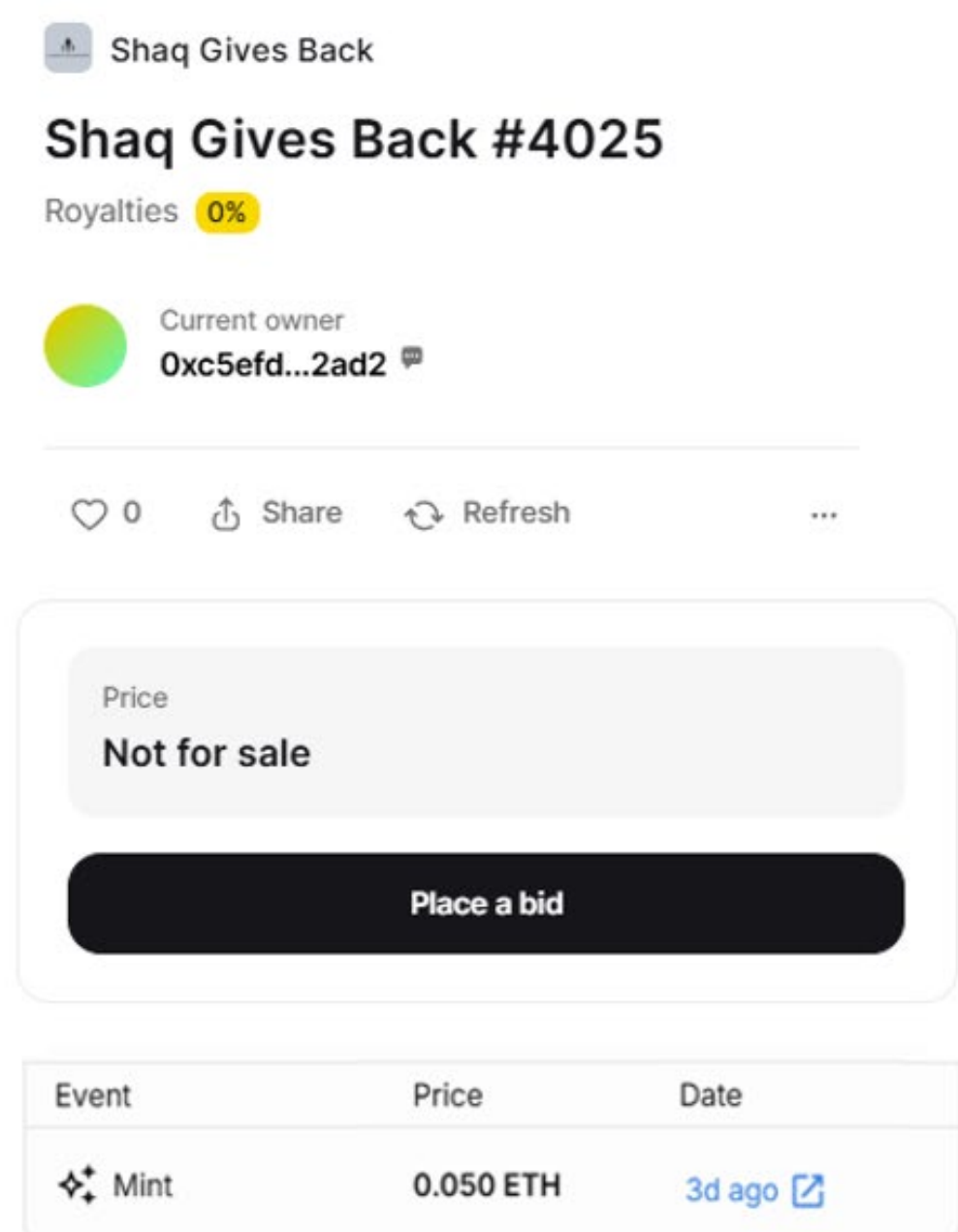

### *Step 2B. Verifying Treatment Comprehension*

- Is this a charity NFT?

Yes No

- Which of the following statements is true?

The owner relisted the charity NFT at a price lower than its original price.
The owner relisted the charity NFT at a price higher than its original price.
The owner has not relisted the charity NFT for sale.

**[Clarification information shown to participants in the HighRelistingPrice group]**
Please take some time to carefully review the explanation for the correct statement.

*The current listing price of this SHAQ GIVES BACK* ***charity NFT is 0.5 ETH ($1,333)****, indicating that the owner has relisted it at a price* ***higher*** *than the original.* ***Any sales money from this resale*** *will go directly and exclusively to the owner (****Oxc5efd..2ad2****).*

**[Clarification information shown to participants in the LowRelistingPrice group]**
Please take some time to carefully review the explanation for the correct statement.

*The current listing price of this SHAQ GIVES BACK* ***charity NFT is 0.02 ETH ($52)****, indicating that the owner has relisted it at a price* ***lower*** *than the original.* ***Any sales money from this resale*** *will go directly and exclusively to the owner (****Oxc5efd..2ad2****).*

**[Clarification information shown to participants in the NoRelisting group]**
Please take some time to carefully review the explanation for the correct statement.

*This SHAQ GIVES BACK* ***charity NFT is not listed for resale****.*

## *Step 3. Seller Preference Evaluation for the Non-Charity NFT*

You now decide to make an offer (place a bid) for the NFT that initially attracted your interest. You may bid a price lower than, equal to, or higher than the stated listing price. Assume you have a budget of **0.1 ETH (approximately $268)**.

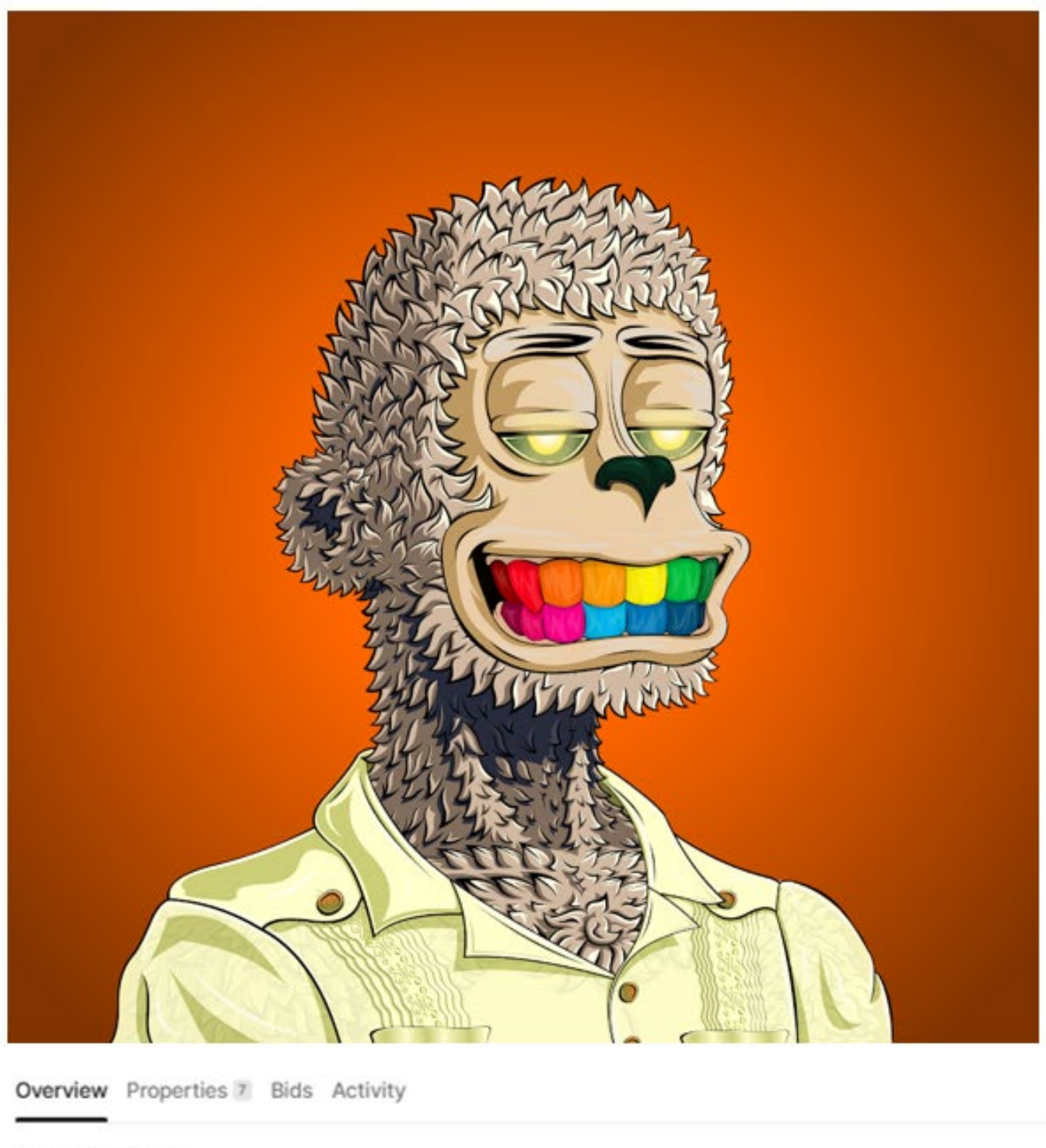

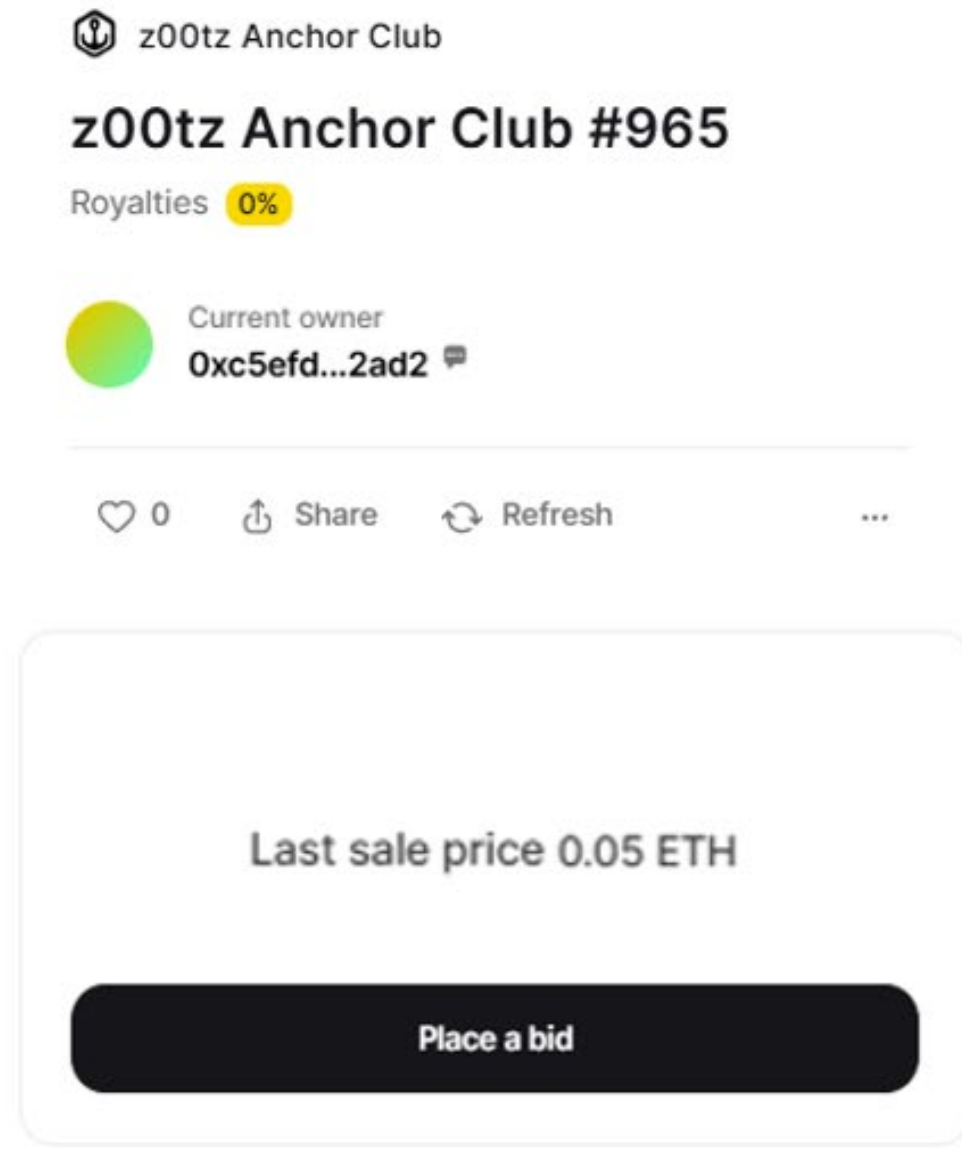

- How much would you choose to bid on this NFT?
  Less than 0.05 ETH
  Exactly 0.05 ETH
  More than 0.05 ETH

While browsing the collection page, you discover a copy of the same NFT that initially drew your interest (NFT name: **z00tz Anchor Club #965**). You notice that the copy is held by another owner with the same last sale price.

**[Question shown to participants in the HighRelistingPrice and LowRelistingPrice groups]**

- Would you prefer to purchase the NFT from the other owner, or the first owner you considered?
  Second owner (does not own a SHAQ NFT)
  First owner (is re-selling a SHAQ NFT)

**[Question shown to participants in the NoRelisting group]**

- Would you prefer to purchase the NFT from the other owner, or the first owner you considered?
  Second owner (does not own a SHAQ NFT)
  First owner (owns a SHAQ NFT)

## *Step 4. Perception Assessment (Strategic Generosity, Immorality)*

**[Questions about participants' perception of strategic generosity]**

Please answer the following questions regarding your perceptions of the actions taken by the first seller (Oxc5efd..2ad2), whom you discovered had re-listed a charity NFT for resale (the SHAQ NFT):

- To what extent do you believe the owner acted out of self-interest when they purchased the SHAQ charity NFT? (Note: Higher numbers indicate a higher level of self-interest.)
  (1 = not self-interested at all and 7 = highly self-interested)

**[Questions about participants' perceptions of the owner's relisting behavior]**

- To what extent do you feel it is unethical for the owner to re-list the charity NFT for sale? (Note: Higher numbers indicate a higher level of unethicality.)
  (1 = completely ethical, 7 = completely unethical)

- To what extent do you feel it is immoral for the owner to re-list the charity NFT for sale? (Note: Higher numbers indicate a higher level of immorality.)
  (1 = completely moral, 7 = completely immoral)

- To what extent do you approve or disapprove of the owner relisting the charity NFT for sale? (Note: Higher numbers indicate a higher level of approval.)
  (1 = Strongly disapprove, 7 = Strongly approve)

## *Step 5. Background Information Collection*

- What is your gender?
  Male
  Female
  Other
  Prefer not to say

- What is your race?
  White
  Black or African American
  Hispanic or Latino
  Asian
  American Indian or Alaska Native
  Native Hawaiian or Other Pacific Islander

- What is your highest level of education?
  Some high school or lower
  High school
  Some college
  Bachelor degree
  Graduate school or higher

- Have you ever owned any NFTs?
  Yes No

- On a scale of 1-7, how familiar are you with NFT marketplace?
(1 = Not familiar at all, 7 = Very familiar)

- If you saw someone attempting to resell a charity NFT for profit, how would you respond?

  Avoid buying from this seller
  Feel disappointed but take no action
  Feel neutral; reselling is the owner's choice
  Support; reselling is common in NFT markets
  Other (please specify): _______

[Note: All texts with shaded backgrounds are not visible to participants.]

## Appendix U. Randomization Checks and Additional Results from the Scenario-based Experiment

We conducted randomization checks, as presented in Table U1, which indicate that the four groups are well-balanced with respect to demographic characteristics.

**Table U1. Randomization Checks for the Experiment**

(Panel A)

| Gender | Female | |
|---|---|---|
| | Mean diff | *p*-value |
| LowRelistingPrice vs. NoRelisting | 0.076 | 0.435 |
| HighRelistingPrice vs. NoRelisting | 0.036 | 0.836 |
| HighRelistingPrice vs. LowRelistingPrice | -0.040 | 0.792 |

(Panel B)

| Race | Asian | | Black or African American | | Native Hawaiian or Pacific Islander | | White | |
|---|---|---|---|---|---|---|---|---|
| | Mean diff | *p*-value | Mean diff | *p*-value | Mean diff | *p*-value | Mean diff | *p*-value |
| LowRelistingPrice vs. NoRelisting | 0.025 | 0.823 | -0.038 | 0.833 | -0.012 | 0.408 | 0.019 | 0.962 |
| HighRelistingPrice vs. NoRelisting | 0.022 | 0.862 | -0.050 | 0.737 | -0.012 | 0.424 | 0.098 | 0.389 |
| HighRelistingPrice vs. LowRelistingPrice | -0.003 | 0.998 | -0.012 | 0.981 | 0.000 | 1.000 | 0.079 | 0.526 |

| Race | Other | |
|---|---|---|
| | Mean diff | *p*-value |
| LowRelistingPrice vs. NoRelisting | 0.005 | 0.984 |
| HighRelistingPrice vs. NoRelisting | -0.059 | 0.130 |
| HighRelistingPrice vs. LowRelistingPrice | -0.064 | 0.081 |

(Panel C)

| Education | High School | | Some College | | Bachelor's Degree | | Graduate School or Higher | |
|---|---|---|---|---|---|---|---|---|
| | Mean diff | *p*-value | Mean diff | *p*-value | Mean diff | *p*-value | Mean diff | *p*-value |
| LowRelistingPrice vs. NoRelisting | 0.046 | 0.553 | 0.070 | 0.474 | 0.029 | 0.918 | -0.146 | 0.057 |
| HighRelistingPrice vs. NoRelisting | 0.034 | 0.736 | 0.080 | 0.402 | -0.064 | 0.683 | -0.050 | 0.720 |
| HighRelistingPrice vs. LowRelistingPrice | -0.012 | 0.958 | 0.009 | 0.987 | -0.093 | 0.426 | 0.096 | 0.282 |

Note: Following the procedures in prior studies (Huang et al. 2019), we evaluate the efficacy of our randomization procedure by conducting pairwise Tukey's Honestly Significant Difference (HSD) tests to detect if there is any significant demographic difference between groups. Values represent the differences between groups, with corresponding *p*-values provided alongside.

## Appendix V. Additional Results from the Scenario-based Experiment

Regarding perceived immorality, as shown in Figure V1, we find that participants in the *HighRelistingPrice* group reported significantly higher levels of perceived immorality ($p$-value < 0.05) and a greater likelihood of disapproving of the relisting behavior ($p$-value < 0.05) compared to the other two groups. Additionally, as shown in Tables V1-V2, participants in the *HighRelistingPrice* group were more inclined to avoid purchasing NFTs from the owner ($p$-value < 0.10) compared to those in the *LowRelistingPrice* group. Note that participants in the *NoRelisting* group answered these questions based on a hypothetical scenario of relisting a charity NFT, as no information about the actual relisting price was provided. As a result, their responses likely reflect their expected relisting price, and the mean response for the *NoRelisting* group tends to fall between the averages observed in the *LowRelistingPrice* and *HighRelistingPrice* scenarios.

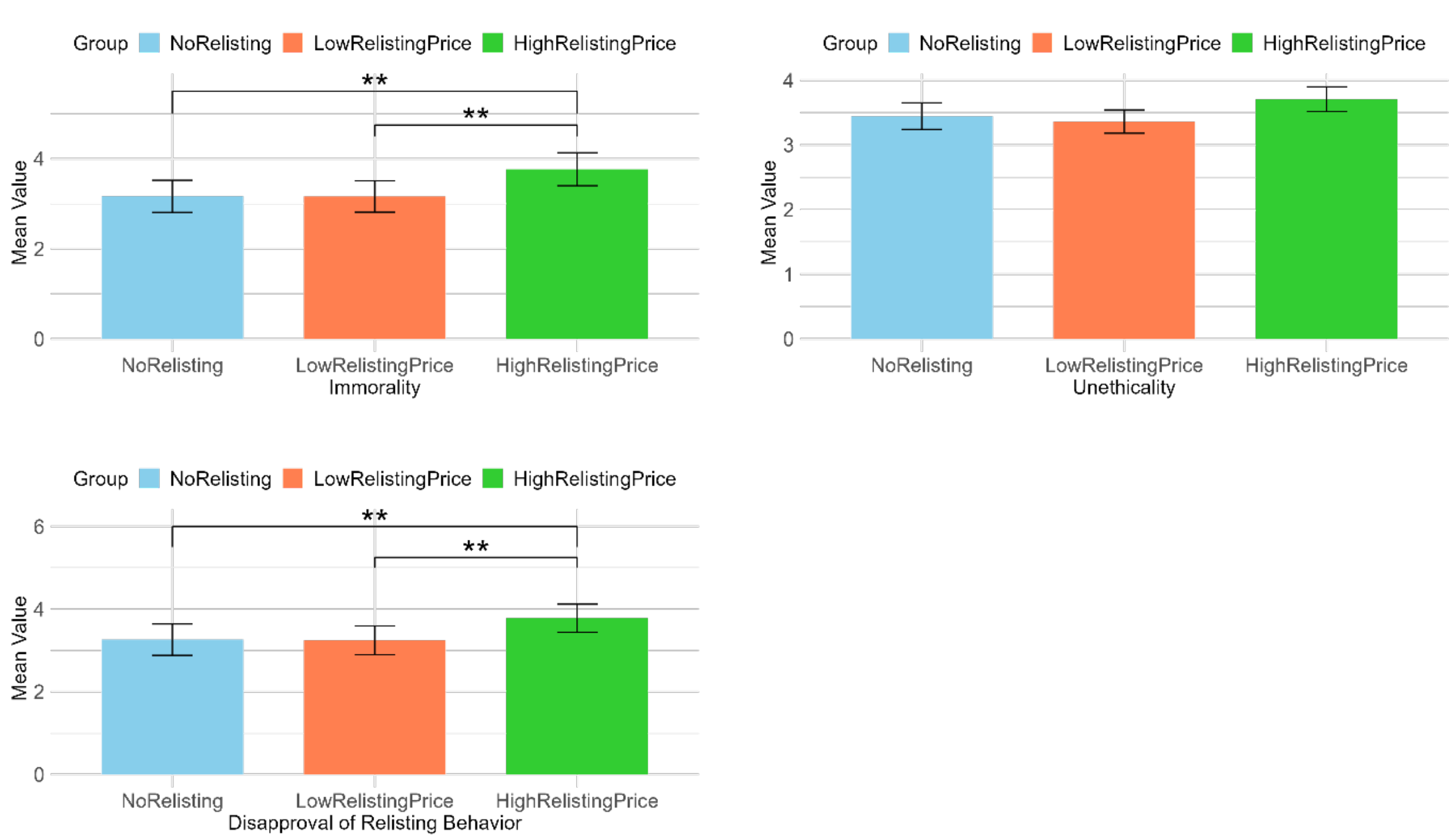

**Figure V1. Treatment Effects on Immorality Perception Measures**

*Note: Error bars denote 95% confidence intervals.*

**Table V1. Pairwise Comparison in Response to the Seller's Resale of a Charity NFT for Profit**

| Choosing "Avoid buying from this seller" | Mean diff | *p*-value |
|---|---|---|
| LowRelistingPrice vs. NoRelisting | -0.049 | 1.000 |
| HighRelistingPrice vs. NoRelisting | 0.079 | 0.543 |
| HighRelistingPrice vs. LowRelistingPrice | 0.128 | 0.082 |
| **Choosing "Feel disappointed but take no action"** | **Mean diff** | ***p*-value** |
| LowRelistingPrice vs. NoRelisting | -0.030 | 1.000 |
| HighRelistingPrice vs. NoRelisting | 0.022 | 1.000 |
| HighRelistingPrice vs. LowRelistingPrice | 0.052 | 0.669 |

| Choosing “Feel neutral; reselling is the owner’s choice” | Mean diff | *p*-value |
|---|---|---|
| LowRelistingPrice vs. NoRelisting | 0.079 | 0.872 |
| HighRelistingPrice vs. NoRelisting | -0.041 | 1.000 |
| HighRelistingPrice vs. LowRelistingPrice | -0.120 | 0.327 |
| **Choosing “Support; reselling is common in NFT markets”** | **Mean diff** | ***p*-value** |
| LowRelistingPrice vs. NoRelisting | -0.021 | 1.000 |
| HighRelistingPrice vs. NoRelisting | -0.061 | 0.960 |
| HighRelistingPrice vs. LowRelistingPrice | -0.039 | 1.000 |
| **Choosing “Other (please specify)”** | **Mean diff** | ***p*-value** |
| LowRelistingPrice vs. NoRelisting | 0.021 | 0.304 |
| HighRelistingPrice vs. NoRelisting | 0.000 | 1.000 |
| HighRelistingPrice vs. LowRelistingPrice | -0.021 | 0.301 |

**Table V2. Choice Preferences in Response to the Seller’s Resale of a Charity NFT for Profit**

| Group | Selected Choice | Observations | Ratio |
|---|---|---|---|
| HighRelistingPrice | Avoid buying from this seller | 22 | 25.58% |
| | Feel disappointed but take no action | 10 | 11.63% |
| | Feel neutral; reselling is the owner’s choice | 40 | 46.51% |
| | Support; reselling is common in NFT markets | 14 | 16.28% |
| | Other (please specify) | 0 | 0.00% |
| LowRelistingPrice | Avoid buying from this seller | 12 | 12.77% |
| | Feel disappointed but take no action | 6 | 6.38% |
| | Feel neutral; reselling is the owner’s choice | 55 | 58.51% |
| | Support; reselling is common in NFT markets | 19 | 20.21% |
| | Other (please specify) | 2 | 2.13% |
| NoRelisting | Avoid buying from this seller | 15 | 17.65% |
| | Feel disappointed but take no action | 8 | 9.41% |
| | Feel neutral; reselling is the owner’s choice | 43 | 50.59% |
| | Support; reselling is common in NFT markets | 19 | 22.35% |
| | Other (please specify) | 0 | 0.00% |

## References

Abadie, A., Diamond, A., Hainmueller, J. (2015). Comparative politics and the synthetic control method. *American Journal of Political Science*, 59(2), 495-510.

Brynjolfsson, E., Li, D., & Raymond, L. (2025). Generative AI at work. *The Quarterly Journal of Economics*, *140*(2), 889-942.

Burtch, G., Carnahan, S., & Greenwood, B. N. (2018). Can you gig it? An empirical examination of the gig economy and entrepreneurial activity. *Management Science, 64*(12), 5497-5520.

Cox, D. R. (1972). Regression models and life-tables. *Journal of the Royal Statistical Society: Series B (Methodological)*, 34(2), 187-202.

Heckman JJ (1976). The common structure of statistical models of truncation, sample selection and limited dependent variables and a simple estimator for such models. J. Econ. Soc. Meas. 5(4): 475-492.

Huang, J., Xia, P., Li, J., Ma, K., Tyson, G., Luo, X., ... & Wang, H. (2024, May). Unveiling the paradox of NFT prosperity. In *Proceedings of the ACM Web Conference 2024* (pp. 167-177).

Huang N, Burtch G, Gu B, Hong Y, Liang C, Wang K, Fu D, Yang B (2019) Motivating user-generated content with performance feedback: Evidence from randomized field experiments. *Management Science,* 65(1): 327-345.

Mundlak, Y. (1978). On the pooling of time series and cross section data. *Econometrica: journal of the Econometric Society*, 69-85.

Neyman, J., & Scott, E. L. (1948). Consistent estimates based on partially consistent observations. *Econometrica: journal of the Econometric Society*, 1-32.

Rabe-Hesketh, S., & Skrondal, A. (2013). Avoiding biased versions of Wooldridge's simple solution to the initial conditions problem. *Economics Letters*, 120(2), 346-349.

Rosenbaum, P. R., & Rubin, D. B. (1983). Assessing sensitivity to an unobserved binary covariate in an observational study with binary outcome. *Journal of the Royal Statistical Society: Series B (Methodological)*, *45*(2), 212-218.

Shaver JM (1998). Accounting for endogeneity when assessing strategy performance: Does entry mode choice affect FDI survival?. Manag. Sci. 44(4): 571-585.

Wooldridge, J. M. (2005). Simple solutions to the initial conditions problem in dynamic, nonlinear panel data models with unobserved heterogeneity. *Journal of applied econometrics*, 20(1), 39-54.